\documentclass[aps,floats,superscriptaddress,onecolumn,showpacs]{revtex4-1}

\usepackage{graphicx}
\usepackage{graphics,dcolumn,bm,epic, eepic,fleqn,float}
\usepackage{amssymb,amsmath,amsfonts,multirow,rotate,color,array}
\usepackage{soul}
\usepackage[normalem]{ulem}
\usepackage{wasysym}
\usepackage{dcolumn}
\usepackage{bm}
\usepackage{subfigure}
\usepackage{comment}
\usepackage{mathtools}
\usepackage{soul}

\usepackage{xcolor}
\usepackage{hyperref}
\hypersetup{	
    colorlinks=true,
    linkcolor={blue!45!black},
    citecolor={blue!45!black},
    urlcolor={blue!45!black}
}

\usepackage{setspace}

\definecolor{Myorange}{cmyk}{0,0.42,1,0}
\definecolor{brown}{rgb}{0.59, 0.29, 0.0}

\newcommand{\vito}[1]{{#1}}
\newcommand{\mattia}[1]{{#1}}

\newcommand{\lay}[1]{^{[#1]}}
\newcommand{\laysup}[1]{^{#1}}
\newcommand{\low}[1]{_{[#1]}}
\newcommand{\newtext}[1]{{#1}}

\begin{document}

\title{Dynamical processes and emergent behaviors in multiplex networks}


\author{Federico Battiston}
\email{battistonf@ceu.edu}
\affiliation{Department of Network and Data Science, Central European University, Vienna 1100, Austria}
\affiliation{Department of AI, Data and Decision Sciences, Luiss University of Rome, Viale Romania 32, 00197, Rome, Italy}

\author{Mattia Frasca}
\affiliation{Department of Electrical, Electronics and Computer Science Engineering, University of Catania, 95125 Catania, Italy}

\author{Jesus G{\'o}mez-Garde{\~n}es}
\affiliation{GOTHAM laboratory, Institute for Biocomputation and Physics of Complex Systems (BIFI), University of Zaragoza, 50018 Zaragoza, Spain}
\affiliation{Department of Condensed Matter Physics, University of Zaragoza, 50009 Zaragoza, Spain}

\author{Byungjoon Min}
\affiliation{Department of Physics, Chungbuk National University, Cheongju,  Chungbuk 28644, Korea}


\author{Filippo Radicchi}
\affiliation{Center for Complex Networks and Systems Research,
Luddy School of Informatics, Computing, and Engineering,
Indiana University, Bloomington, Indiana 47408, USA}

\author{Andrea Santoro}
\affiliation{ISI Foundation, Turin, Italy}
\affiliation{Neuro-X Institute, \'{E}cole Polytechnique F\'{e}d\'{e}rale de Lausanne (EPFL), Geneva, Switzerland}

\author{Vito Latora}
\email{v.latora@qmul.ac.uk}
\affiliation{School of Mathematical Sciences, Queen Mary University of London, London E1 4NS, United Kingdom}
\affiliation{Dipartimento di Fisica ed Astronomia, Universit\`a di Catania and INFN Sezione Catania, I-95123 Catania, Italy}
\affiliation{Complexity Science Hub Vienna (CSHV), Vienna, Austria}


\begin{abstract}
Over the last two decades, network science has greatly 
\vito{advanced our understanding of} how the collective behaviors of
a complex system emerge from the interactions among its basic units.
\vito{Multiplex networks, i.e.~networks with many layers, whose nodes 
  are in one-to-one correspondence, provide a more realistic description
  for social, biological and ecological systems where multiple types of
interactions coexist.}
%
%
After a brief introduction on how \vito{to model} the architecture of
multiplex networks, we present a complete overview of the different
dynamics which can unfold over these structures.
We
\vito{present} 
a unified framework to describe dynamical processes such
as percolation, reaction-diffusion, synchronization, epidemic spreading,
social dynamics and games on multiplex networks, as
well
\vito{as the coupled evolution of different dynamical processes},  
and the 
coevolution of a process with the network structure.
Our focus is
on truly-multiplex collective behaviors, i.e., all those phenomena
which cannot emerge on the corresponding aggregated networks,
or when the different layers of these systems are considered in
isolation.
\vito{We identify three main mechanisms leading to new collective behaviors:
the existence of structural correlations across layers, the presence
of dynamical correlations in the processes taking place at the
different layers, and the dynamical interplay of inter- and
intra-layer interactions.}
We conclude with a summary of the main takeaways from a
decade of work in the field. 
%

\end{abstract}

{
    \let\clearpage\relax
    \maketitle
}

\begin{spacing}{0.9} 
\tableofcontents
\end{spacing}



\section{Introduction}
\label{sec:intro}

Over the last two decades, network science has become an
invaluable tool to study complex systems~\cite{albert2002statistical,dorogovtsev2002evolution,newman2003structure,boccaletti2006complex}, greatly enhancing our
ability to understand and predict their collective
behaviors~\cite{acebron2005kuramoto, dorogovtsev2008critical, castellano2009statistical, pastor2015epidemic}.  Yet, traditional network
approaches often fail to describe the dynamics of many natural or
man-made complex systems when the units of such systems have interactions of
radically different types.  Examples of systems with many levels (or layers)
of interactions are
ubiquitous.  For instance, the individuals of a social network are often 
connected through different types of social relations, such as
kinships, friendships, work collaborations, co-locations, online or
offline communication~\cite{szell2010multirelational, battiston2014structural}.  Large metropolitan areas are increasingly
characterised by complex multimodal transportation systems,
involving means of transportation with different temporal and spatial
scales, which range from buses to underground rail, trains, riverboat
networks and airplanes~\cite{alessandretti2022multimodal}. Similarly, due to recent progresses in brain
imaging, different modalities of data acquisition from DTI to EEG and
fMRI, allow us to map even the human connectome at
different structural and functional levels~\cite{de2017multilayer,lim2019discordant,presigny2022multiscale}. 

Multiplex networks, namely networks with many layers, with the
links at each layer standing for a different type of interaction
among the same set of nodes, allow to better describe the architecture
of all such and many other real-world systems. They are very handy to 
extract and quantify the richness associated to interconnected systems,
and they also provide 
mathematically grounded tools to reduce the structural complexity of 
the data. Indeed, the multiplex paradigm has become common in a
wide variety of domains, from network biology and network medicine to
the social sciences, and has been the topic of thousands of
theoretical and applied papers in the last few years~\cite{hervias-parejo2024structure,hackett2024multihabitat,timoteo2018multilayer,dedomenico2015muxviz}
Despite the explosion of interest in the topic, the multiplex analysis
of a network is more complicated and requires longer time to run and more
computer memory than traditional network analysis. 
It is therefore crucial and timely to understand when and how a
multiplex description truly matters to
characterize the dynamics of a complex system and to
unveil its fundamental functions. 
For such a reason, in this review we present 
%
a complete and systematic discussion of {\em dynamical processes on
multiplex networks}. We will cover processes ranging from percolation 
to diffusion and epidemic spreading, synchronization and linear
control of coupled dynamical systems, social dynamics and evolutionary
game theory. Our attention will always be on the
emergent collective behaviors induced by multiplexity,
namely on all the dynamical behaviours 
that can not be observed nor predicted by considering the 
layers of a multiplex network in isolation, or by studying the graph
obtained by aggregating them.

The review is organized in four 
parts. 
%
In the first part \vito{(Sections \ref{sec:multiplex} and
  \ref{sec:structure_bis})} \vito{we introduce a general mathematical
framework to describe the structure of multiplex networks and the
dynamical processes that will be treated in the following sections.
We then summarise the main measures
and models that have been proposed to characterize and reproduce
the structural properties of real-world multiplex systems.}
The second part (Sections \ref{sec:percolation} to \ref{sec:games})
is the core of the review and provides a complete picture of the effect
of multiplexity on a large variety of dynamical processes. Here we 
discuss how multiple layers of interactions can lead to larger and more
abrupt cascading failures than those observed in single-layer networks.
We then focus on the emergence of superdiffusion in multiplex networks,
on novel mechanisms of pattern formation in reaction-diffusion processes,
and on abrupt transitions and explosive synchronization in ensembles of
oscillators coupled via multilayered interactions. Finally we discuss
strategic behaviors in multiplex networks where individuals may be involved
in different games at the different layers, highlighting how a layered
structure can boost prosocial behavior, inducing spontaneous symmetry
breaking and novel spatial patterns in the formation of cooperative
clusters.
The third part (Section \ref{sec:intertwined}) considers the case of
intertwined dynamical processes. Here we discuss collective phenomena
emerging when two or more dynamics of distinct nature operate at the
different layers, and interact through mutual couplings and feedbacks.
The fourth part (Section \ref{sec:coevolution}) deals with coevolution
of networks and processes, namely with all those cases in which 
the structure of a multiplex network changes in time together with
the dynamical states of the network nodes. Here, we focus on the novel
and rich phenomenology appearing when the structure of a network 
and a dynamical process over the network evolve together under mutual
feedbacks.
The review concludes with a summary of the main takeaways from a
decade of work in the field, and an outlook of open questions for the
future.

\begin{figure*}[t!]
	\begin{center}
		\includegraphics[width=0.98\textwidth]{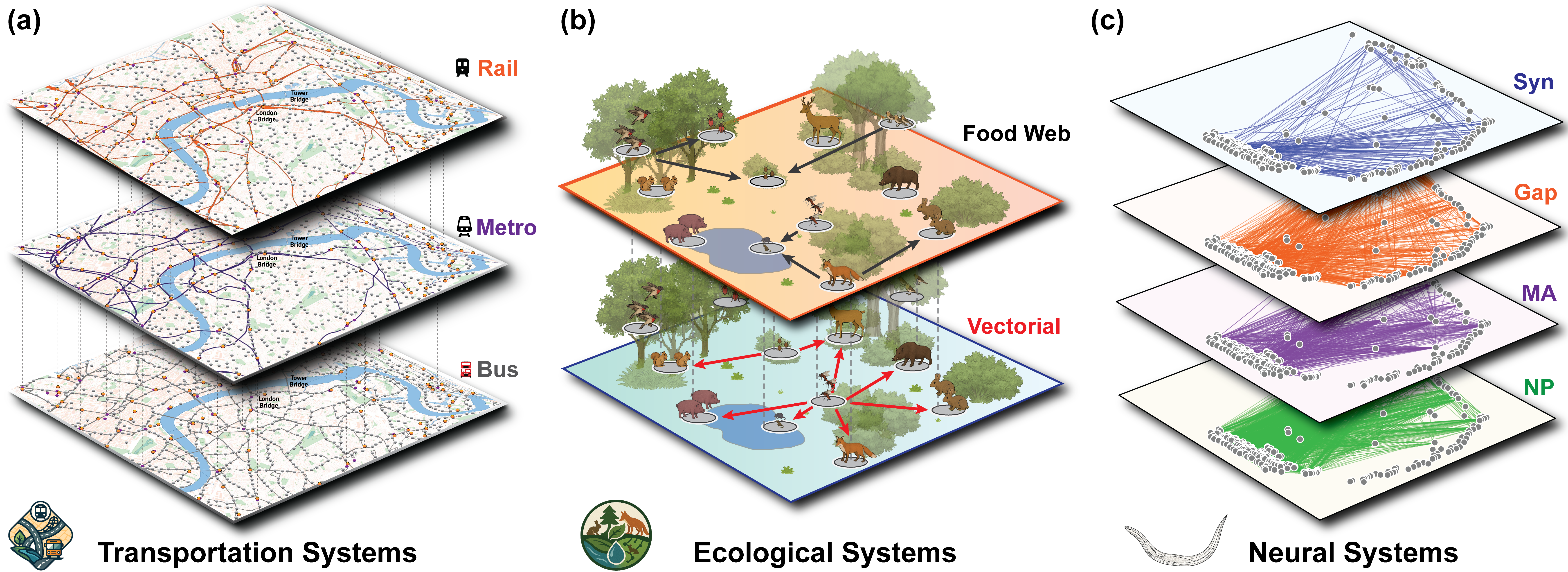}
		\caption{\vito{Real-world examples of multiplex networks. (a) The London 
				transportation system consists of multiple layers, corresponding to 
				different travel modes. Inter-layer connections describe transfer points where 
				passengers can switch from one mode to another, e.g. from a
				train to a bus.
				(b) An ecosystem with two interaction layers, respectively describing
				predator-prey relations (food web) and host-parasite exchanges
				(vectorial). Here, inter-layer connections capture how parasitism
				affects trophic dynamics.
				(c) The C. elegans nervous system integrates different types of communication
				between neurons, namely chemical synapses (Syn), directed electrical gap junctions
				(Gap), monoamine (MA) signalling and neuropeptides (NP).  } }
		\label{fig:figure_real}
	\end{center}
\end{figure*}

\section{Multiplex networks} 
\label{sec:multiplex}

\subsection{\vito{Structure}}

\label{subsec:structure}

\vito{Let us consider the multimodal transportation system 
in Fig.~\ref{fig:figure_real}(a), where travellers can move from a
location to another by a combination of different transportation
modes, e.g. trains, metro and buses
\cite{gallotti2015multilayer}. 
The structure of such a system can be modelled as a multiplex network
$\mathcal M$ with $N$ nodes and $M=3$ layers. 
The nodes represent city locations, while each layer describes the network 
associated with a transportation mode.  
A multiplex network is characterized by two types of links: {\em intra-layer}
links describing, in our example, connections between two nodes 
within a given
transportation mode, and {\em inter-layer} links representing the possibility to
switch from one transportation mode to another at some given node.  
More in general, a multiplex network $\mathcal M$ is useful to describe the
structure of complex systems whose units are coupled through $M$
different types or classes of interactions, each represented by the
edges of one of the $M$ layers of the network.
Other real-world examples include ecological and neural systems. 
  Fig.~\ref{fig:figure_real}(b) is an example of an ecosystem 
  with $M=2$ layers, where
different species interact at the trophic predator-prey layer (food web),
and through the exchange of parasites over the host-parasite layer (vectorial). 
Here, inter-layer links encode how parasitism modulates species activity in the
food web layer, capturing heterogeneous, species-specific effects
(e.g. by altering mortality or feeding rates)~\cite{stella2018ecological}.  
%
%
Fig.~\ref{fig:figure_real}(c) shows the neural network of the
Caenorhabditis elegans, where nodes are neurons and the $M=4$ layers represent
different types of wired transmissions among neurons, namely chemical
interactions through synapses (Syn) and directed electrical
interactions through gap junctions (Gap), but also non-wired transmissions
based on non-synaptic monoamine (MA) signalling and on neuropeptides
(NP) \cite{bentley2016caenorhabditis}.
Notice, that each layer of a multiplex network contains the
same number of nodes, $N$,
while the number of edges per layer, or between layers, can be arbitrary.
An important feature of a multiplex network is the one-to-one
correspondence of nodes across layers. Namely, each node $i$ in layer
$\alpha$ has its own corresponding nodes $i$, also called replica
nodes, in all the other layers, so that we have a total of $N \cdot M$
replica nodes.  Two replica nodes, $i$ at layer $\alpha$ and $i$ at
layer $\beta$, can then be connected by an inter-layer link, while  
intra-layer links describe interactions among nodes of the same layer. 
\begin{figure}
    \begin{center}
        \includegraphics[width=0.48\textwidth]{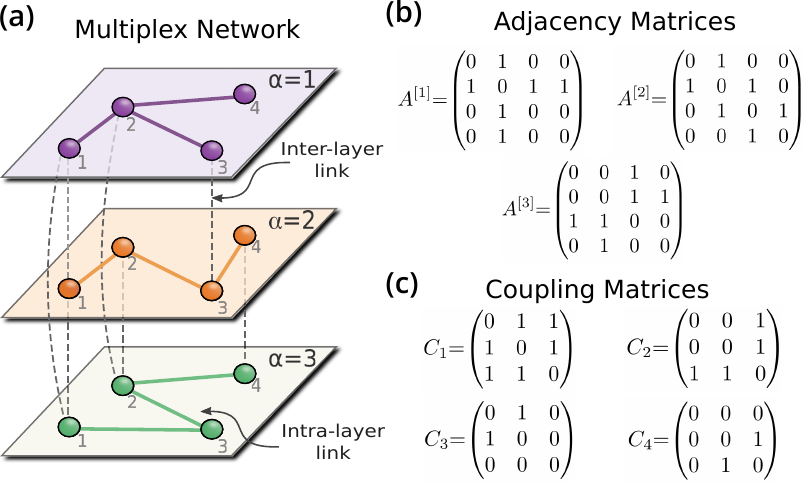}
        \caption{ \vito{ Formal representation of a multiplex network.
            (a) An example of 
           a multiplex network with $N=4$
           nodes and $M=3$ layers, modelling a system with three different
           types of relations among the nodes. Solid lines represent
            intra-layer links, i.e. connections between pairs of nodes 
            at the same layer, while dashed lines represent
            inter-layer links between 
            replica nodes across layers.
          (b) Adjacency matrices $A^{[\alpha]}~(\alpha=1,2,3)$
          encoding the intra-layer connections for each of the $M$ layers.
          (c) Coupling matrices $C_i~(i=1,\dots,4)$ describing the
          inter-layer connectivity patterns of each node $i$.
          }
        }
        \label{fig:figure_interlinks}
    \end{center}
\end{figure}
%
%
Formally, the intra-layer links of 
layer $\alpha$, with $\alpha=1,2, \ldots,M$,
can be described by an $N \times N$ adjacency matrix
$\mathrm{A}\lay{\alpha}=\{a_{ij}\lay{\alpha}\}$, where the entry $a_{ij}\lay{\alpha}$ for $i,j=1,2,\ldots,N$ is either 1 or 0, depending on the existence or
absence of a connection from node $i$ to node $j$ in layer $\alpha$.
Notice that, in the adopted notation, subscripts in Roman letters indicate
nodes, while superscripts in Greek letters indicate layers. More
generally, the entries of the adjacency matrix for any layer $\alpha$ can be 
real numbers describing the weight of the link from
node $i$ to node $j$, $\mathrm{A}\lay{\alpha} \in \mathbb{R}^{ N \times
  N} $.
For instance, in the example of the multimodal transportation
system, the weights may represent the costs or the times required to move from location $i$ to location $j$ using the trasportation mode $\alpha$.
Moreover, matrix $\mathrm{A}\lay{\alpha}$ can be asymmetric 
as in the ecological system and in the gap junction layer of the C. elegans
in Fig.~\ref{fig:figure_real}(b-c), where the links are directed.  
%
%
All the intra-layer links of the system
are encoded in the set $\bm A$ of $M$ adjacency matrices, i.e., $\bm
A=\{\mathrm{A}\lay{1},\mathrm{A}\lay{2}, \ldots, \mathrm{A}\lay{M} \}$.
%
The inter-layer links of node $i$ can be described by 
a $M \times M$ coupling matrix
$\mathrm{C}_i=\{c_{i}\lay{\alpha \beta}\}$, where the entry $c_{i}\lay{\alpha \beta}$
for $\alpha,\beta=1,2,\ldots,M$ is either 1 or 0, depending on whether
or not  replicas of node $i$ in layers $\alpha$ and $\beta$
interact. More generally, $\mathrm{C}_i$ can be a weighted directed matrix.  
E.g. the weight of an inter-layer link can represent the 
time to move from one transportation mode  
to another at a given location $i$. 
All the inter-layer links are stored in the set 
of $N$ coupling matrices, one for each node, i.e., $\bm
C=\{\mathrm{C}_1,\mathrm{C}_2,\ldots,\mathrm{C}_N \}$. 
%
In summary, a multiplex network $\mathcal M$ is described
by the two sets 
$\bm A$ and $\bm C$:
\begin{equation} 
  \mathcal M  \equiv (\bm A, \bm C) =   \{\mathrm{A}\lay{1},\mathrm{A}\lay{2}, \ldots, \mathrm{A}\lay{M}, \mathrm{C}_1,\mathrm{C}_2,\ldots,\mathrm{C}_N \}
  \label{multiplex2}
\end{equation} 
where $\bm A$ and $\bm C$ account, respectively, for the intra-layer
and inter-layer connectivity. This means that to define 
$\mathcal M$ we need to specify 
$M$ adjacency matrices of dimension $N \times N$ for the 
connections between nodes at each of the layers, as well as  
$N$ coupling matrices of dimension $M \times M$
describing the connections between layers at each of the nodes. 
Fig.~\ref{fig:figure_interlinks} reports an  example of a multiplex network
and its adjacency and coupling matrices.} 

\vito{Eq.~(\ref{multiplex2}) describes the most general multiplex network.
However, there are cases 
where the coupling among layers assumes very specific 
forms.
%
For example, in some multiplex networks the inter-layer connectivity
is all-to-all: the replicas of each node 
are all coupled. In this
maximally coupled regime, every node coupling matrix $\mathrm{C}_i$
has all off-diagonal entries nonzero.
In general, $\mathrm{C}_i$ can vary from node to node, as in
the ecological multiplex network of Fig.~\ref{fig:figure_real}(b), 
where all
inter-layer links are present,
but their weights differ across nodes, reflecting species-specific effects
of the parasitism. 
Furthermore, if the coupling between layers is a node-independent property, 
i.e., $\mathrm{C}_i=\mathrm{C} ~\forall i$, then the
inter-layer connectivity $\bm C$ is
described by a single matrix $\mathrm{C}=\{c\lay{\alpha \beta}\}$.
This is the setting considered in some of the
diffusion models of Section~\ref{sec:diffusion} and in 
in most of the synchronization examples in
Section \ref{sec:synchro}. 
}

\vito{In other cases, it is the structure of $\bm C$ that 
simplifies because of the specific ways in which the layers are
coupled.  This fact is reflected in
the number and positions of zeros in the matrices $\mathrm{C}_i$.
For instance, this is the case 
%
of temporal 
networks, i.e. networks whose links 
can fluctuate over time ~\cite{holme2012temporal}.
A temporal network can be described by an ordered sequence 
of adjacency matrices 
$\mathrm{A}\lay{1},\mathrm{A}\lay{2},\ldots,\mathrm{A}\lay{T}$, 
where $A\lay{t}$ is the adjacency matrix of the network at
time step $t=1,2,\ldots,T$ \cite{tang2010smallworld}. 
A temporal network can therefore also be viewed as a
multiplex network $\mathcal M$ as in Eq.~(\ref{multiplex2}),
with $M=T$ layers, and where each layer $\alpha$
describes the system connectivity at time step $t=\alpha$. However, 
in this case, $\bm C$ has a very special structure 
since, for each node $i$, we have $c_i\lay{\alpha \beta}= \delta_{\alpha+1, \beta}$ where $\delta_{\alpha \beta}$ is the Kronecker delta.
%
%
%
Because their inter-layer couplings encode causality, temporal networks have structural and dynamical features that require a
dedicated treatment, which can be found in recent reviews~\cite{holme2012temporal, holme2015modern} and books~\cite{masuda2016guide,holme2019temporal}.
\\
}
\begin{figure}
    \begin{center}
        \includegraphics[width=0.55\textwidth]{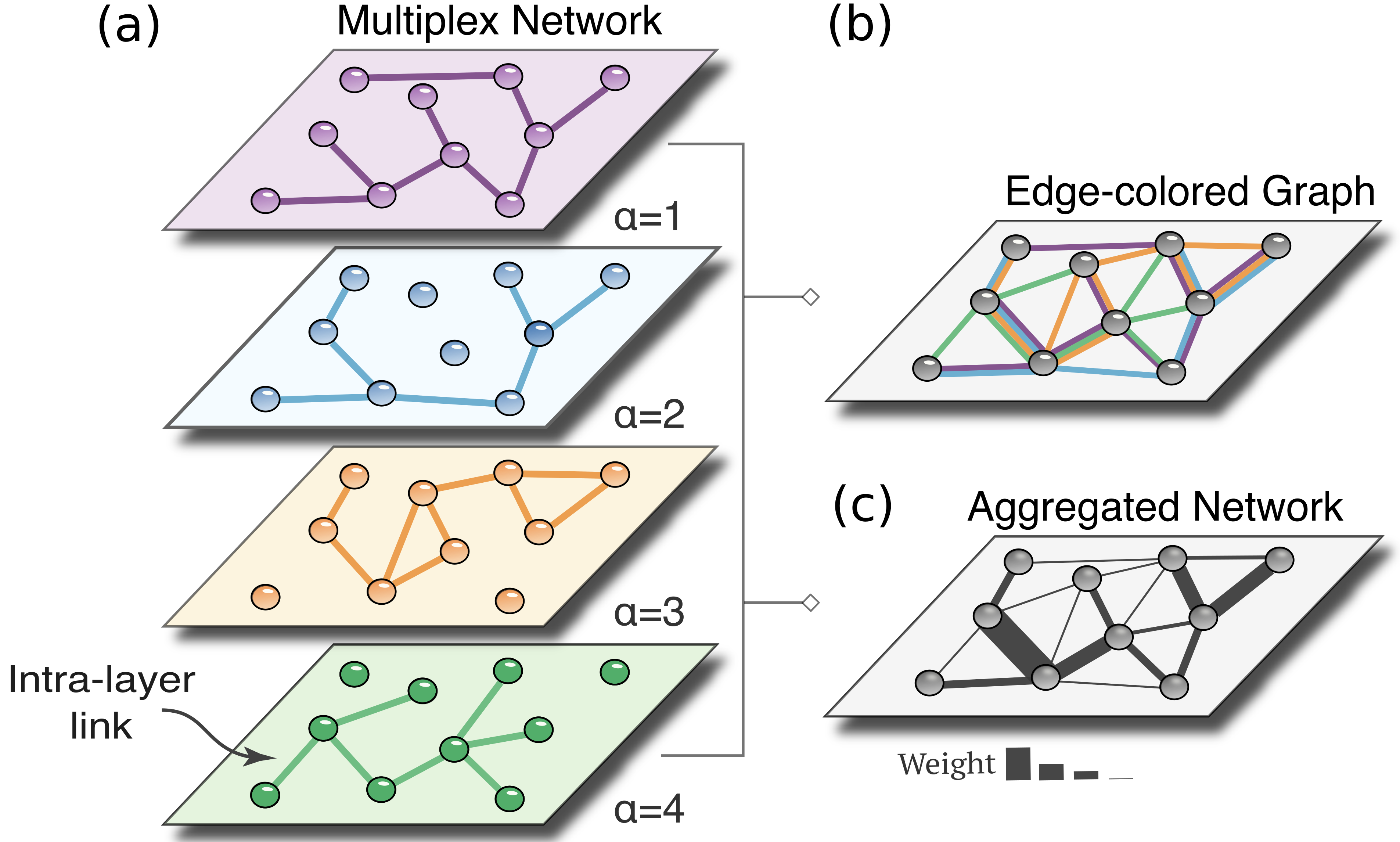}
        \caption{    \vito{   Special cases of multiplex networks.  
            (a) An example with $N=10$ and $M=4$ of a multiplex network whose 
            replica nodes represent the same physical entity across layers. Here the 
            system is solely characterized by the intra-layer links, shown as 
            solid lines. (b) In this case, the multiplex network 
            is fully equivalent to an edge-colored graph with
          $M=4$ colors. (c) By collapsing all the layers of the multiplex
          network, it is possible to obtain the corresponding
          aggregated weighted network,
          which instead disregards the information on 
          the different nature of the interactions. }
          }
        \label{fig:figure_mult}
    \end{center}
\end{figure}
%

\vito{Finally in some multiplex networks, such as the 
  neural network of the C. elegans in Fig.~\ref{fig:figure_real} (c), 
  the replica nodes represent the same physical entity across layers, and 
  there is no need to specify the coupling matrices of nodes
  across layers. In all such cases the distinction of the 
  layers is made at the edge level, but not at the node level.
  In this case, the multiplex network $\mathcal M$ is fully-defined
    by specifying
$\bm A$, i.e.:    
\begin{equation}
\mathcal M \equiv \bm A = \{\mathrm{A}\lay{1},\mathrm{A}\lay{2}, \ldots, \mathrm{A}\lay{M} \}.
\label{multiplex1}
\end{equation}
This special case of multiplex networks} 
is known in the mathematical
literature under the name of edge-colored graphs~\cite{ramsey2009problem},
i.e., multigraphs where each type of links is denoted by a
different color. 
An example of a multiplex network of this type, with $N=10$ and $M=4$ layers, 
is shown in Fig.~\ref{fig:figure_mult}, together with its equivalent
representation as an edge-colored graph. Also reported is the corresponding 
aggregated network, a weighted graph obtained from the
multiplex network by disregarding the information on the nature
of the interactions.
\vito{This special case of multiplex networks with replica nodes representing the
  same physical entity across layers,
}
are used in the description of dynamical models
of  
social agents  involved in different types of interactions, such as those in 
Sec.~\ref{sec:SD}, or 
playing different types of games and/or with
different types of strategies as in Sec.~\ref{sec:games}.

To conclude this part on the structure of a multiplex network, we
observe that Eq.~(\ref{multiplex2}) can be seen as a special case
of a more general 
formalism encoding all possible connections of a system with many
layers in a rank-four tensor ${\mathcal T}= \{ \tau\lay{\alpha
  \beta}_{i j} \}$~\cite{de2013mathematical}.  Also here,
superscripts indicate layers while subscripts indicate nodes,
and entry $\tau\lay{\alpha \beta}_{i j}$ is a non-negative real
number representing the weight of a link from node $i$ at layer
$\alpha$ to node $j$ at layer $\beta$. Tensor ${\mathcal T}$ thus
describes the most general system with many layers, and can also
represent interdependent and
interconnected networks, i.e., systems allowing for inter-layer
interactions between
different and multiple entities~\cite{boccaletti2014structure,
  kivela2014multilayer, aleta2026multilayer}.
In this review we will focus on
multiplex networks. In the tensorial formalism, multiplex networks 
can be described as $\mathcal M \equiv \mathcal T$, with the
supplementary requirement on the tensor $\mathcal T$ 
that, for $\alpha \neq \beta$, only entries
$\tau\lay{\alpha \beta}_{i j}$ with $i=j$ can be different from zero.

%
\subsection{\vito{Dynamics}}  
\label{subsec:dynamics}

\vito{To model the dynamics of a complex system, whose structure is
described by a multiplex network $\mathcal M$ as in
Eq.~(\ref{multiplex2}), we need first to define the set of variables
that characterize the state of the nodes, and then to write the 
equations governing the time evolution of the state variables.
Let us assume that the state of the replica node $i$ at layer $\alpha$
is a column vector ${\bm \sigma_i\lay{\alpha}}=
{\bm \sigma_i\lay{\alpha}}(t) \in \mathbb{R}^{d\lay{\alpha}}$,  
whose $d\lay{\alpha}$ components are real numbers. Notice
that the same set of state variables is used to describe each node $i$ at
layer $\alpha$, that is $d\lay{\alpha}$ does not depend on $i$. 
We can then define the state of node $i$ as:  
\begin{equation}
  \bm \sigma_i = [(\bm \sigma_i \lay1)^{T}, (\bm \sigma_i \lay2)^{T}, 
                  \ldots, (\bm \sigma_i \lay{M})^{T} ]^T 
 \label{dyn_state_node}
\end{equation}
which is a vector with $D=\sum_{\alpha=1}^{M} d\lay{\alpha}$ components, 
or $D=M \cdot d$ when the state of the node at each layer is a vector
in $d$ dimensions. 
Analogously, we can define the state of a layer $\alpha$ as: 
\begin{equation}
  \mathcal{S}\lay{\alpha}=[(\bm \sigma_1 \lay \alpha)^{T},
                           (\bm \sigma_2 \lay \alpha)^{T},
    \ldots, (\bm
\sigma_N\lay \alpha)^{T} ]^T 
 \label{dyn_state_layer}
\end{equation}
which is a vector of $N \cdot d\lay{\alpha}$ components.
Finally, the dynamical state of the multiplex network is fully
determined by the vector:  
\begin{equation}
  \mathcal{S} =  
[ ( \mathcal{S}\lay1  )^{T},
  ( \mathcal{S}\lay2)^{T},
  \ldots,
  ( \mathcal{S}\lay{M})^{T} ]^T =
[(\bm \sigma_1 \lay1)^{T}, \ldots, (\bm \sigma_N\lay1)^{T},  
\ldots, 
(\bm \sigma_1 \lay{M})^{T}, \ldots, (\bm \sigma_N\lay{M})^{T} ]^T
%
\nonumber
 \label{dyn_statemult}
\end{equation}
 with $N \cdot D$ components.
For instance, in most of the systems of $N$ coupled R\"ossler
oscillators studied in Section \ref{sec:synchro}, we have 
$d\lay{\alpha} =d=3$ for each $\alpha$, and $ \mathcal{S}$ is a vector
of $3 N M$ components. Instead, in the majority of diffusion and reaction
processes of Section \ref{sec:diffusion}, the state of a node
at each layer is a scalar, namely we have $d\lay{\alpha}=1 ~\forall \alpha$,
and $\mathcal{S}$ is a vector
of $N M$ components.
\\
Finally, in the special case when the replica 
nodes of the multiplex correspond to the same physical entity, and the 
multiplex network is equivalent to an edge-colored graph,
the state ${\bm \sigma_i}$ of node $i$ cannot be decomposed anymore over
the layers as in Eq.~(\ref{dyn_state_node}). 
In such a situation, the state of the multiplex network is described by
the vector: 
\begin{equation}
 \mathcal S = [ {\bm \sigma_1}^{T}, {\bm \sigma_2}^{T} \ldots,
          {\bm \sigma_N}^{T} ]^{T}
 \label{dyn_statemult-edgecolored}          
\end{equation}
where, for each node $i$, ${\bm \sigma_i}$ is a vector with $D=d$ components.
For instance, $D=3$ in the case of coupled 3-dimensional R\"ossler oscillators
on edge colored graphs considered in Eq.~(\ref{dyn_oscill_mult}) of
Section \ref{sec:synchro}.
In some simpler cases we have 
$D=d=1$ and the state of each node is a scalar, $\sigma_i \in \mathbb{R}$.
This is what happens in the basic multiplex voter model of Section
\ref{sec:SD},
where the scalar $\sigma_i$ represents the
opinion of individual $i$, independently on the number of different types
of interactions (layers), and in some particular random walk discussed in 
Section \ref{sec:diffusion}.}
\vito{However, we may have
$D=1$ also in multiplex networks with non-trivial inter-layer
couplings, as for example in some of the percolation models of Section
\ref{sec:percolation}.}

\begin{figure*}[t!]
 	\centering
 	\includegraphics[width=0.98\textwidth]{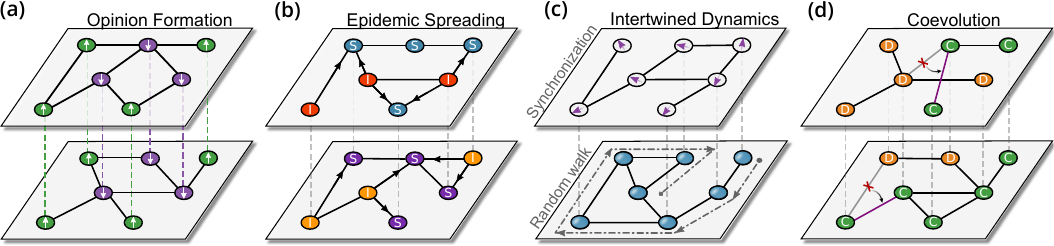}
 	
 	\caption{Multiplex dynamical processes. (a) A voter model, where each
 		agent is characterized by a single state (node color) describing
 		its opinion, and imitation occurs by selecting a neighbor in one of
 		$M=2$ social layers.  (b) Disease spreading, where each
 		node can be susceptible (S) or infected (I) by two different
 		viruses diffusing over two distinct layers, and the infection from
 		one disease also increases the probability of getting the other
 		one. (c) Intertwined synchronization and transport dynamics, where
 		two different types of processes, namely a Kuramoto model and a
 		biased random walk evolve together under mutual feedback.  (d)
 		\vito{Coevolution of network and processes, where cooperators (C) involved
 			in two distinct games on the two layers of a multiplex network can
 			rewire their old (grey) links into new (purple) links to avoid playing
 			with defectors (D). For simplicity, rewiring is independent across layers, though dynamics may couple in more complex settings.}}
 	\label{fig:figure_section2_dynamics}
 \end{figure*}

\vito{Now that we have formalized how to define the
  state of a multiplex network at time $t$, we need to model how such a state
  evolves in time.} 
In the most general form, the equations describing the dynamics of a
multiplex network can be written as:
\begin{equation}
    \frac{d {\mathcal S}} {dt}  = F( {\mathcal M}, {\mathcal S}(t)  ) 
\label{eq:multev_cont}
\end{equation}
for a continuous-time process, or
\begin{equation}
    \mathcal S(t+1) = F( {\mathcal M}, {\mathcal S(t)}      ) 
\label{eq:multev_discr}
\end{equation}
for a discrete-time process. Function $F$ governing the time-evolution of
the multiplex in general depends both on the structure
$\mathcal M$ and on the dynamical state $\mathcal S$ of the multiplex.
Such a function can couple the states of two nodes linearly, as in the
diffusion processes considered in Section \ref{sec:diffusion}, or non-linearly, 
as in the coupled dynamical systems of Section \ref{sec:synchro}.  
%
Figure \ref{fig:figure_section2_dynamics} shows some examples of
multiplex dynamics that will be covered in this review.  Panel (a) and
(b) respectively sketch opinion formation and disease spreading dynamics
occurring over two different layers of interactions. These will be
discussed in Section \ref{sec:SD} and in Section \ref{sec:spreading}.
Panel (c) describes instead a case of intertwined dynamics as those
covered in Section \ref{sec:intertwined}, where two different types of
processes, namely a transport dynamics at the first layer and a
synchronization model at the second layer evolve under mutual
feedback.

In Eqs.~\eqref{eq:multev_cont} and \eqref{eq:multev_discr} 
the structure of the multiplex network is fixed in time,  
and
only the node state $\mathcal S$ is a dynamical variable.
However, in many cases multiplex networks also evolve over
time,
i.e., $d {\mathcal M} / {dt} = G
( {\mathcal M}(t))$, where $G$ is the function governing the time
evolution of the multiplex ${\mathcal M} (t)$. 
%
%
When the evolution of a multiplex structure is influenced not only
by the topology of the network, but also by the dynamical states of
nodes, the previous equation can be generalized to $d {\mathcal M} / {dt} = G ( {\mathcal
   M}(t), {\mathcal S}(t)   )$. This naturally leads to the most general case 
 of the coevolution of multiplex structures with the dynamical processes
 running on the networks, which can be described by the coupled equations
 \cite{thurner2018introduction}:
%
 \begin{equation}
\left\{ \begin{array}{c}
    d {\mathcal S} / {dt}  =   F( {\mathcal M}(t), {\mathcal S}(t)      ) 
    \\
    \\
     d {\mathcal M} / {dt}  = G ( {\mathcal M}(t), {\mathcal S}(t)      )   
\end{array}
\right.
\label{eq:coev_cont}
\end{equation}
 for continuous-time systems, and by
 \begin{equation}
\left\{ \begin{array}{c}
     \mathcal S(t+1)  =   F( {\mathcal M(t)}, {\mathcal S(t)}      ) 
    \\
    \\ \mathcal M(t+1) = G ( {\mathcal M(t)}, {\mathcal S(t)}      )   
\end{array}
\right.
\label{eq:coev_discr}
\end{equation}
%
if time is discrete. Figure \ref{fig:figure_section2_dynamics}(d)
illustrates an example of such coevolution, where cooperators involved in two distinct games on the two layers of a multiplex network can strategically
rewire their links to avoid interacting with defectors in order
to maximize their payoffs. This and other cases of multiplex coevolution will be considered in Section \ref{sec:coevolution}.

%




\section{Structural properties and models}
\label{sec:structure_bis}
\newtext{In this section, we briefly overview key measures and models used to characterize the topology of multiplex networks~\cite{boccaletti2014structure,kivela2014multilayer,bianconi2018multilayer,cozzo2018multiplex,artime2022multilayer,de2023more}.  Such measures concisely capture intra- and inter-layer connectivity and quantify layer importance (e.g., activity, overlap, correlations), providing the minimal notation needed to relate multiplex structure to dynamics.}

\subsection{\vito{Edge properties}}
\label{subsec:edge} 
Due to the presence of multiple layers, the crucial signature of a
multiplex network is the diverse configuration of links at the
different layers. \newtext{That is, the same pair of nodes can be 
connected in distinct ways across layers, giving rise to a rich variety of 
connectivity patterns. While the measures we introduce primarily apply to the common case of undirected, unweighted multiplexes, most of them can be naturally extended to weighted or directed networks}. To simplify this complexity, the
layers of the multiplex can be collapsed into a single-layer network,
the so-called \emph{aggregated network}. \newtext{This often serves
  as a useful baseline for comparing and characterizing the properties of the 
  full multiplex  structure.  Several variants have been proposed for this purpose, including the average
  network~\cite{sole2013spectral}, the overlay or projected
  monoplex network~\cite{de2013mathematical}, and the quotient
  network~\cite{sanchez2014dimensionality,cozzo2018multiplex}. In the following,
  we consider two widely used types of aggregated networks.} First, the
topological aggregated network, $\mathcal A=\{a_{ij}\}$, whose generic
element is defined as
\begin{equation}
    a_{ij} = \begin{cases}1 & \text{if} \>\>\>\exists\> \alpha:
        a\lay{\alpha}_{ij}=1 \\ 0 & \text{otherwise,} \end{cases}
\end{equation}
\newtext{which simply encodes the existence of a connection between nodes $i$ and $j$ in at least one layer, but disregards both the frequency and the specific nature of such connections across layers.}

A second aggregate,
represented by the overlapping matrix $O=\{o_{ij}\}$ and shown in
Fig.~\ref{fig:figure_mult},
accounts instead for the amount of \emph{edge overlap}
in the multiplex~\cite{battiston2014structural,boccaletti2014structure},
\begin{equation}
    o_{ij}=\sum_{\alpha=1}^{M} a_{ij}\lay{\alpha}, 
    \label{eq:chapter3_overlapping_matrix}
\end{equation}
namely the number of layers at which each link is present, \newtext{which 
aligns with the overlay network defined in \cite{de2013mathematical} in the case of multiplex networks, or link multiplicity~\cite{bianconi2013statistical,bianconi2018multilayer}}.  \newtext{For weighted multiplexes, $o_{ij}$ can be generalized to sum edge weights across layers, capturing the cumulative interaction intensity rather than a binary count. For directed networks, overlap can be defined separately for in- and out-edges.} To get a
global indicator of edge overlap in a multiplex network, one can
average the overlapping matrix over all possible pairs of nodes,
i.e., $\bar{o}= 2/[N(N-1)] \sum_i \sum_{j < i} o_{ij}$
~\cite{battiston2014structural}, or over all the links of the
topological aggregated network $\mathcal A$, i.e., 
\begin{equation}
  o= \frac{ \sum_i \sum_{j < i} o_{ij}} { \sum_i \sum_{j < i} a_{ij}},
  \label{eq:sec2_edgeoverlap}
\end{equation}
\newtext{with $o=1/M$ when every edge appears in exactly one layer, and $o=1$ when all edges are present in all layers} ~\cite{lacasa2015network}.
Among several definitions of edge overlap between
layers~\cite{boccaletti2014structure, cellai2013percolation,
  cellai2016message}, \newtext{we mention the global overlap between layers $\alpha$
and $\beta$, which quantifies the total number of links in common as
$\hat{o}\lay{\alpha,\beta}= \sum_i \sum_{j < i} a_{ij}\lay{\alpha}
a_{ij}\lay{\beta}$~\cite{szell2010multirelational,bianconi2013statistical}.} By contrast, the local edge overlap of a node $i$
counts the number of overlapping edges incident to node $i$ in both
layers $\alpha$ and $\beta$, i.e.,  $\tilde{o}_i\lay{\alpha \, \beta} =
\sum_{j=1}^{N} a_{ij}\lay{\alpha}
a_{ij}\lay{\beta}$~\cite{bianconi2013statistical}.
Extensions of these metrics include the \textit{multiplexity index}, introduced by \textcite{gemmetto2015multiplexity}, which provides a normalized version of the global overlap for both unweighted and weighted multiplex networks.  In a similar way, the edge intersection index introduced by~\textcite{de2015structural} quantifies the probability of finding a pair of nodes that is connected by an edge at all the $M$ layers of the multiplex.
\newtext{However, all the aforementioned metrics of overlap provide only a raw count of the number of overlapping links, overlooking the precise pattern of connections across different layers.} One possible way to solve this issue is to consider the $2^M$ different configurations, using the so-called \textit{multilinks}~\cite{bianconi2013statistical,menichetti2014weighted}, which keep track of the precise layers at which a pair of nodes is connected.
\newtext{\subsection{\vito{Node properties}}}
\label{subsec:node} 
Given the layered structure of a multiplex network, any property 
of node $i$ needs to be described as a
vector
\begin{equation}
    \bm p_{i} = \left\{p_i\lay 1, \ldots, p_i\lay{\alpha},\ldots, p_i \lay M \right\},\quad
    i=1,\ldots,N,
    \label{eq:node_property}
\end{equation}
whose component $p_i\lay \alpha$ represents the node property 
at layer $\alpha$. \newtext{While this full vector contains maximal information, it is often advantageous --- both for analytical tractability and for intuitive interpretation --- to compress it into a small number of meaningful scalars.}

For instance, the most basic property of a node $i$, namely its degree, 
can be expressed in a multiplex network
as ${\bm k_i} = \{ k_i\lay{1}, \ldots, k_i\lay{M} \}$, where $k_i\lay{\alpha} =
\sum_{j\neq i}^{N}a_{ij}\lay{\alpha}$ is the number of connections of 
node $i$ at layer $\alpha$~\cite{battiston2014structural}.  \newtext{For weighted layers the formula gives node strength (sum of incident edge weights), while in directed networks one distinguishes in-degree and out-degree (or strengths), defined by restricting the summation to incoming or outgoing neighbors.}
A first compact way to describe the information contained in 
vector ${\bm k_i}$ is to evaluate the sum of its components,
usually denoted in the
literature as the total or 
\emph{overlapping degree}~\cite{battiston2014structural}
\begin{equation}
    o_i = \sum_{\alpha=1}^M k_i\lay{\alpha} 
\end{equation}
of node $i$, \newtext{or referred to as \textit{multidegree centrality} when considering the tensorial formalism~\cite{de2013mathematical,artime2022multilayer}}. \newtext{This notion extends directly to weighted and directed cases, where it corresponds to total node strength (in, out, or both) accumulated across layers.}
\newtext{To capture how heterogeneously the links of node $i$ are distributed across layers, one defines the multiplex \emph{participation coefficient} 
}
\begin{equation}
    \label{participationcoefficient}
    P_i=1- \sum_{\alpha=1}^M\biggl(\frac{k_i\lay{\alpha}}{o_i}\biggr)^2 \in [0,1],
\end{equation}
\newtext{with $P_i=0$ if all of node $i$'s edges lie in a single layer, and $P_i=1$ when they are equally spread across the M layers, while intermediate values quantify partial mixing.   Equivalently,  the Shannon entropy of the normalized degree vector ${\bm k_i}/o_i$ can be used to capture the same heterogeneity.}
\newtext{Moreover, since nodes in real-world multiplex networks
may lack connections on certain layers,} we can characterize the \newtext{structural} activity-pattern of node $i$ by means of vector
${\bf b}_i = \left\{b_i\lay{1}, \ldots, b_i\lay{M} \right\}$,
%
where $b_i\lay{\alpha}=\delta_{0,k_i\lay{\alpha}}$ is equal to 1 if
node $i$ has at least one connection at layer $\alpha$.   \newtext{This notion extends directly to weighted and directed cases, where it corresponds to total node strength (in, out, or both) accumulated across layers.}
\newtext{We emphasize that this vector is purely structural, indicating edge presence per layer,  independent of any dynamics. }
The total  \newtext{structural} activity of node $i$, $B_i = \sum_{\alpha} b_i\lay{\alpha}$, \newtext{with $B_i \in [0,M]$,}
counts the number of layers where node $i$ is active, and  \newtext{exhibits heterogeneous distributions across many real-world multiplex systems}~\cite{nicosia2015measuring}.

Measures of node centrality beyond the degree~\cite{newman2018networks}
have been proposed to capture the relative importance of the
nodes of a multiplex network. 
\newtext{The most direct generalization, in line with Eq.~(\ref{eq:node_property}), defines the multiplex centrality of a node $i$ as the vector of its centrality scores in each layer,  ${\bf c}_i =\left\{c_i\lay{1}, \ldots, c_i\lay{M} \right\}$. 
While simple, this approach ignores structural correlations between layers that can strongly affect rankings. Computing centralities per layer and aggregating them, or flattening the multiplex, discards these correlations and may create artificial links, distorting flow-based measures and local structure.}
\textcite{sola2013eigenvector} first introduced a genuinely multiplex
centrality measure proposing to evaluate the eigenvector centrality
of nodes on each layer $\alpha$ as the normalized eigenvector relative
to the largest eigenvalue of $\tilde{A}\lay{\alpha}=
\sum_{\beta=1}^{M} i\lay{\alpha,\beta} A\lay{\beta}$. Here, $I =
\left\{ i\lay{\alpha,\beta} \right\}$ is the influence matrix, where each entry specifies
\newtext{how much the centrality of layer $\alpha$ affects rankings in layer $\beta$. 
In practice, entries of the matrix $I$ can be fixed from domain knowledge, estimated from data (e.g., via inter-layer correlations), or tuned to explore coupling regimes. Setting $I$ as the identity matrix recovers independent layers, whereas a unitary $I$ yields the aggregate-network eigenvector centrality.}
\newtext{More recent approaches compute node and layer centralities simultaneously through coupled equations, often formalized using multi-homogeneous maps~\cite{rahmede2018centralities,tudisco2018eigenvector,taylor2021tunable}.} The main idea is that nodes have a high centrality if they are active in highly central layers, while layers are ranked as highly influential if highly central nodes are active in them. 
Additionally, an entire class of eigenvector centralities can be obtained using the multilayer tensorial representation~\cite{sole2014centrality,de2015ranking,wu2019tensor,kumar2023multicens,frost2024generalized}, such as the ``versatility'' measure~\cite{de2015ranking}.

\newtext{Random-walk centralities have also been extended to multiplex settings, allowing walkers to move within and between layers, possibly with biases that reflect inter-layer dependencies (see Section~\ref{sec:random_walks}).}	
For example, \textcite{halu2013multiplex},  
among others~\cite{sole2016random,ding2018centrality,bottcher2021classical}, 
introduced a multiplex PageRank where  walkers in a layer are  biased  by PageRank scores of other layers.  \textcite{iacovacci2016functional} refined  this by considering that different patterns of connections across layers might contribute differently to the centrality of a node.   \newtext{Related adaptations extend betweenness and current-flow centralities to multiplex structures~\cite{sole2016random,bottcher2021classical}.}
\newtext{Other strategies compute multiple centralities} independently in each layer and combined them via consensus ranking to define a single 
index~\cite{posfai2019consensus}, \newtext{outperforming any single-layer measure in predicting behavior in primate multiplex social networks~\cite{beisner2020multiplex}.}

\newtext{Path-based methods also play a central role. Extensions of the communicability matrix to multiplex networks~\cite{estrada2014communicability,bergermann2021matrix} capture how information or influence can propagate through multi-layered pathways. To  quantify how the presence of many layers affects the reachability
of each node $i$, \textcite{morris2012transport}  introduced the \emph{node interdependence}  
$\lambda_i=\frac{1}{N-1}\sum_{\substack{j \in N}}
\frac{\psi_{ij}}{\sigma_{ij}}$ 
where $\psi_{ij}$ is the number of shortest paths between $i$ and $j$ that span across more than one
layer, whilst $\sigma_{ij}$ is the total number of shortest paths between $i$ and $j$ in the multiplex. Hence, $\lambda_i = 1$ when all shortest paths make use of edges laying at least on two layers and equal to 0 when each path lies in the same layer. \\}

\newtext{\subsection{\vito{Layer properties}}}
\label{subsec:layer} 

Similarly to the \newtext{structural node activity, one can define the
\emph{structural layer-activity vector} of layer $\alpha$,
as $\mathbf{d}\lay{\alpha} = \left\{b_1\lay{\alpha},b_2\lay{\alpha}
, \ldots,b_N\lay{\alpha}\right\}$, which describes the patterns of
node activities at that layer, where again
$b_i\lay{\alpha}=\delta_{0,k_i\lay{\alpha}}$. The \emph{structural layer activity} is $N\lay{\alpha} = \sum_i b_i\lay{\alpha} \in [0,N]$, equal to $0$ when all nodes are isolated and to $N$ when none are. This quantity} captures the number of nodes with at least one
connection in layer $\alpha$, and its distribution has been found to
be quite broad in many real-world
systems~\cite{nicosia2015measuring}. Since many
multiplex systems exhibit highly correlated structures between layers, a number of
different similarity indicators have been
proposed~\cite{kao2018layer}. For instance, the pairwise
multiplexity~\cite{nicosia2015measuring} characterizes the similarity
between the activity-vectors of two layers $\alpha$ and $\beta$,
i.e., $Q\lay{\alpha,\beta} = \frac{1}{N}\sum_i b_i\lay{\alpha}
b_i\lay{\beta}$, in terms of the fraction of nodes that are
simultaneously active in both layers, \newtext{with $0 \leq Q^{[\alpha,\beta]}\leq \min\{N^{[\alpha]},N^{[\beta]}\}/N$. Such a quantity is zero for disjoint activity sets, or equal to one only if all nodes are active in both layers.} A more fine-grained metric, the normalized Hamming distance, quantifies instead the similarity among
the patterns of activity in two layers~\cite{nicosia2015measuring}.

More generally, \newtext{to compare a node property 
$\mathbf{P}$ between two layers $\alpha$ and $\beta$, one may use Pearson, Spearman, or Kendall correlations, each bounded in $[-1,1]$. } 
%
%

When the node property of interest $\mathbf{P}$ is the
degree, a quantity extensively used to quantify
\emph{inter-layer degree correlations} is
$\overline{k\lay{\beta}}\left(k\lay{\alpha}\right)=
\sum_{k\lay{\beta}}k\lay{\beta}P\left(k\lay{\beta} \vert
k\lay{\alpha}\right)$,which quantify the average degree at layer $\beta$ of 
nodes with degree $k\lay{\alpha}$ at layer $\alpha$. 
\newtext{
An increasing trend indicates positive inter-layer assortativity; a decreasing trend indicates disassortativity; a flat curve at $\langle k^{[\beta]}\rangle$ mirrors no dependence. If $k_i^{[\alpha]}=k_i^{[\beta]}$ for all $i$, then $\overline{k^{[\beta]}}(k^{[\alpha]})=k^{[\alpha]}$; if degrees are independent$, \overline{k^{[\beta]}}(k^{[\alpha]})=\langle k^{[\beta]}\rangle$.  Scalar summaries such as $\mathrm{corr}(k^{[\alpha]},k^{[\beta]})\in[-1,1]$ (and rank analogues) are also common~\cite{nicosia2015measuring}. Notice that }this is the multiplex generalization of the nearest-neighbours average degree function
$k_{nn}(k)$, traditionally used to quantify degree-degree correlations
in single-layer graphs~\cite{pastor2001dynamical}. Other measures of
inter-layer degree correlations rely on pairwise
mutual information between the degree sequences of the two
layers~\cite{lacasa2015network}, or on the tensorial
formalism~\cite{de2016degree}.


%
\begin{figure*}[t!]
	\centering
	\includegraphics[width=0.95\textwidth]{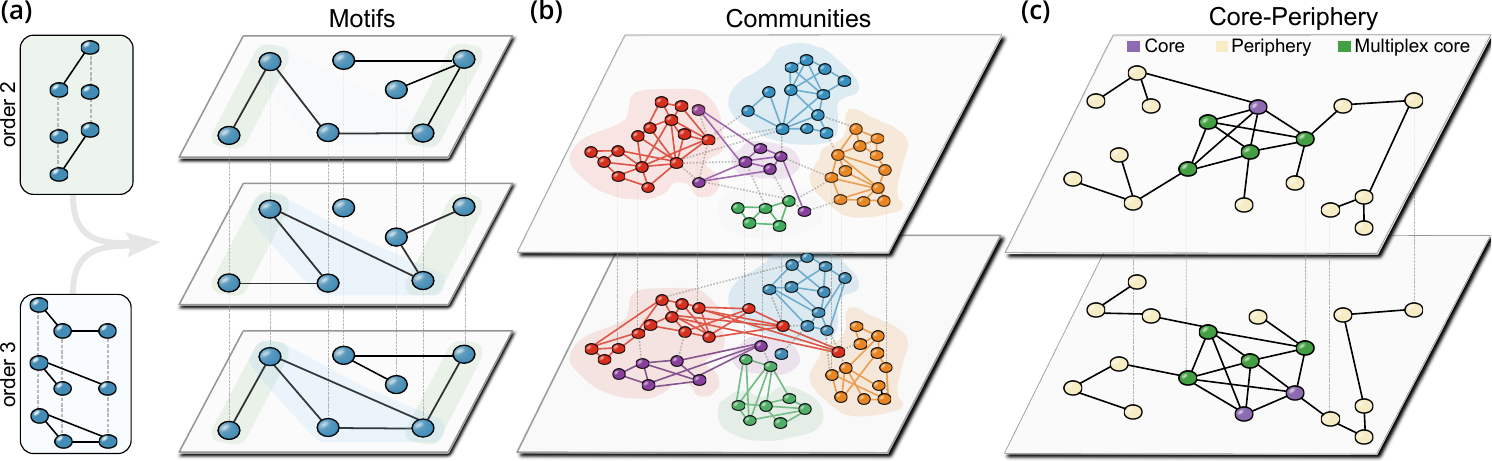}
	\caption{Micro- and meso-scale patterns in multiplex
		networks.
		\newtext{(a) Examples of multiplex motifs: a 2-node multilink (edge configuration across layers) and a 3-node motif in a system with three layers, where identical aggregate topology can correspond to distinct layered patterns.}
		(b) A multiplex community structure
		in a network with two layers, reported as colored shaded areas, which often 
		differs from the community partitions of the two 
		single layers, shown by colored nodes.  
		\newtext{(c) Core-periphery organization in multiplex networks can also depart from that of single layers, reflecting roles that may align across layers, vary by layer, or emerge only jointly.} }
	\label{fig:section2_mesoscale}
\end{figure*}

\newtext{\subsection{\vito{Mesoscale properties}}}

\newtext{Non-trivial mesoscopic structures are a key feature of many real-world networks. These structural patterns, which bridge local and global scales, include well-known features such as motifs~\cite{milo2002network}, communities~\cite{porter2009communities,fortunato2016community}, and core-periphery structures~\cite{csermely2013structure,rombach2017core}. While these concepts have been extensively studied in single-layer networks~\cite{newman2018networks}, their analysis in multiplex networks requires a more nuanced approach. The distribution of edges across different layers fundamentally alters the form and interpretation of these structures (see Fig.~\ref{fig:section2_mesoscale})}.

\newtext{In single-layer networks, a motif is defined as a small, statistically over-represented subgraph~\cite{alon2007network}.  In multiplex networks, the definition must also consider which layers host the edges. Indeed, subgraphs that are identical in their aggregated structure can represent different patterns depending on their layer assignment (see Fig.~\ref{fig:section2_mesoscale}).  For example, in edge-colored graphs,  a 2-motif with a pair present on two of three layers (e.g., layers 1 and 3) differs from a pair present in only one layer (e.g., layer 1), corresponding to two instances of the $2^{M}-1$ nontrivial multilinks~\cite{bianconi2013statistical,menichetti2014weighted}. Likewise, for connected  3-motifs, both the triad (three nodes, two edges) and the triangle (three nodes, three edges) admit multiple layered patterns. The classification of such motifs on edge-colored or uncoupled multiplex graphs has clear precedents in the colored-graph literature~\cite{wernicke2005faster,wernicke2006fanmod}, which addresses the same underlying combinatorial problem on the same class of architectures. Building on this line of work, the multiplex interpretation further specifies the edge-to-layer assignment within a richer structural formalism: in general, a multiplex motif is determined by its size, the induced aggregate topology, and the edge–to–layer assignment~\cite{battiston2017multilayer}. The presence of layers motivates extensions of other concepts, such as clustering or cycles~\cite{baxter2016cycles}. A multiplex clustering coefficient distinguishes between within-layer and inter-layer closures of triangles~\cite{battiston2014structural}, with similar generalizations using the tensorial formalism~\cite{cozzo2015structure,bartesaghi2021clustering}. Multilayer isomorphisms extend standard graph isomorphisms to layer labels, enabling compact classification of motifs~\cite{kivela2017isomorphisms}. At the computational level, counting motifs in multiplex networks is challenging due to the combinatorial complexity introduced by the layers; algorithms designed for multiplex and temporal motifs have made it possible to study these structures in larger systems~\cite{boekhout2019efficiently}. Recently, \textcite{dimitrova2020graphlets} extended graphlet analysis to the case of multiplex networks.}

\newtext{Nevertheless, the most ubiquitous meso-scale feature observed in complex systems is community structure, the tendency of a network's nodes to cluster into densely connected groups with only sparser connections between them \cite{fortunato2022years}. In a multiplex network, this concept is considerably more subtle than in a single layer. A multiplex community is a set of nodes that exhibits high internal cohesion when connections are considered across one or multiple layers. The core challenge is that this structure may not be discernible from any single layer alone or from a simple aggregation; instead, it often emerges from the combined, and sometimes complementary, patterns of intralayer connectivity. Consequently, multiplex communities may (i) persist coherently across all layers, (ii) be specific to a single layer, or (iii) arise only through inter-layer dependence, where no individual layer shows strong community structure, but their combination does \cite{mucha2010community}. Given this complexity, no single, universally agreed-upon definition of a multiplex community exists; rather, the precise definition is often implicit in the detection methodology employed. For comprehensive reviews of the field, we refer the reader to dedicated surveys \cite{kim2015community, huang2021survey, magnani2021community}.}

Loosely speaking, multiplex community detection strategies can be
classified into three broad categories: flattening, aggregation, and
direct methods~\cite{tagarelli2017ensemble}. Most of the early
approaches either flatten all the layers of the multiplex network
into a single weighted network~\cite{berlingerio2011finding,tang2012community,taylor2016enhanced,taylor2017super},
or rely on existing algorithms for each layer and then merge the
partitions via consensus clustering~\cite{lancichinetti2012consensus,papalexakis2013more,cantini2015detection}.
Since most of these approaches have the drawback of ignoring 
inter-layer links, methods directly tailored on the layered
structure have become increasingly
popular~\cite{de2015identifying,mondragon2018multilink}. A large
fraction of these methods are based on the optimization of
objective functions such as the 
modularity~\cite{mucha2010community,bennett2015detection,weir2020multilayer}, \newtext{with inter-layer coupling tuning layer dependencies. Infomap is an information-theoretic method that finds communities by minimizing the description length of a random walker’s trajectory \cite{rosvall2008maps}. In multiplex form, the walker moves within and across layers, and the partition that best compresses these paths reveals modules where flow is preferentially retained~\cite{de2015identifying}. Similarly, related random-walk formulations reveal functional modules by allowing walkers to switch layers under tunable diffusion parameters ~\cite{kuncheva2015community,jeub2017local,de2015identifying,bertagnolli2021diffusion}. 
Additionally, stochastic block model (SBM) approaches fit generative models via Bayesian inference or maximum likelihood~ \cite{peixoto2015inferring,stanley2016clustering,subhadeep2016consistent,de2017community,pamfil2019relating}, defining communities probabilistically, enabling explicit inter-layer coupling and degree correction, and supporting principled model selection and uncertainty quantification (see Sec.~\ref{subsec:models}).
}


Alternative approaches to analyze the
mesoscale structure of multiplex networks rely on different tensor 
factorizations~\cite{gauvin2014detecting,chen2019tensor,aguiar2024tensor}, which are
often computationally efficient thanks to their closed-form
solution. In addition to the above-mentioned methods, a part of
recent literature has focused on discovering overlapping
communities~\cite{liu2018finding}, or extended existing method by
including also node
attributes~\cite{contisciani2020community}. Finally, other approaches
are based on various similarity metrics between nodes, group
of nodes, or layers of the layered system~\cite{brodka2018quantifying,iacovacci2015mesoscopic,iacovacci2016extracting,de2016spectral,kao2018layer}.

\newtext{Core–periphery structure partitions a network into a dense, cohesive core and a sparser periphery whose nodes connect mainly to the core \cite{borgatti2000models,rombach2017core}. In multiplex systems, roles must be assessed across layers: the core may (i) align consistently across layers (a global core), (ii) vary by layer (core in one, peripheral in another), or (iii) emerge only jointly when layers reinforce one another, so capturing both cross-layer consistency and complementarity is essential. Early analyses relied on single-layer proxies such as k-core variants \cite{corominas2016weak,galimberti2017core}, but general multiplex treatments now exist: a nonparametric criteria based on local node information have been extended to layered settings \cite{ma2015rich,battiston2018multiplex}, nonlinear spectral formulations jointly inferring node and layer coreness~\cite{bergermann2024nonlinear}, and recent refinements to identify dense multiplex cores~\cite{nie2025effective}. Empirically, multiplex core-periphery differs from single-layer cores and is altered in the brain by neurodegenerative disease \cite{pontillo2025more,guillon2019disrupted} and by learning~\cite{corsi2021bci}.}
%

\medskip

\newtext{\subsection{\vito{Reducibility}}}

\begin{figure}[tbh!]
		\centering
		\includegraphics[width=0.55\textwidth]{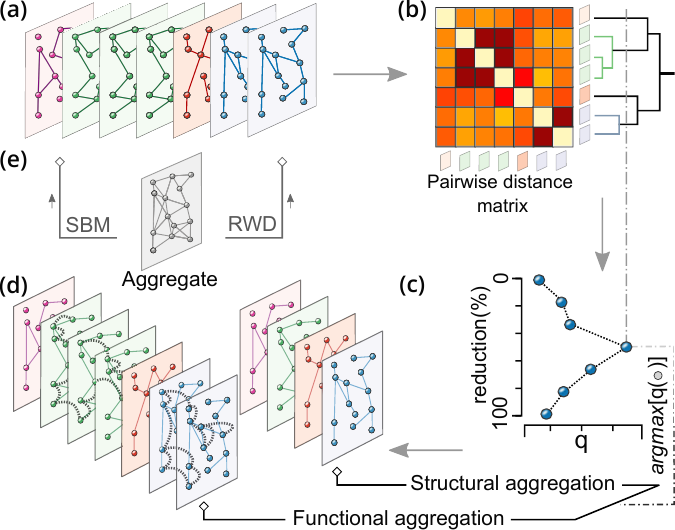}
		\caption{Reducibility of multiplex networks. (a)
			Starting from a multiplex network with $M$ layers, (b) a general
			reducibility procedure usually relies on a distance measure
			to quantify the similarity/dissimilarity between all the
			possible pairs of layers. The resulting $M \times M$ matrix of 
			pairwise distances allows to obtain a hierarchical 
			diagram (dendrogram) whose leaves represent the initial
			layers, while the internal nodes denote layer merging. (c) A
			structural or functional quality function q($\bullet$)
			is then considered to
			determine an optimal partition of the layers, (d) and obtain
			\newtext{a structurally or functionally aggregated multiplex network.}
			(e) Vice versa,
			starting from a single-layer (aggregated) network, it is possible
			to reconstruct an underlying multiplex structure,  
			either by relying on a stochastic block model (SBM)
			formulation~\cite{valles2016multilayer} or by considering
			Markovian random walk dynamics
			(RWD)~\cite{lacasa2018multiplex}.}
		\label{fig:figure_section2_reducibility}
	\end{figure}

\newtext{One of the fundamental issues with a multiplex network description of a real-world system is that of {\em reducibility}~\cite{de2015structural}: determining whether the layered representation can be simplified without significant loss of information. Given a multiplex with $M$ layers, the task is to construct a reduced representation --- by merging, reweighting, or removing layers --- that best satisfies a clearly defined objective. This objective specifies what counts as ``essential'' information and is typically evaluated through a quality function that measures how well the reduced network preserves the desired properties. Structural objectives aim to preserve key topological features, while functional objectives seek to maintain the performance of target dynamical processes, such as diffusion, synchronization, or epidemics.  In principle, the optimal reduction requires testing all partitions of the $M$ layers, a number given by the $M$-th Bell number $B_M$ that grows super-exponentially, making exhaustive search intractable. In practice, methods rely on heuristics, typically hierarchical clustering or greedy algorithms, to navigate the search space efficiently. The general workflow of a reducibility approach is sketched in 
	Fig.~\ref{fig:figure_section2_reducibility}.}

	%

\newtext{A first structural approach was proposed by \textcite{de2015structural}, drawing a formal parallel between graph Laplacians and density operators in quantum systems: each network layer is represented by a normalized Laplacian, analogous to a quantum density matrix whose spectrum captures structural features. This representation allows one to quantify layer-to-layer dissimilarity using the quantum Jensen–Shannon divergence, and to measure the information retained by a given partition through changes in the Von Neumann entropy. A greedy hierarchical clustering iteratively merges the most similar layers, producing a reduced multiplex that maximizes entropy-based distinguishability from the fully aggregated network, thereby removing redundancy while retaining the most informative structures. Many later methods follow this principle but use alternative similarity measures (e.g., cosine, spectral, embedding-based) or quality functions~\cite{de2016spectral, wang2017layer,santoro2020algorithmic, aguiar2022factor, baccini2022similarity,nan2025assessing}. In particular, \textcite{santoro2020algorithmic} further refine this approach via algorithmic information theory, yielding reduced multiplexes that better preserve the structure and some aspects of the dynamics of the original multiplex. Other structural reducibility methods target mesoscopic organization, particularly community structure, by clustering layers according to shared stochastic block model parameters or community-similarity measures, yielding reduced representations tailored for community detection~\cite{stanley2016clustering, taylor2017super,kao2018layer}.}
	
\newtext{In contrast, functional reducibility defines its objective in terms of the performance of a dynamical process on the network. A representative example is \textcite{de2020enhancing}, who use random walks as a proxy for information flow, grouping or coupling layers to reduce redundant diffusion pathways and improve navigability without altering intralayer topology. Here, the quality function is explicitly dynamical, capturing properties such as reachability, return probabilities, or diffusion times.}

\newtext{Beyond reduction, the inverse problem has also been explored: reconstructing hidden multiplex structure from partial or aggregated data. \textcite{valles2016multilayer} addressed this using a multilayer stochastic block model to infer the most likely division of an observed aggregate into multiple layers. \textcite{lacasa2018multiplex} instead exploited non-Markovian signatures in random-walk trajectories to detect whether a hidden multiplex underlies the observed network and to estimate its number of layers. More recent works~\cite{ma2020data,zhang2021detangling,bagrow2021recovering,wu2022discrimination,kaiser2023multiplex,kaiser2024reconstruction} have framed this as a classification problem, developing methods to discriminate between genuine single-layer networks and aggregates of multiplex structures.}\\

\newtext{\subsection{\vito{Models}}}
\label{subsec:models}

We briefly survey here the main approaches to \newtext{generate multiplex networks with certain structural characteristics, a task often referred to as generative modeling. These models can be broadly categorized as:} equilibrium models 
(i.e., the canonical and microcanonical ensembles), models of growth,
and stochastic block models. Extensive discussion can be found in dedicated
reviews~\cite{boccaletti2014structure,kivela2014multilayer} and
books~\cite{bianconi2018multilayer,artime2022multilayer}.

%
Tools coming from equilibrium statistical mechanics are well known to provide a
principled way to construct null models that satisfy a set of
constraints and are the least possible
biased~\cite{park2004statistical,cimini2019statistical}. 
The main idea is to construct the best ensemble of random graphs
constrained by a number of structural properties, such as the degree sequence,
or the community structure of a given empirical network. 
Any set of constraints gives rise to a
microcanonical network ensemble, when each graph of the ensemble
satisfies all the 
constraints exactly (hard constraints), or to a canonical network
ensemble when the constraints are satisfied on average (soft
constraints). \textcite{bianconi2013statistical} laid 
down the foundations of a statistical mechanic approach to 
multiplex networks, starting from the simplest hypothesis that the
various layers are uncorrelated, and then gradually considering more
realistic models of correlated or weighted
multiplexes~\cite{menichetti2014correlations}.
The approach has been generalised to a wide variety of
structures, including spatial multiplex
networks~\cite{halu2014emergence} and multiplex networks with
heterogeneous activities of the nodes~\cite{cellai2016multiplex}. For
instance,~\textcite{sagarra2015role} studied the canonical ensemble of
networks generated by the aggregation of multiplex networks, where
information on the connection between nodes is only accessible at the
aggregated level.

The purpose of models of growth, in which the number of nodes, links, but also
layers, may vary over time, is to find simple dynamical
mechanisms responsible for the emergence of structural properties (observed
in real-world multiplex networks) such as heterogenous degree distributions, 
positive and negative degree correlations~\cite{kim2013coevolution} and
community structures.
\newtext{A natural starting point is the single-layer preferential attachment rule~\cite{barabasi1999emergence}, where new nodes connect with probability proportional to degree, producing scale-free networks. In multiplex settings, the key question is how to define a node's ``attractiveness'' across layers. \textcite{nicosia2013growing} proposed an attachment rule where connection probability on a layer depends linearly on neighbors’ degrees across layers, later extended beyond two layers~\cite{momemi2015growing}. However, linear kernels only yield positive inter-layer correlations. To reproduce negative correlations, as seen in transportation systems, nonlinear kernels were introduced~\cite{nicosia2014nonlinear}, allowing high degree in one layer to reduce attractiveness in another.}
More sophisticated models are able to produce tunable
intra- and inter-layer community structure, for instance, when
considering triadic closure mechanisms~\cite{battiston2016emergence}.
Another class of growth models describes systems in which new layers
are sequentially added. The main purpose of these models is to explain the
fat-tail distributions of node and layer activity observed in
real-world systems~\cite{nicosia2015measuring}, the exponential
distribution of edge overlap in transportation
systems~\cite{santoro2018pareto}, or the scale-free degree
distribution observed on the aggregated networks when distinct
mesoscale structures are present at the different layers of the
system~\cite{criado2012mathematical}.

\newtext{Lastly, we note that stochastic block models (SBMs) provide a unified way to generate and recover multiplex community structure: they partition nodes into latent groups with statistically similar connectivity patterns and let edge tendencies depend on the groups involved \cite{holland1983stochastic,karrer2011stochastic}. In multiplex settings, this idea extends by specifying how group structure relates across layers—from a shared partition whose between-group interactions vary by layer to layer-specific partitions coupled by priors that encourage alignment without enforcing it.  
This flexibility allows SBMs to express not only assortative (positively correlated) communities but also disassortative (negatively correlated) patterns and core–periphery organization within the same framework. A crucial refinement for real data is degree correction, which separates heterogeneous node activity from genuine grouping and prevents hubs from being misidentified as communities \cite{ball2011efficient,peixoto2015inferring}. Model families span shared-partition, degree-corrected, mixed-membership formulations that plant the same groups across layers while letting each layer realize them differently \cite{de2017community,subhadeep2016consistent}; strata models that cluster layers and fit a common SBM within each stratum \cite{stanley2016clustering}; and fully flexible constructions that tune mesoscale structure arbitrarily across layers \cite{bazzi2020framework}. Additional realism comes from incorporating inter-layer edge correlations or broader layer interdependence beyond communities \cite{pamfil2020inference,aguiar2022factor}. On the inferential side, likelihood-based and Bayesian methods estimate group assignments and between-group interactions --- often with scalable variational or message-passing approximations --- while minimum-description-length and related evidence-based criteria select the number of groups and prevent against overfitting \cite{peixoto2015inferring}. In short,  SBMs furnish a generative–inferential toolkit that can plant multiplex communities (assortative, disassortative, core–periphery) and retrieve them from data, with layer dependencies made explicit and tunable.\\}

\newcommand{\change}[1]{{#1}}

\section{Percolation}
\label{sec:percolation}

\change{
Percolation theory studies how the macroscopic
connectedness of a network is affected by
the deletion of some of its microscopic
elements~\cite{stauffer1992introduction,  araujo2014recent}.
As being part of the same connected component is a necessary condition
for two nodes in a network to interact, percolation is key to understand
many other processes on networks, e.g., the spreading of diseases and opinions,
and applications of percolation theory to real-world problems are numerous
~\cite{albert2002statistical, li2021percolation}.

On a multiplex network, percolation is naturally interpreted as a dynamic process, where
an initial perturbation caused
by the microscopic failure of some of the network's individual components
propagates within the individual network layers
and/or across coupled layers potentially leading
to the macroscopic failure of the entire system~\cite{buldyrev2010catastrophic}.
Such a cascade of failures due to the
interplay between within- and cross-layer
interactions
gives rise to
a physics of
percolation which has no analogue when the process is studied in
single-layer networks.
}

The goal of this section is to
provide a concise overview of the main results for percolation
models on multiplex networks.
\change{Although they 
have been reviewed extensively
by~\textcite{boccaletti2014structure},
~\textcite{kivela2014multilayer},
and~\textcite{bianconi2018multilayer}, among others,
our overview will include some classical results as they are necessary to grasp
  the main features of the physics of percolation on multiplex networks. However, 
  we will dedicate a significant portion of the section to highlight recent literature.}

\subsection{Cascades of failures}
\label{perc1}

Percolation models assume the presence of an underlying network
structure where either nodes (site percolation) or edges (bond
percolation) are deleted according to a predefined
protocol~\cite{dorogovtsev2008critical,li2021percolation}.
\change{The most studied protocol is the one prescribed by the
  so-called classical or ordinary percolation model where the microscopic elements to be deleted are selected randomly with uniform probability $0 \leq 1 - p \leq 1$~\cite{cohen2000resilience,  newman2001random,  dorogovtsev2008critical}.
  Another popular deletion protocol is the one of targeted attacks, where the elements to be removed are selected according to some topological criterion, e.g., their centrality in the network~\cite{albert2000error, cohen2001breakdown}.}
\change{Once some elements of the network are deleted, the connectedness of the network is studied in terms of
nearest-neighbor non-deleted elements forming connected components or clusters.} Depending
on the size of such components, the system can be found in two
different phases: (i) the non-percolating phase, where the network is
fragmented in many non-extensive clusters; (ii) the percolating
regime, where a giant connected component encompasses a finite
fraction of the network. By switching from one phase to the other, the
network undergoes a percolation transition, whose characteristics
solely depend on
the structure of the network and/or
the protocol according to which microscopic elements are deleted from the network.

\change{
  On single-layer networks, percolation is a static process where the connectedness of the system is affected instantaneously by the failure of some its components. Models of percolation are static too, meaning that all failing elements are removed at the same time. 
  Kinetic models where edges and/or nodes are removed according to some dynamic protocol are sometimes used to enhance real-world interpretability/applicability of the theory and/or
facilitate its mathematical analysis~\cite{ben-naim2005kinetic, karrer2014percolation, dsouza2015anomalous}.}

\change{On multiplex networks, however, dynamics is a key feature of percolation, as the process describes the unfolding of failures within and across layers started from a given initial damage of the network's components. The protocol used to determine the initial failure of the network components corresponds to the specific percolation model under scrutiny.  Such a dynamic perspective of the percolation process on multiplex networks has been used since the seminal paper by~\textcite{buldyrev2010catastrophic}, where the authors introduced such a process with the goal of
  explaining the sudden electrical blackout that affected much of Italy in
  2003~\cite{rosato2008modelling} as a cascade of failures propagating in
  the interdependent infrastructures serving energy supply and communication.
}

The multiplex descriptor $\mathcal M = (\bm A, \bm C)$ of
Eq.~(\ref{multiplex2}) is perfectly suited for
describing
cascading failures in one-to-one interdependent
networks~\cite{buldyrev2010catastrophic, gao2012robustness}. 
Each interdependent network corresponds to a
layer of the multiplex, and one-to-one
interdependent nodes are uniquely identified
by their labels.
We remind that the topology of layer $\alpha$ is fully
encoded in the adjacency matrix $A\lay{\alpha}$.
The coupling matrix $C_i$
is used here to specify one-to-one interdependencies of node $i$ between pairs
of layers, i.e., $c_i\lay{\alpha \beta} = 1$ if such interdependency exists
among layers $\alpha$ and $\beta$, and $c_i\lay{\alpha \beta} = 0$
otherwise.  At the node level,
configurations of the system are denoted by 
$\mathcal S$, see Eq.(\ref{dyn_statemult}).
The state vector ${\bm \sigma_i}$ associated with node $i$
has $D=M$ components, each
corresponding to a layer of the multiplex. 
In particular, the state of node $i$ in layer $\alpha$
is  $\sigma_i\lay{\alpha} =1$ if the node belongs to the so-called
mutually connected giant component (MCGC) of the graph, whereas
$\sigma_i\lay{\alpha} =0$ otherwise.
\change{Mutual connectedness serves as a proxy for network function in an inter-dependent
 multiplex system; it replaces the notion of connectedness valid for single-layer networks.}
According to the formulation by~\textcite{son2012percolation},
the MCGC is defined in a self-consistent
manner so that the state of node $i$ in layer $\alpha$
is $\sigma_i\lay{\alpha}=1$ if:

\begin{itemize}
  
\item[(i)] node $i$ is connected on layer $\alpha$ to at least
  another node $j$ that is in the MCGC, i.e., $\exists  j$
  such that $a_{ij}\lay{\alpha} =  \sigma_j\lay{\alpha}=1$.
  
  \item[(ii)] node $i$ is simultaneously in the MCGC in all
    layers that are coupled with layer $\alpha$, i.e.,
    $\sigma_i\lay{\beta} = 1$, $\forall \, \beta$ such that
    $c_i\lay{\alpha \beta} = 1$.
    
\end{itemize}

\change{The dynamics of the cascading failure is started by setting} the initial conditions for
\change{the nodes $\sigma_i\lay{\alpha} =0, 1$} for $i =1, \ldots, N$ and $\alpha = 1,
\ldots, M$,
\change{as well as for the edges $a_{ij}\lay{\alpha} =0, 1$ for $i,j =1, \ldots, N$ and $\alpha = 1,
\ldots, M$,} 
and then implemented by iterating rules (i) and (ii) until
convergence. During the dynamics, only the change of state
$\sigma_i\lay{\alpha} = 1 \to \sigma_i\lay{\alpha} = 0$ is allowed,
i.e., a failed node can not recover. 
\change{The above-mentioned irreversibility for the state of the nodes
  makes the dynamic process very sensitive to the initial configuration of the dynamics,
  meaning that initial failures of one or more elements may
 trigger an avalanche of failures unfolding within and
 across the layers of the multiplex, respectively
 represented by rules (i) and (ii).
}
\change{In site-percolation models, the initial condition is such that
  $\sigma_i\lay{\alpha} =0$ for all nodes $i \in \mathcal{R}$
  and $\sigma_i\lay{\alpha} =1$ otherwise.
  Here, $\mathcal{R}$ represents the set of initially removed nodes.}
\change{
In bond percolation models, initial failures are implemented at the level
of the edges of the multiplex by setting $a_{ij}\lay{\alpha} = 0$ if the corresponding edge
in layer $\alpha$ is in the set $\mathcal{R}$ of edges to be initially removed~\cite{hackett2016bond, kryven2019bond}.}
\change{These initial edge deletions do not cause additional removal of edges, however, they may induce changes in the state of some nodes, eventually
  triggering a cascade of failures.}

\change{We stress that a specific percolation model consists in the protocol used to
  set the initial configuration of the dynamic percolation process, which corresponds to determining
  the composition of the set $\mathcal{R}$. Several percolation models considered
  for multiplex networks are the same as those used for single-layer networks,
  for example the ordinary model and the target attack protocol.
}
\change{Irrespective of the specific percolation model considered, }
the relative size of the MCGC, i.e., $P_\infty\lay{\alpha} = \frac{1}{N} \,
\sum_{i=1}^{N} \sigma_i\lay{\alpha}$, is
used to monitor the mutual connectedness of the layer $\alpha$ in the
multiplex.
In particular, the steady-state value of the relative size of the MCGC averaged over all layers,
namely $P_\infty = \frac{1}{M} \, \sum_{\alpha = 1}^M
P_\infty\lay{\alpha} $, represents a natural quantity to monitor the
extent of the cascade of failures.
$P_\infty$  is a function  of the topology of the multiplex $\mathcal M$,
and of the initial microscopic failures that are present in the
system.

We note that the self-consistent definition of the MCGC
is valid for all mutually
connected clusters that may be present in the
multiplex. Indeed, the MCGC is only the largest among
all mutually connected components of the
graphs~\cite{buldyrev2010catastrophic}.
At the same time, we note that, except for pathological cases, only one
MCGC may exist in a multiplex network therefore the
use of the binary state $\sigma_i\lay{\alpha} =0,1$ is sufficient
to distinguish the MCGC from all other microscopic clusters.
Eventual pathological cases may be treated by imposing the
initial condition $\sigma_i\lay{\alpha} =1$ only if node $i$ belongs to
the giant connected component of layer $\alpha$, and
$\sigma_i\lay{\alpha} =0$ otherwise.

\change{\subsection{\change{Message-passing approximation}}}
\label{perc1a}

As illustrated by several authors including
~\textcite{bianconi2014multiple}, 
an immediate
\change{approximation}
of the dynamical system that
describes the propagation of cascading failures in multiplex networks is obtained
by writing the following set of message-passing (MP) equations:

\begin{equation}
\eta_{i \to j}\lay{\alpha \alpha} = p_i\lay{\alpha} \, \prod_{\beta =
  1}^M \,    \left( \eta_{i \to i} \lay{\beta \alpha}
\right)^{c_{i}\lay{\alpha  \beta} } \, \left[  1 - \prod_{\substack{\ell =1 \\ \ell \neq j}}^N \,
  \left( 1 -  \eta_{\ell \to i}\lay{\alpha \alpha} \right)^{a_{\ell
      i}\lay{\alpha}} \right] \; ,
\label{eq:section3_message_passing_1}
\end{equation}
and
\begin{equation}
\eta_{i \to i}\lay{\alpha \beta} = p_i\lay{\alpha} \, \prod_{\substack{\gamma =1 \\ \gamma \neq \beta}}^M \,    \left( \eta_{i \to i} \lay{\gamma \alpha}
\right)^{c_{i}\lay{\alpha  \gamma} } \, \left[  1 - \prod_{\substack{\ell =1}}^N \,
  \left( 1 -  \eta_{\ell \to i}\lay{\alpha \alpha} \right)^{a_{\ell
      i}\lay{\alpha}} \right] \; .
\label{eq:section3_message_passing_2}
\end{equation}

In the above equations, we explicitly considered the case of
the site-percolation model. The binary variable $p_i\lay{\alpha}
$ serves to set the initial condition of the dynamical system, i.e., $p_i\lay{\alpha}
=0$ if node $i$ in layer $\alpha$ is initially failing, and $p_i\lay{\alpha}
=1$ otherwise.
\change{In the MP approximation, the values of the variables $p_i\lay{\alpha}$ define the specific percolation model at hand.
  For example, for the ordinary site-percolation model where a fraction $1-p$ of nodes is deleted
 uniformly at random, one sets $p_i\lay{\alpha} = 0$ with probability $1-p$ and $p_i\lay{\alpha} = 1$ otherwise, for all $i =1, \ldots, N$ and  $\alpha = 1,
\ldots, M$.
  }
Messages are used by pairs of connected/interdependent
nodes to
exchange dynamical information about the
propagation of the cascade of failures in the 
multiplex. The generic message $\eta_{i
  \to j}\lay{\alpha \beta}$ informs the copy of node $j$ in layer
$\beta$ if the copy of node $i$ in layer $\alpha$ 
currently belongs or not to the MCGC.  
 Messages have binary values: a
 message equal to one indicates that the sender surely belongs to
 the MCGC; by contrast, a message equal to zero indicates that the sender
 has been potentially involved in the cascade of failures.
 Given the details of the model, only two types of messages are
 effectively exchanged: (i)
within-layer messages exchanged between pairs of connected nodes on the same
layer, i.e.,  $\eta_{i \to j}\lay{\alpha \alpha}$ such that
$a_{ij}\lay{\alpha} =1$; (ii) cross-layer messages exchanged by 
nodes with identical labels that are coupled across the layers,
i.e., $\eta_{i \to i}\lay{\alpha \beta}$ such that $c_{i}\lay{\alpha
  \beta} =1$. We note that the two types of messages mirror the conditions at the
basis of the self-consistent definition of the MCGC. 
The dynamics of the within-layer messages is described in
Eq.~(\ref{eq:section3_message_passing_1}). Each edge $(i,j)$ of layer
$\alpha$ is associated with two distinct messages, i.e., $\eta_{i \to
  j}\lay{\alpha \alpha}$ and $\eta_{j \to i}\lay{\alpha \alpha}$,
depending on the direction of propagation of the message.
The value of the message is automatically equal to zero if node $i$ is such
that $p_i\lay{\alpha} =0$. Otherwise, the message is not zero if node $i$
is receiving at least a non-null message from another neighbor $\ell \neq j$, i.e.,
$ 1 - \prod_{\substack{\ell =1 \\ \ell \neq j}}^N \,
  \left( 1 -  \eta_{\ell \to i}\lay{\alpha \alpha} \right)^{a_{\ell
      i}\lay{\alpha}}$, in all layers which the node is coupled to,
  hence the product $ \prod_{\beta =
  1}^M \,    \left( \eta_{i \to i} \lay{\beta \alpha}
\right)^{c_{i}\lay{\alpha  \beta} }$.
The constraint $\ell \neq j$ is imposed to avoid messages to immediately
traverse the same edge in the opposite direction.
This constraint is rather standard in MP algorithms
developed for the solution of various graph problems~\cite{gallager1962lowdensity,
  mezard2009information, karrer2014percolation}.
Cross-layer messages are defined
in Eq.~(\ref{eq:section3_message_passing_2}) in a similar fashion as
within-layer messages. 
The only difference is that the layer receiving
the message is excluded from the product appearing  on
the right hand side of Eq.~(\ref{eq:section3_message_passing_2}).

\change{The MP approximation of the percolation process consists in starting from the initial messages
$\eta_{i \to j}\lay{\alpha \beta} =1$ for all $i, j = 1, \ldots, N$
and $\alpha, \beta = 1, \ldots, M$,} and iterating the MP equations until convergence to the steady state. At each stage of the iteration, it is possible to
estimate the state of the system using the system of equations
\begin{equation}
\sigma_i\lay{\alpha}= p_i\lay{\alpha} \, \prod_{\beta=1}^M \,    \left( \eta_{i \to i} \lay{\beta \alpha}
\right)^{c_{i}\lay{\alpha  \beta} } \, \left[  1 - \prod_{\substack{\ell =1}}^N \,
  \left( 1 -  \eta_{\ell \to i}\lay{\alpha \alpha} \right)^{a_{\ell
      i}\lay{\alpha}} \right] \; .
\label{eq:section3_message_passing_3}
\end{equation}

In essence, the copy of node $i$ in layer $\alpha$ is part of the MCGC
if it receives a positive message from at least one of its neighbor in
all layers which the node is coupled to.

The MP Eqs.~(\ref{eq:section3_message_passing_1}-\ref{eq:section3_message_passing_3})
rely on
two strong
assumptions about the topology of the multiplex:
(i) no edge overlap exists between network layers, and
(ii) the network topology of the individual layers is
locally tree-like~\cite{dorogovtsev2008critical}.
Under such assumptions, it is possible to
generalize the MP equations to multilayer networks, thus accounting for
arbitrary many-to-many interdependencies among
layers~\cite{boccaletti2014structure, bianconi2015mutually,
  shao2011cascade, gao2013percolation}.
Assumption (i) may be avoided either by
decomposing the links of the multiplex in overlapping vs. exclusive
links~\cite{radicchi2015percolation, min2015link}
or by developing MP equations 
for multilinks~\cite{cellai2016message,
  bianconi2016percolation}.
Whereas MP approaches that do not require the locally tree-like ansatz
for percolation in single-layer
networks exist~\cite{radicchi2016beyond, cantwell2019message},
we are not aware of similar attempts for percolation on multiplex networks.

\change{The MP approximation is useful to derive 
  analytically some properties of the percolation process on multiplex
  networks~\cite{boccaletti2014structure}.
This is already apparent in the simplest scenario
of the ordinary percolation model.}

\bigskip
\subsection{Ordinary percolation}
\label{perc2}

\change{As for the case of single-layer networks, the most studied percolation model on multiplex networks is the one of ordinary site percolation, where each node is deleted with probability $0 \leq 1 - p \leq 1$.} For simplicity, let us focus on
multiplex networks composed of only $M=2$ layers, namely $\alpha$ and $\beta$. We also assume that the coupling between layers is
such that $c_i\lay{\alpha \beta}= 1$ , $\forall \, i = 1, \ldots,
N$. 
\change{This is the setting that was considered in the seminal paper by ~\textcite{buldyrev2010catastrophic}.
  We stress that in such a setting there is still freedom to pick arbitrary network structures for defining the two layers of the multiplex.
\textcite{buldyrev2010catastrophic} considered
multiplex networks whose layers are random graphs generated according to the standard
configuration model~\cite{molloy1995critical}.  Those networks are
further assumed to be generated independently one from the other. Although its apparent simplicity, the setting
studied by~\textcite{buldyrev2010catastrophic} captures the main physics of percolation on multiplex networks.
}

The finite-size scaling analysis performed by~\textcite{buldyrev2010catastrophic}
on multiplex formed by Poisson random graphs with average degree
$\langle k \rangle$ reveals that the model undegoes a discontinuous
percolation transition for $p_c \langle k \rangle \simeq 2.45541 $.
We remind that the ordinary percolation model applied to 
single-layer Poisson graphs displays a continuous phase transition
at the critical threshold $ p_c  \, \langle k \rangle  = 1$~\cite{molloy1995critical,
 cohen2000resilience, newman2001random}.
\textcite{buldyrev2010catastrophic} further show that,
if the degree distribution of the layers is a power law, the
transition is discontinuous at $p_c >0$, at
odds with ordinary percolation in single-layer SF graphs which exhibits
a continuous phase transition at
$ p_c = 0$~\cite{molloy1995critical,
 cohen2000resilience, newman2001random}.

\begin{figure}[t!]
  \begin{center}
        \includegraphics[width=0.55\textwidth]{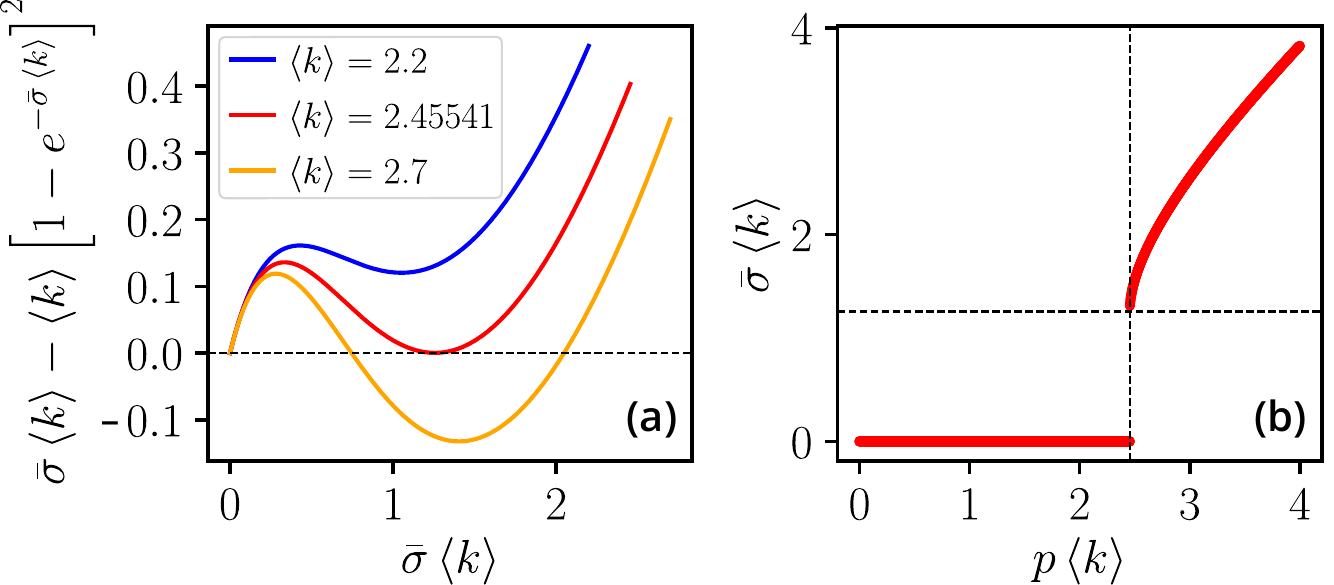}
        \caption[]{Theoretical prediction of the percolation phase
                diagram of synthetic multiplex networks. 
            The multiplex is composed of
            two uncorrelated Poisson network layers, and
            individual nodes are randomly deleted with
            uniform probability $1-p$, as considered
            by~\textcite{buldyrev2010catastrophic} and ~\textcite{son2012percolation},
            among others.
            (a) Graphical solutions of
            Eq.~(\ref{eq:section3_message_passing_rnd}) are obtained
            by setting $p=1$ and using the
            known generating function of the Poisson distribution. Different curves
            correspond to different choices of the average degree $\langle k
            \rangle$. All points of intersection of the curves with
            the horizontal dashed line represent mathematical solutions of Eq.~(\ref{eq:section3_message_passing_rnd}).
            However, the solution with physical meaning for the percolation
            problem is the one with the largest magnitude.
            (b) Relative size of mutually connected giant component
            as a function of $p$. Both quantities are
            multiplied by $\langle k \rangle$. The vertical dashed line identifies the percolation threshold
            $  p_c \, \langle k \rangle  =  2.45541\ldots \,$. The horizontal dashed line is placed at
            $\bar{\sigma}  \, \langle k  \rangle = 1.25643\dots$  and corresponds to
            the magnitude
            of the discontinuous jump of the MCGC at criticality.
            \label{fig:section3_figure1}}
    \end{center}
\end{figure}

A theoretical explanation of the numerical findings
can be obtained directly from
the
MP Eqs.~(\ref{eq:section3_message_passing_1}-\ref{eq:section3_message_passing_3})
under the reasonable
assumption of sparsity.
When layers are
generated independently one from the other, 
both the hypotheses of locally tree-like structure and absence of
overlap between multiplex layers at the basis of the MP equations
are satisfied to a very good extent. A great simplification to the MP
equations arises from replacing all binary
variables by their expectation values over the ensemble
of random graphs with prescribed degree distributions, and over the
ensemble of initial random failures for the individual nodes. As a
matter of fact, the
entire system of MP equations reduces to
\begin{equation}
\bar{\sigma} = p \,  \left[ 1 - G_0 (1 - \bar{\sigma}) \right]^2 \; ,
\label{eq:section3_message_passing_rnd}
\end{equation}
which differs from the analogous equation for ordinary percolation in
single-layer random networks only for the presence of a square on
the right hand side~\cite{cohen2000resilience, newman2001random}.
Here, $\bar{\sigma}$ is the expectation value of the state of a
randomly chosen node. No explicit dependence on the labels of the multiplex layers
appears in the equation as the two layers are assumed to be
generated according to the same configuration model.
We note that the relative size of the MCGC is
simply $P_\infty = \bar{\sigma}$ so that the robustness of the
multiplex is directly studied by monitoring how $\bar{\sigma}$  changes
as a function of $p$. In Eq.~(\ref{eq:section3_message_passing_rnd}),
$G_0 (x) = \sum_{k} P(k) \, x^k$ is the generating function of the
degree distribution of the network
layers~\cite{newman2001random}.
For a Poisson
degree distribution,
we have
$G_0(1 - \bar{\sigma}) = (1 - e^{- \bar{\sigma} \, \langle k \rangle })$, and 
Eq.~(\ref{eq:section3_message_passing_rnd}) 
can be rewritten as $ \bar{\sigma}  =
p\, (1 - e^{- \langle k \rangle \,
  \bar{\sigma}})^2$. 
The equation has a trivial solution
$\bar{\sigma}=0$ corresponding to the non-percolating phase
of the multiplex; non-trivial solutions, denoting the percolating
phase, can be
obtained graphically as shown in Fig.~\ref{fig:section3_figure1}.
As long as $p \, \langle k \rangle <
2.45541\ldots$, the equation displays only the trivial solution.
At the critical point $p_c \, \langle k \rangle  =
2.45541\ldots$, the equation develops the non-trivial solution
$\bar{\sigma} \, \langle k \rangle = 1.25643\ldots$, indicating
that the system exhibits a discontinuous phase transition.
The transition is hybrid in nature, having a discontinuity, but
exhibiting critical behavior, only above the transition,
like a continuous transition~\cite{baxter2012avalanche}.

Eq.~(\ref{eq:section3_message_passing_rnd}) can be
readily generalized to deal with an arbitrary number of
layers~\cite{gao2011robustness, gao2012networks, yagan2012analysis}
as well as to deal with layers obeying different degree
distributions~\cite{boccaletti2014structure, buldyrev2010catastrophic,
  son2012percolation, baxter2012avalanche}.

To describe partially one-to-one interdependent networks one can assume
that the generic entry of the coupling matrix is
$c_i\lay{\alpha \beta} = 1$ with probability $q$, and $c_i\lay{\alpha
  \beta} = 0$ otherwise ~\cite{parshani2010interdependent,
  bashan2011percolation, hu2011percolation,
  son2012percolation}. Eq.~(\ref{eq:section3_message_passing_rnd}) 
becomes $\bar{\sigma} = p \,  \left[ 1 - G_0 (1 - q\, \bar{\sigma})
\right] \left[ 1 - G_0 (1 - \bar{\sigma}) \right]$, which 
reduces to the previous case for $q=1$. For low $q$ values,
thus when the
coupling between layers is sufficiently weak, the multiplex acts as it
is composed of two non-interdependent network layers, and the
percolation transition is continuous;  the model displays a
tricritical point
where the percolation
transition changes its nature from continuous to discontinuous.
If the multiplex network is composed of more than $M=2$ layers, then
different types of partial interdependencies can be
considered, and the scenario becomes even richer displaying
multiple transition points~\cite{bianconi2014multiple}.

Structural correlation among 
multiplex layers
generally
leads to systems that are more robust to the random deletion of nodes
than their uncorrelated counterparts.
An increase of robustness means that the network can tolerate
a larger number of node failures before collapsing. Also,
the actual value of the discontinuous jump is lower compared to the
one observed in the uncorrelated case.
Different types of structural correlations can be considered.
Some authors study correlation of the
degree, or other centrality metrics, between
interdependent nodes~\cite{watanabe2014cavity,
  buldyrev2011interdependent, parshani2010intersimilarity,
  lee2012correlated, valdez2013triple}.
Other papers focus on the edge overlap among network layers
~\cite{min2014network, min2015link,
  cellai2016message, hu2013percolation, li2013critical}.

\begin{figure}
  \begin{center}
   \includegraphics[width=0.6\textwidth]{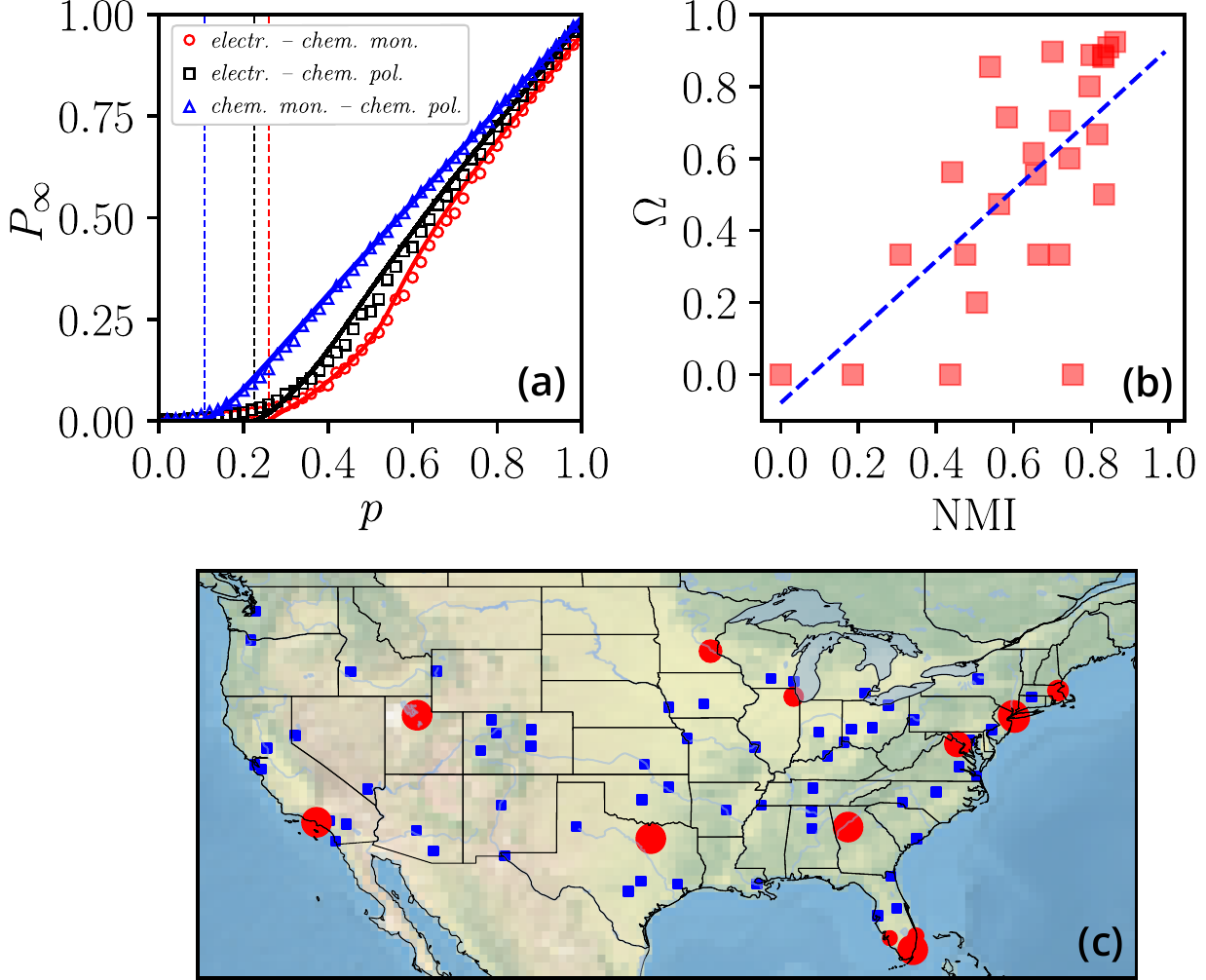}
\caption[]{Percolation 
    on real-world multiplex networks.
  (a) Ordinary percolation on the multiplex network of the {\it
    C. Elegans} connectome~\cite{dedomenico2015muxviz}. The analysis is performed on
  the three different duplex networks that can be constructed
  depending on the flavor of the interactions among neurons.  The relative size
  of the MCGC is computed via numerical simulations (large
  symbols)  and MP equations (small symbols). The vertical lines
  are located at $p=1/\mu_I$, with $\mu_I$ largest eigenvalue of the
  non-backtracking matrix of the
  graph composed of edges shared by both
  layers of the duplex.
  (b) Targeted attacks on real multiplex networks. The robustness
  $\Omega$ of a real multiplex, as defined by~\textcite{kleineberg2017geometric}, is well
  predicted by the similarity of the community
  structure, identified
  with the InfoMap algorithm~\cite{rosvall2008maps}, of the network
  layers and measured in terms of normalized
  mutual information (NMI)~\cite{danon2005comparing}. The dashed line
  corresponds to the best linear fit (correlation coefficient $r = 0.68$).
  Results reproduced from~\textcite{faqeeh2018characterizing}.
  (c) Optimal percolation on the multiplex network of US domestic
  flights operated in January 2014 by American Airlines and Delta. Red
  circles represent airports that
   \change{are part of solutions to the optimal percolation
    problem, i.e., if removed, they lead to the collapse of the network.
    Note that multiple solutions exist, meaning that
    different sets of removed nodes lead to network collapse.
  }
  The size of each circle is proportional to
  \change{number of solutions the corresponding node is part of.} 
  All other airports in the multiplex are
  represented as blue squares.
  Figures adapted from~\textcite{osat2017optimal}.
 \label{fig:section3_figure2}}
\end{center}
\end{figure}

When applied to real-world
multiplex networks, the ordinary percolation model generally undergoes
a smooth transition~\cite{radicchi2015percolation, bianconi2016percolation}.
Rigorously speaking no phase transition can be observed in a real,
thus finite, network~\cite{coghi2018controlling}.
\change{However, empirical results from numerical
simulations do not highlight any abrupt behaviour in the percolation
diagram [Fig.~\ref{fig:section3_figure2}(a)].}
\change{The finding can be understood by first classifying the
  edges of a real multiplex network with two layers into three mutually exclusive sets:
  one set is given by the edges shared by both layers, and the other two sets contain layer-exclusive edges.
  One can then realize that the mutual connectedness in the multiplex network
  is effectively determined
  by the standard connectdness of the single-layer graph composed of shared edges only, hence the emergence of a smooth
  percolation transition.}
The MP approach is able to accurately reproduce the results
of numerical simulations; it
further allows to approximate the transition point of the smooth
percolation transition observed in real-world multiplex networks as
$p_c \simeq 1 / \mu_I$, with $\mu_I$ largest eigenvalue of the
non-backtracking matrix \change{of the
graph composed of edges shared by both
  layers of the multiplex~\cite{radicchi2015percolation}.}

\change{So far, we focused our attention on the average behavior, over a large number of realizations,
  of the ordinary percolation model. We neglected, however, potential changes in the behavior of the
  model from realization to realization. Large fluctuations around the average are in fact possible due to the extreme
  sensitivity of the model depending on the specific initial configuration of the damaged nodes. Neglecting
the presence of such fluctuations is}
a serious shortcoming in the
  assessment of risk given the abruptness of the transition and
  the fact that real-world
  multiplex networks considered in this context have generally
  a small size.   Only a few papers consider explicitly this issue.
  ~\textcite{bianconi2019large} develops a MP approach able to
 capture the large deviation of percolation in interdependent
 multiplex networks with a locally tree-like structure.~\textcite{coghi2018controlling}
 introduce a metric, named safeguard centrality, able to single out
 the nodes that control the response of the entire multiplex network
 to random damage. Safeguarding the function
 of top-scoring nodes is  sufficient to prevent system collapse.

 As far as it concerns strategies for improving
 robustness or facilitate recovery, it is known that
 reinforcing a fraction of nodes
 \change{, so that they can not be removed,}
 may help
 to mitigate the abruptness of the percolation
 transition~\cite{yuan2017eradicating}, and that
simple heuristic strategies may be used to allow system recovery
during  and/or after collapses~\cite{dimuro2016recovery, danziger2020recovery}.

\subsection{Targeted attacks and optimal percolation}
\label{perc3}

Several percolation models are based on deletion protocols that are
correlated with the structure
of the network where they are applied
to. For example in targeted attacks, nodes are preferentially
deleted on the basis of the value of a network
centrality metric, e.g., node degree~\cite{albert2000error, cohen2001breakdown}.
The model shows that heterogenous networks,
whose connectedness heavily relies on hubs, 
can be easily dismantled by the removal of a small portion of their
most central nodes. On multiplex networks, targeted attacks 
  can be performed in different ways, depending on how
  centrality metrics in the various layers are combined together to
  define the order of deletion of the individual nodes.

  ~\textcite{huang2011robustness} study the problem on
  multiplexes composed of two random uncorrelated network layers
  where nodes are deleted according to their degree centrality
  as measured in only one of the layers. Studies about targeted
  attacks on random multiplex networks
  concern also arbitrary number of layers~\cite{dong2013robustness},
  structural correlation among network layers~\cite{wang2019targeted}, and 
  partially interdependency~\cite{zhou2020dependency}.
  
~\textcite{kleineberg2017geometric} consider targeted attacks on
real-world multiplex networks composed on two
layers, namely $\alpha$ and $\beta$, where the centrality
of each node $i$ is $K_i = \max \{
k_i\lay{\alpha}, k_i\lay{\beta} \}$, and the percolation protocol
prescribes nodes to be deleted, in descending order,
according to this metric. Eventual ties are randomly broken.
They analyze the behavior of the model in many real-world
networks, noting that network collapses are milder
than those observed on random networks with
degree sequences identical to those of the real multiplexes.
They leverage the empirical finding to define a metric of robustness named
$\Omega$, and show that $\Omega$ positively correlates with the
geometric similarity between the layers of the
multiplex~\cite{kleineberg2016hidden}. The finding holds even in
the case of vanishing edge overlap between network layers, meaning
that higher-order structural correlations 
mitigate the vulnerability of multiplex networks under targeted attacks.
The finding is confirmed by~\textcite{faqeeh2018characterizing} who
quantify higher-order structural correlations of network layers in
terms of similarity among their community structure
[Fig.~\ref{fig:section3_figure2}(b)]. By means of
finite-size-analyses performed on
multiplex networks composed of instances of the
Lancichinetti-Fortunato-Radicchi (LFR)
benchmark graphs~\cite{lancichinetti2008benchmark},
~\textcite{faqeeh2018characterizing} further show that strong and
correlated modular structure between
the layers of a multiplex leads to a smooth percolation
transition; if the community structure of the layers is not correlated, or if
communities are too fuzzy, then the transition is abrupt.

The spirit of the model for targeted attacks is extremized by
the so-called optimal percolation problem, i.e., the
NP-hard problem aimed at finding 
the minimal set of nodes whose removal from a network
fragments the system into non-extensive
disconnected clusters~\cite{morone2015influence}.
\textcite{osat2017optimal} generalize the problem of optimal percolation
to multiplex networks by 
adapting approximate algorithms for solving the problem,
as for example those developed by~\textcite{morone2015influence}, ~\textcite{clusella2016immunization},
and ~\textcite{braunstein2016network}, 
from single-layer to multiplex networks, and by characterizing
solutions of the problem on both synthetic and real multiplex networks
[Fig.~\ref{fig:section3_figure2}(c)].
\change{
  \textcite{osat2017optimal} systematically compare solutions to the optimal percolation problem
  that can be obtained by solving the problem on multiplex networks or their single-layer aggregated versions. The main
  finding is that neglecting the ground-truth multiplex nature of a network may result in signiﬁcant inaccuracies about its
robustness: a multiplex network can be fragmented by removing considerably smaller sets of nodes than its single-layer based network
representation; the error committed when relying on single-layer
representations of the multiplex does not regard only the number of nodes, but also the identity of those nodes.
}
In a follow-up study, 
~\textcite{baxter2018targeted} show that the optimization problem can be
effectively solved with an heuristic based on network decycling,
similar to the one already considered by~\textcite{zdeborova2016fast} in
single-layer networks. Also,
~\textcite{santoro2020optimal}
propose novel heuristic methods for multiplex dismantling
able to outperform all methods existing on the market in case of
multiplexes displaying strong correlation in the structure of
their network layers.

\subsection{\change{Alternative percolation processes}}
\label{perc4}

\change{This final subsection is
devoted to  brief descriptions of variants of the percolation process that are intended to address real-world  scenarios where the classical version of the process falls short.}

Geographical embedding
is a quite relevant, and basically universal, feature of real-world
critical infrastructures, and a series of papers consider percolation
on spatially embedded multiplex networks, \change{where both the within and cross-layer propagations of the cascade of failure may have a space dependence}. 
Whether or not the ordinary percolation model displays an abrupt
transition on space-embedded multiplex models
is a topic of debate of some early publications on the topic
~\cite{son2011percolation, berezin2013comment, son2013reply}.
The conundrum is clarified in some follow-up papers.
\change{If interdependent nodes are randomly
selected without accounting for spatial constraints, lattice networks
display a mixed-order percolation transition,
where the MCGC 
is characterized by a discontinuity, but also
  self-similar fluctuctions and a well-defined fractal dimension~\cite{bashan2013extreme, gross2022fractal}.
}
However, if interdependent nodes
are sufficiently close in space, then the
percolation transition is continuous~\cite{li2012cascading,
  danziger2014percolation, danziger2016effect,shekhtman2014robustness}.
Recent efforts focus also on space-based strategies for
targeted attacks and network recovery,
e.g.,~\textcite{berezin2015localized} and ~\textcite{stippinger2014enhancing}.

In all percolation models based on
mutual connectedness, two nodes are part
of the same mutually connected cluster if all copies
of these two nodes in the various interdependent layers
are also part of such a cluster.
Hence,
adding a new layer to an existing multiplex can only
decrease its robustness. The scenario is not realistic for
infrastructures where instead the addition of
new layers is performed with the purpose of
improving system robustness. For example, the function of
a multimodal transportation network
should benefit from the addition of a new mode of
transportation.  In their redundant percolation process,
\textcite{radicchi2017redundant} address this issue by
redefining the mutual connectedness of two
nodes as a condition that must be satisfied in at least
two interdependent layers. The process is identical to the standard one
by~\textcite{buldyrev2010catastrophic} for $M=2$ layers, but for $M>2$ it
correctly describes a scenario where redundant interdependencies among layers
boost system robustness. The authors develop
a MP theory that is able to characterize redundant in both
synthetic and real-world multiplex graphs.

A $\bm k$-core, with $\bm k = \{k\lay{1}, \ldots, k\lay{M}\}$, in a multiplex
network is defined as the maximal set of nodes such that each node complies
with the corresponding degree threshold condition
in each layer of the multiplex~\cite{azimi-tafreshi2014core}.
Cores are identified with a iterative procedure where all nodes with
degree below threshold are removed, and the degree of the remaining
nodes is recomputed discounting for the eventual removal of their
neighbors. The procedure is repeated until stable $\bm k$-cores are identified.
$\bm k$-cores are important in the study of spreading processes taking
place on networks, as for example in the identification of influential
spreaders~\cite{kitsak2010identification}
\change{and the emergence of self-sustained spreading localized
  in densely connected portions of networks~\cite{pastor-satorras2018eigenvector}.}
The emergence of a macroscopic
$\bm k$-core structure in random uncorrelated multiplexes is considered by
\textcite{azimi-tafreshi2014core}. The study is extended to
correlated graphs by~\textcite{shang2020generalized}. Finally, 
~\textcite{osat2020core} consider  $\bm k$-core percolation on real multiplex networks. In
particular, they show that the robustness of $\bm k$-core structure
is positively correlated with the geometric similarity of the layers
forming the multiplex.

The notion of viability replaces mutual connectedness in the study
by~\textcite{min2014multiple}. \change{This serves to describe scenarios where} some special
nodes provide resources essential for the function of the other nodes.
A node is said viable if it can reach, via
other viable nodes, to a resource node in each and every layer, and
\change{the viability of a multiplex network is given by the fraction of
  its nodes that are viable.}
Viability is studied by monitoring
cascades of activations or deactivations
\change{that are triggered by removing a fraction $1-p$ of edges
selected uniformly at random from the multiplex. By varying $p$, }
viability
is characterized by discontinuity, bistability,
and hysteresis. A similar process, characterized by analogous features, is
referred as weak or bootstrap percolation by ~\textcite{baxter2014weak}.

The observability process is a variant of percolation
that finds its motivation in the study of some dynamical processes
where the state of the system can be determined
by monitoring or dominating the states of a limited number of nodes in
the network~\cite{liu2011controllability, yang2012network}. The process
is extended to multiplex networks via the definition of so-called
mutually observable clusters by~\textcite{osat2018observability}.
\change{The paper specifically focuses on the case when observable nodes are selected randomly at uniform, as
for the ordinary model; the paper includes the development of a MP
theoretical framework and the analysis of several real multiplex
networks.}

Finally, antagonistic instead of synergistic (a.k.a. interdependent)
interactions are considered by~\textcite{zhao2013percolation}.
    In this percolation process on multiplex networks, the function of a
    node is incompatible with the function of its antagonistic node.
    Percolation can occur only in one layer at time, depending on the initial failure in
  the system. The corresponding MP description of the process
  is characterized by bistable solutions, and the
  percolation diagram displays a hysteresis loop.
  A similar phenomenology is observed in dynamical systems
  on multiplex networks with competing interactions by~\textcite{danziger2019dynamic}.
  



\section{Diffusion and reactions} 
\label{sec:diffusion}

In this section we focus on dynamical behaviors emerging when
a set of agents move over a multiplex network or, in more complex
situations, move over the network and interact at the nodes of the
network. Depending on the information available to the agents, the
movement can range from simple diffusion or unbiased random walks, to
biased random walks or navigation in search of optimal paths
\cite{blanchard2011random}.  Agents can be of the same or of different
types (species, families), while the node state variable
$\sigma_i\lay{\alpha}(t)$ can either represent their concentration, their
number, or the probability of finding them at node $i$ at layer
$\alpha$. The layers of the network can account for the
diverse mobility patterns of different species, or for various
channels of communication or transportation modes.  
The agents can move from a layer to another when this can be
beneficial to them according to a given utility function, for instance
a transport cost. 
Additionally, the situation can be made richer when agents are allowed to
interact at the nodes of the network. Interactions can be of different nature,
and mimic chemical or biological nonlinear (for instance prey-predator)
reactions, but can also describe the tendency for agents to avoid each
others to prevent congestion in multimodal transportation systems.

\subsection{\vito{Diffusion}}
\label{diff1}

Diffusion is the physical process by which atoms and molecules move
from regions of high concentration to regions of low concentration.
In a diffusive process on a graph, $\sigma_i(t)$ denotes the
concentration of the quantity of interest at node $i$ at time $t$, and
the flow from a neighbouring node $j$ to node $i$ is proportional to the
difference in concentrations $\sigma_j(t) - \sigma_i(t)$.
We can then write the rate of change of $\sigma_i(t)$ as:   
\begin{equation}
  \frac{ d \sigma_i}{dt} =
   D \sum_{j=1}^N  a_{ji} \left( \sigma_j -  \sigma_i  \right) =
  - D \sum_{j=1}^N  l_{ji} \sigma_j                
\label{eq:diff} 
\end{equation}
where $D$ is the so-called diffusion constant, $a_{ij} $ are the
entries of the adjacency matrix $A$ of the graph, and $l_{ij} = k_i
\delta_{ij} - a_{ij} $ are the entries of the
      Laplacian matrix $L= K - A$
      (where $K$ is the diagonal matrix of node degrees). 
%
%
%
Eq.~(\ref{eq:diff}) is the graph analogous of the standard diffusion
equation for a gas (or of the heat equation) in the Euclidean space.
Notice that this equation is linear, but as we will see in the
following sections, it can also be a good approximation for different
types of non-linear dynamical processes, such as synchronization.
%
%

\vito{The most general} way to extend the diffusion equation to a
multiplex network is to consider the state ${ \bm \sigma}_i = {\bm
  \sigma}_i(t)$ of node $i$, with $i=1,2,\ldots N$, as a vector with
$M$ components $\sigma_i\lay{\alpha}$, representing the concentrations
at layer $\alpha$, with $\alpha=1,2,\ldots M$. \vito{We can then
write~\cite{gomez2013diffusion,buldu2018frequency}:}
\begin{equation}
{\small \frac{ d \sigma_i\lay{\alpha}}{dt} = D\laysup{\alpha}
  \sum_{j=1}^N a_{ji}\lay{\alpha} \left( \sigma_j\lay{\alpha} -
  \sigma_i\lay{\alpha} \right) + \sum_{\beta=1}^M D\laysup{\beta \alpha}_{\vito{i}}
  \vito{c_i\lay{\beta \alpha}}
  \left( \sigma_i\lay{\beta} - \sigma_i\lay{\alpha} \right) }
\normalsize
\label{eq:diff_A}
\end{equation}
where the first term on the right hand side accounts for  
the intralayer diffusion, while the second term accounts 
for the diffusion between layers.
\vito{The topology of each layer $\alpha$ is fully
encoded in the adjacency matrix $A\lay{\alpha}$, while the 
matrix $\mathrm{C}_i=\{c_{i}\lay{\alpha \beta}\}$
describes the topology of the coupling among layers at node $i$.} 
$D\laysup{\alpha}$ is the diffusion constant in layer $\alpha$, while  
$D\laysup{\alpha \beta}_{\vito{i}}$
denotes the interlayer diffusion constant 
  between layer $\alpha$ and layer $\beta$, \vito{which in general can differ
    from node to node.} 
\vito{ Eqs.~(\ref{eq:diff_A}) indicate
  that the concentration} at node $i$ of layer $\alpha$ depends on
the \vito{concentrations} of all 
nodes $j$ connected to $i$ at layer $\alpha$, but also on
the \vito{concentrations} at the same node $i$ at the other layers
$\beta$, with $\beta \neq \alpha$. 
\\  
In the case of two layers only, we have two
equations for each node, namely: 
\begin{equation}
   {\small
   \left\{  \begin{array}{c}
  { d \sigma_i\lay{1}}/{dt} =   D\laysup{1}   \sum_{j=1}^N  a_{ji}\lay{1} \left( \sigma_j\lay{1} -  \sigma_i\lay{1}  \right)  

  +        D_{\vito{i}} 
 ~ \vito{c_i\lay{12}} \left( \sigma_i\lay{2} -  \sigma_i\lay{1}  \right)  
\\
\\
 { d \sigma_i\lay{2}}/{dt} =   D\laysup{2}  \sum_{j=1}^N  a_{ji}\lay{2} \left( \sigma_j\lay{2} -  \sigma_i\lay{2}  \right)  
 +       D_{\vito{i}}
~ \vito{c_i\lay{1 2}} \left( \sigma_i\lay{1} -  \sigma_i\lay{2}  \right)  
\end{array}
\right.
}\normalsize
\label{eq:diff_A2}
\end{equation}
where $i=1,2,\ldots N$.
Notice that $A\lay{1}= \{ a\lay{1}_{ij}\}$, $A\lay{2}= \{ a\lay{2}_{ij}\}$, $D\laysup{1}$ and $D\laysup{2}$
are respectively the adjacency matrices and the diffusion constants
of the two layers, and, for simplicity, we have
set \vito{$c_i\lay{12}= c_i \lay{21}$ and} 
$D\laysup{1 2}_{\vito{i}} =  D\laysup{2 1}_{\vito{i}} =   D_{\vito{i}} ~ {\vito{\forall i}}   $.  
These equations can be rewritten 
in the form of Eq.~(\ref{eq:multev_cont}) in terms of the $2N$-dimensional (column) vector
%
%
%
$\mathcal S = ( \sigma\lay{1}_1, \sigma\lay{1}_2 \ldots, \sigma\lay{1}_N,
\sigma\lay{2}_1, \sigma\lay{2}_2 \ldots, \sigma\lay{2}_N)^{T}$ as:
\begin{equation}
    \frac{d {\mathcal S}} {dt}  =  - \cal{L} S       
\label{eq:difvect}
\end{equation}
where matrix $\cal L$ is the so-called supra-Laplacian:
\begin{equation}
     \cal{L}  =\left( \begin{array}{cc}
D\laysup{1} L\lay{1} + \vito{W\lay{12}}  & - \vito{W\lay{12}}
       \\
       \\
-\vito{W\lay{12}}  & D\laysup{2} L\lay{2} + \vito{W\lay{12}}
       \\
    \end{array}
    \right)
\label{eq:supraL2}
\end{equation}
where $L\lay{1}$ and $L\lay{2}$ are the Laplacian matrices of the two
layers, \vito{and
  $W\lay{12}= \mathrm{diag}(c_1\lay{12} D_1, c_2\lay{12} D_2,\ldots c_N\lay{12}D_N)$
  is a diagonal matrix whose entries contain information about the existence,
  $c_i\lay{12}$, and weights, $D_i$, of the 
  interlayer links at each node. 
}

\begin{figure}[t]
\begin{center}
\centering
\includegraphics[width=0.6\textwidth]{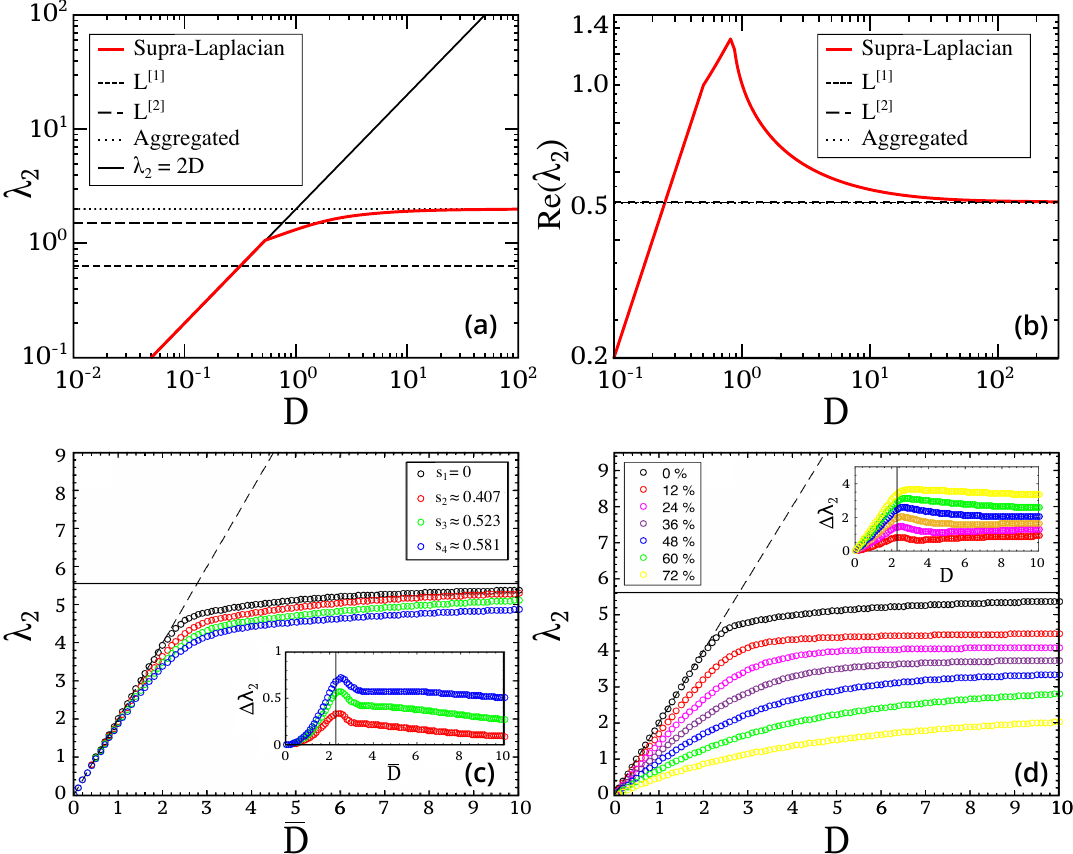}
\caption[]{
  The second smallest eigenvalue $\lambda_2$ of the supra-Laplacian
   matrix $\cal L$ is compared to the second smallest eigenvalues
   of the two layers and of the aggregate for an undirected (a) and a
   directed (b) multiplex network. In both cases, $D\laysup{1} = D\laysup{2}=1$, 
   \vito{$D_i\laysup{1 2} = D_i\laysup{2 1} =
     D$ and $c_i \lay{12}= c_i \lay{21} =1 ~\forall i$},
   so that there is one tuning parameter only, namely the
   interlayer diffusion constant $D$.  
   \vito{Consequences of heterogeneity and of missing interlayer
     links: (c) $\lambda_2$ is plotted as a function of the average
     interlayer diffusion constant $\bar{D}$, for different values of
     the standard deviation $s$;
     (d) $\lambda_2$ is plotted 
     as a function of $D$ for different percentages of missing
     interlayer edges in the case in which the interlayer diffusion constants
     are equal for all nodes, but not all the
     interlayer edges are present.}
     Figures adapted from
     \textcite{gomez2013diffusion}, \textcite{tejedor2018diffusion}, 
     \vito{and \textcite{buldu2018frequency}}.     
   } 
 \label{fig:diffusion_diff}
 \end{center}
\end{figure}

\textcite{gomez2013diffusion} have \vito{considered the particular
  case in which all the interlayer links are present ($c_i\lay{1 2} =
  1 ~\forall i$) and diffuse with the same constant ($D_i = D~\forall
  i$), so that the matrix $W\lay{12}$ in Eq.~(\ref{eq:supraL2}) 
reads $W\lay{12}= D I $, where $I$ is the $N
  \times N$ identity matrix.  They have} used a perturbative approach
to study the spectral properties of  
the supra-Laplacian matrix 
$\cal{L}$ of a multiplex \vito{with two undirected layers}  
in terms of eigenvalues and eigenvectors of $L\lay{1}$ and $L\lay{2}$. 
Fig.~\ref{fig:diffusion_diff}(a) shows the 
second smallest eigenvalue $\lambda_2$ of $\cal L$ as a function of
the interlayer diffusion constant $D$. 
%
%
This eigenvalue, also known as algebraic connectivity, determines  
the diffusion relaxation time scale 
$\tau$, namely $\tau = 1 /\lambda_2$. 
We observe a monotonic increase of $\lambda_2$ as a function of 
$D$, with two regimes of qualitatively distinct dynamics. 
\vito{When $D \ll D^*$, the algebraic connectivity $\lambda_2$ 
follows the linear relation $\lambda_2 = 2D$, while for $D \gg D^*$
the value of $\lambda_2$ approximates that of the aggregated network.
The striking feature} is the emergence \vito{at $D^*$} 
of {\em multiplex \vito{fast} diffusion}, 
where diffusive processes in the multiplex are faster than in any
of its individual layers. 
%
\textcite{sole2013spectral} have extended this result to the case of $M>2$
layers and also derived analytical expressions for the full
spectrum of eigenvalues of the supra-Laplacian. 
The dynamics of a diffusion process is reacher when the multiplex
network is directed. \textcite{tejedor2018diffusion}  
have in fact discovered that, when at least one layer consists of a directed
graph, the relaxation time is a nonmonotonic function of the interlayer
diffusion constant $D$ \vito{as shown in 
  Fig.~\ref{fig:diffusion_diff}(b)}. 
The position of the maximum of $\lambda_2$ indicates
that the diffusion in directed multiplex networks is faster 
at an intermediate value of $D$, which depends on structure of
the system.

\vito{\textcite{buldu2018frequency} have investigated how the speed of the
diffusion is affected by the interlayer topology
and the heterogeneity in the diffusion constants,
an issue that can be quite relevant for applications,
e.g. in neuroscience. 
They have first considered the case where all the interlayer links
are present ($c_i\lay{1 2} =
  1 ~\forall i$), but the diffusion constants vary from node to node.    
Namely, $D_i = \bar{D} h_i$, where the quantities $h_i$ are sampled from a uniform
distribution with unit mean value and standard deviation $s$. 
Fig.~\ref{fig:diffusion_diff}(c) shows $\lambda_2$ as a function
of the average interlayer diffusion constant $\bar{D}$, for different
values of the standard deviation $s$. Notice that $s=0$ 
corresponds to homogeneous interlayer diffusion as in panel (a), 
%
and increasing the heterogeneity of $D_i$ leads to a non-negligible decrease
of $\lambda_2$. Furthermore, the maximum discrepancy $\Delta \lambda_2$
from the homogeneous case is observed for values
of $\bar{D}$ close to the transition between the two regimes.  
Fig.~\ref{fig:diffusion_diff}(d) explores instead how diffusion over a
multiplex is affected by the topology of its interlayer links.  
It shows that removing even a small percentage of interlayer links
can cause a drastic reduction of the values of $\lambda_2$. 
When $D \ll D^*$, the value of $\lambda_2$ increases with a slope that is
smaller than $2D$, while, for  $D \gg D^*$, the value of $\lambda_2$ never reaches
that of the aggregated network. Also in this case, 
the largest discrepancies with respect to the homogeneous case are observed
close to the transition point $D^*$. 
}

The presence of structural correlations can also affect multiplex
diffusion \cite{serrano2017optimizing}. Yet, surprisingly, the critical
point associated to the emergence of \vito{fast diffusion} as a function of
inter-link weights is independent from the value of edge
overlap, and layer dissimilarity only increases the extent of
multiplex \vito{fast diffusion} compared to single-layer diffusion, 
%
without affecting the onset of the phenomenon\cite{cencetti2019diffusive}.
%
    {The findings reported in this section are a direct consequence
      of the emergence, due to multiplexity,  
      of more paths between pairs of nodes \cite{de2016physics}.
      The appearance of a \vito{fast diffusion} regime is a 
      transition \vito{due to the change in the structure of a multiplex
      network that can also arise in other contexts} 
\cite{radicchi2013abrupt}.}

\bigskip
\subsection{
 \vito{Random walks}}  
\label{sec:random_walks} 

Random walks are a versatile way to explore a network by tuning  
the rule of the motion. This is done by opportunely choosing the
so-called transition probabilities $\text{Prob}(i \to j)$ 
for a walker to move from a node $i$ to a node $j$ at each
time step.  We focus here on the simplest possible case of time-invariant
Markov processes, in which the
walkers have no memory and the transition probabilities can be written as
$\text{Prob}(i \to j)= \pi_{ji}$, where $\pi_{ij} $ are the entries of
the {\em transition matrix} $\Pi$ and satisfy the constraint 
$\sum_j \pi_{ji}=1 ~\forall i$ \cite{cover1999elements}. 
If $\sigma_i(t)$ denotes the probability of finding a walker
at node $i$ at time $t$, the time evolution of the process 
reads: 
\begin{equation}
  \sigma_i(t+1)  = \sum_{j=1}^N  \pi_{ij} \sigma_j(t)                
\label{eq:map_rws}
\end{equation}
\vito{Notice that,} if we assume that there are independent, identical Poisson
processes at each node of the graph, such that the walkers jump at
\vito{the same} 
rate from each node, the corresponding continuous-time
process is governed by \cite{lambiotte2008laplacian,masuda2017random}:
\begin{equation}
  \frac{ d \sigma_i}{dt} =
  \sum_{j=1}^N  \left( \pi_{ij} -  \delta_{ij}\right) \sigma_j
  = - \sum_{j=1}^N  l'_{ij} \sigma_j                   
\label{eq:}
\end{equation}
which has been written in a form in all similar to Eq.~(\ref{eq:diff}) 
thanks to the definition of $L'= I - \Pi$ with entries 
$l'_{ij} = \delta_{ij} - \pi_{ij}$. \vito{This is still a Laplacian matrix, but
  is different from the standard diffusion Laplacian $L= K - A$
introduced in Section \ref{diff1}, as the adjacency
matrix $A$ is now replaced by the stochastic matrix  $\Pi$.}
%

%
%
%

%

\vito{Following 
\textcite{de2014navigability},  
the most general way to extend 
Eq.~(\ref{eq:map_rws}) to describe
the dynamics of random walks on multiplex networks
with $M$ layers is:}
\vito{
\begin{equation}
  \sigma_i\lay{\alpha}(t+1)  =  
                         \sum_{j=1}^N  \pi_{ij}\lay{\alpha} \sigma_j\lay{\alpha}(t)
                         +  \sum_{\beta=1, \beta \neq \alpha}^M  \pi_{i}\lay{\alpha \beta} \sigma_i\lay{\beta}(t)
\label{eq:mrw}
\end{equation}
}
\vito{
  In these equations, $\sigma_i\lay{\alpha}(t)$ denotes the probability
of finding a walker at node $i$ in layer $\alpha$ at time $t$, 
while $\pi_{ij}\lay{\alpha}$ and $\pi_{i}\lay{\alpha \beta}$
  are respectively the probabilities for a walker to move
  from node $j$ to node $i$ at layer $\alpha$, or from layer $\beta$
  to layer $\alpha$ 
  at node $i$.
  }
%
%
%
%
\vito{Notice that,} defining a column vector
$\mathcal S = ( \sigma\lay{1}_1, \sigma\lay{1}_2 \ldots, \sigma\lay{1}_N, \ldots, \sigma\lay{M}_1, \sigma\lay{M}_2 \ldots, \sigma\lay{M}_N )^{T}$
with $MN$ components, Eqs.~(\ref{eq:mrw}) can be rewritten in a 
compact form analogous to Eq.~(\ref{eq:difvect}) with a normalized
supra-Laplacian 
matrix $\cal L'$ which depends on the transition matrices internal to each
layer and on the transition matrices across layers, and 
whose structure is similar to the diffusion supra-Laplacian matrix $\cal L$  
reported in Eq.~(\ref{eq:supraL2}) for the case $M=2$.  
%
%
%
%
\vito{In the standard case of an unbiased random walk, the
  probability of moving from node $i$ to node $j$ within the same
  layer $\alpha$, or to switch to node $i$ at another layer is uniformly
  distributed, so the transition matrices can be written as:
\begin{equation}
  \pi_{ji} \lay{\alpha} = \frac {a^{\lay{\alpha}}_{ij}}
               {  S_i^{\alpha}  } 
  \qquad
  \pi_{i}\lay{\beta \alpha} = \frac {c^{\lay{\alpha \beta}}_{i}}
                             { S_i^{\alpha} }
  \label{eq:PIurw}
\end{equation}
where
$S_i^{\alpha} = \sum_j a^{\lay{\alpha}}_{ij} + \sum_{\beta} c^{\lay{\alpha \beta}}_i$
is the total strength of node $i$ at layer $\alpha$. In this way,
the normalization $ \sum_j \pi_{ji} \lay{\alpha} + 
\sum_{\beta \neq \alpha}  \pi_{i}\lay{\beta \alpha} = 1  ~\forall i$ and 
$\forall \alpha$ is satisfied.
}

In order to quantify the efficiency of a random walk in exploring a
multiplex network, \textcite{de2014navigability} have
\vito{developed an analytical method} to evaluate the \vito{so-called}
graph coverage 
$\rho(t)$\vito{, namely} the average fraction of 
nodes of the multiplex network
visited at least once in a time less than or equal to $t$
(regardless of the layer).
%
\vito{This method} allows to investigate how the efficiency of a walk
depends on the navigation strategy and on the topology of the
multiplex. It has also been used in a practical application to show
that the public transport system of London, consisting of three
different layers \vito{as those shown in Fig.~\ref{fig:figure_real}(a)}, is more
resilient to random failures than its individual layers 
separately. This is because connections between layers 
help finding paths from apparently isolated parts
of single layers, enhancing the resilience of the entire
system. 
\vito{Together with the graph coverage, other relevant metrics of 
  the mixing properties of random walks, such as 
  average first passage times or the entropy rate
  \cite{cover1999elements,gomez2008entropy} have also being 
  studied in the context of multiplex networks 
  \cite{battiston2016efficient}.
}

Various other types of random walkers have been considered.
For instance \textcite{taylor2020multiplex} has
%
\vito{proposed a multiplex generalization of Markov chains to 
describe random walkers that, with a probability 
$(1- \omega) \in [0, 1]$, move according to
layer-specific intralayer Markov chains, and with a probability
$\omega$, move to new layers following node-specific 
interlayer Markov chains.}
When $\omega \to 0$, each intralayer Markov chain approaches 
a stationary solution, and these layer-specific solutions are
balanced by the interlayer Markov chains. In the other limit $\omega \to 1$
the interlayer Markov chains individually approach 
stationary solutions, and these node-specific solutions are balanced
by the intralayer Markov chains. For intermediate values of $\omega$ a
novel multiplexity-induced phenomenon called multiplex convection is
identified, in which convection cycles, similar to those commonly
observed in fluid dynamics, emerge in the network due to imbalances
in the intralayer degrees of nodes in different layers.

Random walks that move over the links of a multiplex network with 
the additional possibility of performing non-local hops to randomly chosen
nodes have been studied by 
\textcite{dipatti2015optimal}, with the purpose of finding the 
optimal combination of local and non-local jumps that 
maximises the efficiency of target search, and by 
\textcite{halu2013multiplex} to measure the centrality of a
node by introducing different versions  
%
%
of a multiplex PageRank. 
%
\textcite{guo2016levy} have instead considered L\'evy random walks whose 
intralayer probability to jump from a node $i$ to a node $j$
has a power-law dependence on the distance between $i$ and $j$, and
have studied how average times depends of the value of power-law
exponent in the search of the most efficient navigation strategies.

\bigskip
\subsection{Pattern formation} 

Richer behaviours emerge when the agents not only diffuse
but also interact. Chemical reactions, in which the molecules diffuse
in space and react when in close proximity, are the classical example.
{\em Reaction-diffusion} models have been extensively used to study
physical and chemical systems, but also to model the dynamics of
biological populations and the spreading of diseases
\cite{van1992stochastic,murray2002mathematical}. More recently,
reaction-diffusion
processes have also been explored in complex networks in order to
account for the patterns of mobility observed in real systems
\cite{colizza2007reaction,nicosia2011impact,gomez2018critical}.
Our focus here is 
on a particular type of reaction-diffusion with two species of agents,
namely activators, which autocatalytically enhance their own
production, and inhibitors which instead suppress the activator
growth. Alan Turing was the first to show, in his pioneering paper in
1952, that differences in the diffusion constants of activators and
inhibitors can create a destabilization of the uniform state leading 
to a spontaneous emergence of periodic spatial patterns of the two
species, today known as {\em Turing patterns}
\cite{turing1952chemical,othmer1971instability,othmer1974non}.
\textcite{nakao2010turing} 
have demonstrated that Turing patterns also emerge when the two
interacting species occupy the nodes of a complex network, for 
instance a scale-free graph, and are diffusively transported across
its links.  Through a linear stability analysis of the uniform
stationary state, they have proven that, when the ratio $r =
D^{\rm{inh}}/D^{\rm{act}}$ between the diffusion constants of inhibitor and
activator is larger than
%
%
critical threshold $r_c > 1$,
an initial perturbation leads to the spontaneous differentiation
of the network nodes into activator-rich and activator-poor groups.

\begin{figure}[t]
\begin{center}
 \includegraphics[width=0.4 \textwidth]{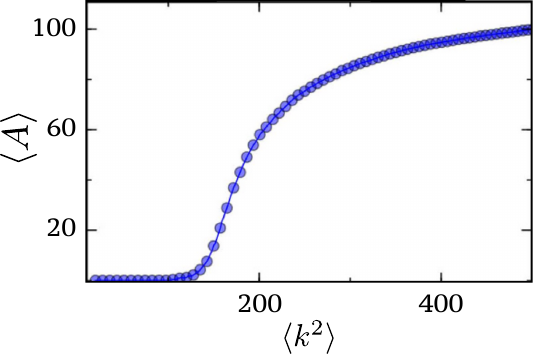}
 \caption[]{Average amplitude of Turing patterns as a function of the average
   degree  $\langle k\lay{2} \rangle$ of the inhibitor layer in multiplex scale-free networks
   with $\langle k\lay{1} \rangle=20$. The diffusion constants are $D\laysup{1}=D\laysup{2}=0.12$, so that
   we have $r=1$. Averages are taken over ten numerical simulations for different
   samplings of the inhibitor network with the same mean degree $\langle k\lay{2} \rangle$. 
   Figure adapted from \textcite{kouvaris2015pattern}. }
 \label{fig:diffusion_tp1}
 \end{center}
 \end{figure}

A natural extension is the case in which the two species diffuse on
two distinct topologies, representing their diverse mobility patterns
\vito{and react across layers.} 
This can be easily implemented in terms
of a multiplex network with two layers, where the quantities 
$\sigma\lay{1}_i$ and $\sigma\lay{2}_i$ are
respectively the concentrations of activators and of inhibitors of a
reaction-diffusion dynamics. The process can be
described by the following equations \cite{kouvaris2015pattern}:
\begin{equation}
   \left\{  \begin{array}{c}
     { d \sigma\lay{1}_i}/{dt} =  f\lay{1}(\sigma\lay{1}_i,\sigma\lay{2}_i)  - D\laysup{1} \sum_{j=1}^N  l_{ij}\lay{1} \sigma\lay{1}_j
\\
\\
{ d \sigma\lay{2}_i}/{dt} =  f\lay{2}(\sigma\lay{1}_i,\sigma\lay{2}_i)  -  D\laysup{2} \sum_{j=1}^N  l_{ij}\lay{2} \sigma\lay{2}_j                   
        \end{array}
\right.
\label{eq:ai}
\end{equation}
where $i=1,2,\ldots N$, $L\lay{1}=\{ l\lay{1}_{ij}\}$, with $l\lay{1}_{ij} =k\lay{1}_i
\delta_{ij} - a\lay{1}_{ij}$ is the Laplacian of the layer accounting for
the diffusion of activators, $L\lay{2} =\{ l\lay{2}_{ij}\}$ is the one for
inhibitors and $D\laysup{1} = D\laysup{\rm act}$, $D\laysup{2}=D\laysup{\rm inh}$ are the diffusion
constants of the two species.
Differently from the equations considered in the previous subsections,
Eqs.~(\ref{eq:ai}) are nonlinear because of the nonlinearity
in the reaction terms. Depending on the specific choice of the two
reaction functions $f\lay{1}$ and $f\lay{2}$, Eq.~(\ref{eq:ai}) can be used to
describe prey-predator systems, neurodynamics, or chemical reactions
where two species interact at the nodes of a network
and diffuse not only with different diffusion constants, but also
on different topologies. As a concrete example let us focus on the
Mimura-Murray ecological model, where the two reaction functions 
read $f\lay{1}(u,v)=\{(a+ bu - u^2)/c - v \}u$ and $f\lay{2}(u,v)=\{u - (1+dv)\}v$,
where $u$ and $v$ correspond to prey (activators) and predator (inhibitors)
densities \cite{mimura1978diffusive,nakao2010turing}.
In particular, the following choice of parameters 
$a=35, b=16, c =9$ and $d =2/5$ yields a fixed point
$(\bar{u},\bar{v}) = (5,10)$ of the reaction dynamics. 
So, in absence of diffusion, a multiplex system with this type of
reaction term is in the uniform state in which
$(\sigma_i\lay{1},  \sigma_i\lay{2}) = (\bar{u},\bar{v})$ for all nodes 
$i=1,2,\ldots, N$. \textcite{kouvaris2015pattern} have shown that
Turing patterns can occur on a multiplex network even when $r=1$, i.e.~when
the two species have the same mobility rate,
condition which can never destabilize single-layer networks.
Figure \ref{fig:diffusion_tp1} reports the stationary amplitude
$A =[\sum_{i=1}^N  [ ( \sigma_i\lay{1} - \bar{u} )^2 +   ( \sigma_i\lay{2} - \bar{v} )^2 ]^{1/2}$  
of a non-uniform pattern in the case in which $r=1$ and
both layers of the multiplex are scale-free networks. The average degree
of the activator layer $\langle k\lay{1} \rangle$ is fixed to 20, while
different values of average degree $\langle k\lay{2} \rangle$ are considered for 
the topology of inhibitors. The numerical results show that the instability occurs
when the network at the activator layer has an average degree $\langle k\lay{2} \rangle$
larger that a certain threshold $\langle k\lay{2} \rangle_c$. The same authors,
under the assumption that both $\langle k\lay{1} \rangle \gg 1$ and  $\langle k\lay{2} \rangle \gg 1$, 
have derived an analytical approximation for $\langle k\lay{2} \rangle_c$ as a function
of $\langle k\lay{1} \rangle$, of the reaction dynamics and of the two diffusion
constants:   
\begin{equation}
  \langle k\lay{2} \rangle_c = \frac {Tr  - f\lay{2}\low{2}  D\laysup{1} \langle k\lay{1} \rangle}
                           {D^2 ( f\lay{1}\low{1} - D\laysup{1} \langle k\lay{1} \rangle) }   
  \label{eq:kouvaris_formula}
\end{equation}
where $f\lay{\alpha}\low{\beta}= \partial f\lay{\alpha}/ \partial \sigma\lay{\beta}$ for
$\alpha, \beta =1,2$, and $Tr$ is the trace of the
Jacobian matrix of the reaction dynamics,   
$Tr= f\lay{1}_{\sigma_1} f\lay{2}_{\sigma_2} - f\lay{1}_{\sigma_2} f\lay{2}_{\sigma_1}$.
This formula clearly indicates that the relevant quantity in
a multiplex network is the product $D^2  \langle k\lay{2} \rangle$.
Hence topology-driven instabilities leading to Turing patterns 
can occur in multiplex networks even if the two species have the
same mobility rates, provided the average node degree of the inhibitor 
topology is large enough. 
%
%
%
%
Within the same framework of Eqs.~(\ref{eq:kouvaris_formula}),  
\textcite{asllani2016tune} have derived analytical formulas to
predict Turing patterns generated or destroyed by small changes 
in a multiplex network, such as the addition of single links to one
of the two layers. 

A different implementation of reaction-diffusion processes 
over multiplex networks has been proposed by
\textcite{asllani2014turing}. In their setting, each of the 
two species is allowed to live on both layers, and to diffuse over the two
layers and also from one layer to the other.
For convenience, we will now indicate as $u$ and $v$ the concentrations 
of activators and inhibitors, respectively. The equations read: 
\begin{equation}
\hspace*{-0.05 cm}
{\small
\left\{  \begin{array}{c}
{d u\lay{1}_i}/{dt} =  f\lay{1}(u\lay{1}_i,v\lay{1}_i)  - D_u\laysup{1} \sum_{j=1}^N  l_{ij}\lay{1} u\lay{1}_j + D\laysup{12}_u (u\lay{2}_i - u\lay{1}_i) 
\\[0.2 em]
{d  v\lay{1}_i}/{dt} = g\lay{1}(u\lay{1}_i,v\lay{1}_i)  - D_v\laysup{1} \sum_{j=1}^N  l_{ij}\lay{1} v\lay{1}_j + D\laysup{12}_v (v\lay{2}_i - v\lay{1}_i) 
\\[0.2 em]
{d u\lay{2}_i}/{dt} =  f\lay{2}(u\lay{2}_i,v\lay{2}_i)  - D_u\laysup{2} \sum_{j=1}^N  l_{ij}\lay{2} u\lay{2}_j + D\laysup{12}_u (u\lay{1}_i - u\lay{2}_i) 
\\[0.2 em]
{d v\lay{2}_i}/{dt} = g\lay{2}(u\lay{2}_i,v\lay{2}_i)  - D_v\laysup{2} \sum_{j=1}^N  l_{ij}\lay{2} v\lay{2}_j + D\laysup{12}_v (v\lay{1}_i - v\lay{2}_i) 
          \end{array}  \right.
}
\label{eq:ai_2}
\end{equation}
where the interactions at the nodes of the first (second) layer are
ruled by the functions $f\lay{1}$ and $g\lay{1}$ ($f\lay{2}$ and $g\lay{2}$), and $D_u\laysup{1},
D_v\laysup{1}, D_u\laysup{2}, D_v\laysup{2}$ are the intralayer, while $D\laysup{12}_u$ and $D\laysup{12}_v$
are the interlayer diffusion constants. 
When $D\laysup{12}_u = D\laysup{12}_v=0$ the two layers are decoupled and we
have two independent pairs of reaction-diffusion equations for
$(u\lay{1}_i, v\lay{1}_i)$ and $(u\lay{2}_i, v\lay{2}_i)$. {\vito In this case  
Turing patterns can eventually set in for each of the two layers. If instead}  
the interlayer diffusion is turned on, a perturbative expansion on the
interlayer diffusion constants shows that the possibility to move  
from one layer to the other can induce Turing patterns even when these
are impossible in the two uncoupled layers.
%
\vito{\textcite{busiello2018homogeneous} have further analyzed the rich dynamics 
of reaction-diffusion processes of the type in Eq.~(\ref{eq:ai_2}) when more
than two layers are considered, including the emergence of patterns
consisting in sequences of different homogeneous states at the
different layers.}

\bigskip
\subsection{Congestion}

\vito{Congestion is a collective phenomenon that occurs in transportation
  or communication networks when nodes or links become overloaded with
  more traffic, agents or data than they can handle, leading to a reduced
  quality of the overall service. This excessive load results in
  delays, as data packets queue up, and can cause packet loss when
  queues overflow and packets are dropped. The effects are a slowdown
  in data transmission, low throughput, and decreased network
  efficiency, impacting transportation or information flow across the
  network.} 
\textcite{sole2016congestion} showed the emergence of congestion
induced by the multiplex structure of multimodal transportation
systems that otherwise would not appear if the individual layers of a
system were not interconnected \cite{strano2015multiplex,chodrow2016demand}.  
Also the effect on congestion of the possibility 
of changing layer has been analyzed as a function of the different
velocities associated to the links of the layers 
\cite{morris2012transport}. 
\textcite{manfredi2018mobility} have proposed a multiplex mobility model
in which the nodes have a limited storage capacity and the agents, instead
of simply following shortest paths, seek for
uncongested paths during their navigation \vito{over a multiplex network}. 
%
%
In such a model, the state $\sigma_i\lay{\alpha}(t)$ represents the queue
length, namely the number of agents being on node $i$ at layer
$\alpha$ and at time $t$, with $\sigma_i\lay{\alpha}(t) \leq B_i\lay{\alpha}~
\forall t$. Here, $B_i\lay{\alpha}$ is the maximum capacity of the node in
storing agents, e.g., the buffer size of a router in a computer
network,  or the maximum allowed  number of passengers in an underground
station. Agents are processed at each node in order of their arrival.
To mimic their propensity to minimise distances but also avoid
congested nodes \cite{scellato2010traffic}, 
the agents 
move from their origins to their destinations by following, at each step, 
minimum-weight paths, where the link weights change in time, depending on the
node queue lengths \cite{crucitti2004cascading}.
Namely, the weight $w_{ij}\lay{\alpha \beta}(t)$ at time $t$ 
of the link from node $i$ at layer $\alpha$ to node $j$ at layer
$\beta$ is defined as:
\begin{equation}
  w_{ij}\lay{\alpha \beta} (t)=c \cdot \gamma_{ij}\lay{\alpha \beta} +
  \frac{ \sigma_{j}\lay{\beta} (t) } { B_j\lay{\beta} -  \sigma_{j}\lay{\beta}(t) }
\label{pesi}
\end{equation}
where $\gamma_{ij}\lay{\alpha \beta}$ represents the link travel 
time from node $i$ at layer $\alpha$ to node $j$ at layer $\beta$,
and $c$ is the equivalent cost per unit time, so that 
$c \cdot \gamma_{ij}\lay{\alpha \beta}$ is the intrinsic cost of traversing the
link.
The second term in the right hand side of Eq.~(\ref{pesi}) 
represents the cost due to the level of congestion of node 
node $j$. The weight $w_{ij}\lay{\alpha \beta}(t)$
%
%
takes the minimum value $c \cdot \gamma_{ij}\lay{\alpha \beta}$ 
when the queue at $j$ is empty,
while it 
diverges when the queue is full, i.e. when $\sigma_{j}\lay{\beta}(t)=B_j\lay{\beta}$.
\vito{Notice that, in a multiplex network, if $\alpha \neq \beta$, we
  have  by definition $\gamma_{ij}\lay{\alpha \beta}=0$ and
  $w_{ij}\lay{\alpha \beta} = 0$  for each $i \neq j$.     
}
Agents are randomly generated at the nodes of a \vito{multiplex}
network, \vito{as the one in Fig.~\ref{fig:figure_real}(a),} and they
are randomly assigned a final node destination.  At each time step,
they move from \vito{the current node to one of its neighbours on a
  given layer, or to the corresponding node at another layer}
mimimising the sum of the weights to their final destination, until
they arrive there and are removed from the network. Due to the form of
the weights in Eq.~(\ref{pesi}), the agents will automatically avoid
nodes when their queues are congested.
%
\begin{figure}[t]
\begin{center}
\includegraphics[width=0.4\textwidth]{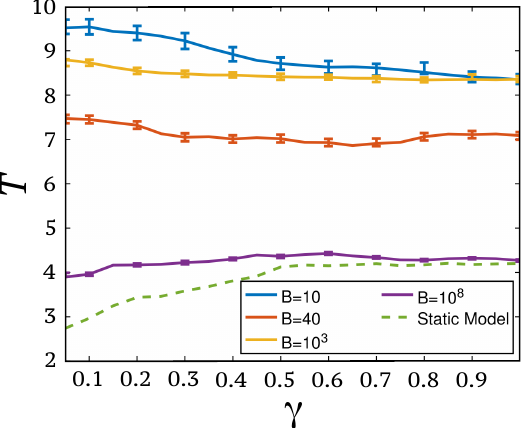}
\caption[]{ Average travel time $T$ of delivered agents as a function
  of the \vito{travel time ratio} $\gamma$ in the multilayer mobility model by
  \textcite{manfredi2018mobility} with three different values of node
  buffer sizes: $B= 10, 40, 10^{3}, 10^{8}$ (continuous lines).
  Numerical results are reported as symbols, with the error bars
  representing fluctuations over random agent generations and 
  different network realizations. Also the case of a static model not
  accounting for the queue dynamics has been considered (dashed line).
  Figure adapted from \textcite{manfredi2018mobility}.
	\label{figs:RGG:gamma:meantime}}
 \label{fig:diffusion_congestion}
 \end{center}
\end{figure}
%
%
Fig.~\ref{figs:RGG:gamma:meantime} reports
%
the average time $T$ taken by the agents to arrive at their 
destinations    
in a multiplex network with two layers. 
Layer $1$ represents a dense but slow transportation system 
with high clustering
coefficient and short-range connections,
while layer $2$ represents a fast transportation system, 
with fewer active nodes but with long-range connections. 
Setting 
$c=1$, $B_i\lay{1}=B_i\lay{2}=B$,  
$\gamma\lay{1 1}_{ij}=\gamma\lay{1 1}=1$, 
$\gamma\lay{1 2}_{ij}=\gamma\lay{1 2}=1$, 
and $0<\gamma\lay{2 2}_{ij}=\gamma\lay{2 2}\leq
\gamma\lay{1 1}$ $\forall i,j$,
it is possible to concentrate on the exploration of different
values of the
ratio $\gamma=\gamma\lay{2 2}/\gamma\lay{1 1}$ in
the range $(0, 1]$.
The plots of $T$ as a function of $\gamma$ for buffer sizes
$B= 10, 40$ and $10^{3}$ (continuous lines) show that the travel time
counterintuitively decreases by increasing $\gamma$, i.e. by
decreasing the velocity of 
the faster layer $2$ while keeping fixed the velocity of layer $1$.  
Such a behaviour, in which an improvement of the system performance is
obtained by reducing the velocity, is a multiplex version of 
the so-called ``slower is faster'' effect
\cite{gershenson2015slower}.  
%
%
Interestingly, this effect disappears for very large values of $B$
(see the curve for $B=10^8$) or in the limit case of a static model that
does not account for the queue dynamics (dashed line): in both
cases $T$ is observed to increase for increasing values of
$\gamma$. 
%
%
Another interesting behaviour shows up in the dependence of $T$ on the
buffer size. When $B$ changes from 10 to 40, the addition of resources to
the nodes leads to a reduction of congestion and consequently to a
drop of the travel time $T$ (for any value of $\gamma$).  However, a
further increase of the node buffer size to $B=10^3$, does not lead to
an additional improvement of the system, but instead to an unexpected
increase of the travel time $T$. This is reminiscent of
the Braess's paradox, in which the addition of 
resources to a network in terms of links  leads to a worsening of
the network performance \cite{braess2005paradox}.


\section{Synchronization}
\label{sec:synchro}

Synchronization is the physical process by which two or more coupled
dynamical systems share a property of their motion, e.g., their phase,
or follow exactly the same trajectory~\cite{boccaletti2018synchronization}.
It has been observed in many real-world systems 
including chemical and biological oscillators, crickets
singing at unison, fireflies flashing at the same rate, and 
electrical power grids~\cite{pikovsky2003synchronization}.
Mathematically, synchronization was
first studied in systems of two coupled oscillators
\cite{pecora1990synchronization,mirollo1990synchronization}, and
then in systems formed by many units with pairwise interactions 
specified by a network \cite{boccaletti2006complex,arenas2008synchronization}.

Here, we focus on synchronization in multiplex networks.
We begin by considering results for dynamical
units whose state is a composition of variables pertaining to the
different layers, and interactions can occur at the intra- and the
inter-layer level. We concentrate on each of the three patterns 
of synchronization that can emerge in such structures, namely
complete, intra-layer and inter-layer synchronization. 
After briefly reviewing them, we comment on the tools
available to study their stability. Then, we discuss how
intra-layer and inter-layer synchronization can be obtained by
dynamical relays. We then move to discuss different phenomena
appearing in multiplex structures, such as explosive synchronization,
cluster synchronization and chimera states. We conclude by considering
the case of systems whose units have a state that cannot
be decomposed into different layer-specific variables and by showing
how this framework naturally leads to multiplex control techniques
where some of the layers can be used to control
the dynamical behavior of the other layers.

\mattia{Examples of applications of synchronization in multiplex networks span different fields~\cite{wu2024synchronization,hu2025inter}. For instance, genetic networks have been modeled as multiplex structures of oscillators~\cite{khalaf2019synchronization}, brain networks have been studied through multiplex models to investigate alterations of brain function under acupuncture stimulation~\cite{yu2021electroencephalographic}, and power grids have been described in terms of multiplex layers, where nodes represent either generators or consumers~\cite{yang2021analysis}.}

\subsection{Complete, inter- and intra-layer synchronization}
\label{sec:secVA}

The structure of a network can affect synchronization in
four main aspects:  
{\em i)} whether or not synchronization can be achieved;
{\em ii)} the synchronization threshold, i.e. the smallest value of
the coupling coefficient to attain synchronization; 
{\em iii)} the path to synchronization~\cite{gomez2007paths}, i.e. the growth
and merging of synchronized clusters; 
{\em iv)} the geometry of the synchronized
clusters~\cite{pecora2014cluster,schaub2016graph}. 


When interactions are of different tpyes, synchronization can be studied 
by considering a multiplex network in which each unit is described by a
$D = M d$ dimensional state, where $d$ is dimensionality of the
node state $\bm \sigma_i\lay{\alpha}$ at layer $\alpha$.
\mattia{For convenience, we restrict here the analysis to the case of diffusive coupling, which is the most commonly adopted formalism due to its solid physical grounding. In this case, the coupling terms can be expressed as the difference of the values of the coupling function at the two interacting nodes, so that the }
dynamics of the quantities $\bm \sigma_i\lay{\alpha}(t)$, for $i=1,\ldots,N$ and
$\alpha=1,\ldots,M$, are governed by the set of coupled equations \cite{tang2019master}: 
\begin{equation}
\label{eq:syncmodel}
\frac{d{\bm \sigma_i\lay{\alpha}}}{dt}
  =   \bm f (\bm \sigma_i\lay{\alpha})  - g  \sum_{j=1}^N l_{ij}\lay{\alpha} \bm h\lay\alpha (\bm \sigma_j\lay{\alpha}) - w \sum\limits_{\beta=1}^M u\lay{\alpha \beta}_{i} \bm v_i (\bm \sigma_i \lay{\beta}),
\end{equation}
\noindent where $\bm f:\mathbb{R}^d\rightarrow\mathbb{R}^d$
describes the local dynamics, here assumed to be equal for
all the nodes, $\bm h\lay\alpha:\mathbb{R}^d\rightarrow\mathbb{R}^d$
represents the intra-layer coupling function at layer $\alpha$, and
$\bm v_i:\mathbb{R}^d\rightarrow\mathbb{R}^d$ are the coupling
functions across layers.
The connectivity within each layer $\alpha$ is encoded by the $N
\times N$ intra-layer Laplacian matrices $L\lay\alpha = \{
l_{ij}\lay{\alpha} \}$, $\alpha=1,\ldots M$, whereas the connectivity
across layers by the $M\times M$ inter-layer Laplacian matrices
$L^I_i = \{ u_{i}\lay{\alpha \beta} \}$, $i=1,2,\ldots N$
(where apex $I$ stands for inter-layer).
When $\alpha \neq \beta$, we have $u\lay{\alpha \beta}_{i}=-1$ 
if the states of node $i$ at layers $\alpha$ and $\beta$ 
are coupled or $u\lay{\alpha \beta}_{i}=0$ otherwise, while
$u\lay{\alpha \alpha}_i=-\sum_{\beta \neq \alpha}u\lay{\alpha,\beta}_i$.
For simplicity, the inter-layer connectivity is assumed to be equal for
each unit $i$ of the system, i.e. $L^I_i = L^I ~ \forall i$.
Finally, $g>0$ and $w>0$ are tunable parameters
controlling the intra-layer and inter-layer coupling strength, respectively.

Eqs.~(\ref{eq:syncmodel}) can be rewritten in a compact form.  Let
$\mathcal{S}\lay{\alpha}=[(\bm \sigma_1 \lay \alpha)^{T}, \ldots, (\bm
  \sigma_N\lay \alpha)^{T}]^T$ be the stack vector of all the state
variables at layer $\alpha$, and let $\mathcal{S}=[(\mathcal{S}\lay
  1)^{T}, \ldots, (\mathcal{S}_N\lay M)^{T}]^T$ indicate the stack
vector of all variables across the whole structure. Correspondingly,
we stack together the local dynamics at each layer $\alpha$, $\bm
\tilde f \lay \alpha= [ \bm f^T(\sigma_1\lay{\alpha}), \dots, \bm
  f^T(\sigma_N\lay{\alpha}) ]^T$, and in the whole structure, $\bm
F(\mathcal S)=[ (\bm \tilde f\lay 1)^T, \ldots, (\bm \tilde f\lay M)^T
]^T$, and the coupling functions: $\bm \tilde h\lay\alpha = [ (\bm
  h\lay \alpha(\sigma_1^{\alpha}))^{T}, \dots, (\bm
  h\lay{\alpha}(\sigma_N^{\alpha}))^T ]^T$, $\bm H(\mathcal S)=[ (\bm
  \tilde h\lay{1})^T, \ldots, (\bm \tilde h\lay{M})^T]^T$, $\bm \tilde
v \lay \alpha = [ \bm v_1^{T}(\sigma_1\lay{\alpha}), \dots, \bm
  v_N^{T}(\sigma_N\lay \alpha) ]^T$, $\bm V(\mathcal S)=[(\bm \tilde
  v\lay 1)^{T}, \ldots, (\bm \tilde v\lay M)^{T}]^T$. With these
positions, the system evolution is described by:
\begin{equation}
\label{eq:syncmodelKron}
 \frac{d{\mathcal S}}{dt}
  =   \bm F (\mathcal S)  - g (\mathcal L^{L} \otimes I_d) \bm H(\mathcal S) - w (\mathcal L^{I} \otimes I_d) \bm V(\mathcal S),
\end{equation}
where 
\begin{equation}
\label{eq:LLLI}
\mathcal L^{L}= \oplus L^{\alpha}, ~~~~~~~  \mathcal
L^{I}= L^{I} \otimes I_{N}
\end{equation}
and the symbols $\oplus$
and $\otimes$ indicate respectively the direct sum and the Kronecker product
of matrices. 

Suppose now that the intra-layer coupling functions are linear and
layer-independent, that is, $\bm h\lay{\alpha}(\bm \sigma_j\lay
\alpha)=\bm h(\bm \sigma_j\lay \alpha)=H\,\bm \sigma_j\lay
\alpha$. Analogously, assume that the inter-layer coupling functions
are also linear and node-independent, i.e., $\bm v_i(\bm \sigma_i\lay
\alpha)=\bm v(\bm \sigma_i\lay \alpha)=V\,\bm \sigma_i\lay
\alpha$. Here, $H$ and $V$ are $d \times d$ matrices
of constant coefficients (for simplicity, they are often
binary). Under these assumptions, Eq.~(\ref{eq:syncmodelKron}) becomes:
\begin{equation}
\label{eq:syncmodelKronLIN}
\frac{d{\mathcal S}}{dt}  =   \bm F (\mathcal S)   -g(\mathcal{L}^L \otimes H)\mathcal S -w(\mathcal{L}^I \otimes V)\mathcal S
\end{equation}

Models~(\ref{eq:syncmodelKron}) and~(\ref{eq:syncmodelKronLIN}) prompt for the occurrence of three patterns of synchronization: intra-layer, inter-layer and complete synchronization.

\begin{figure}
    \begin{center}
        \includegraphics[width=0.6\textwidth]{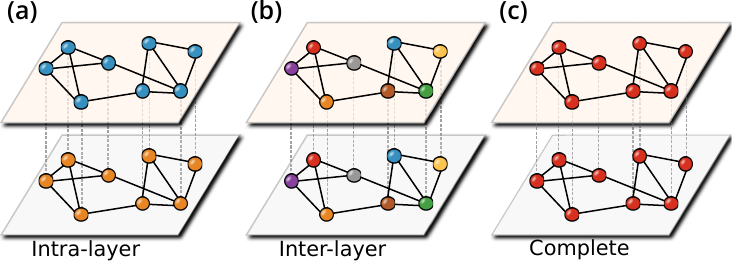}
        \caption[]{Illustration of 
          intra-layer (a), inter-layer (b)
          and complete (c) synchronization in a multiplex network with
          two layers. \label{fig:schemisync}}
    \end{center}
\end{figure}

In \emph{intra-layer synchronization} [Fig.~\ref{fig:schemisync}(a)]
all nodes in a layer $\alpha$ asymptotically follow the same
trajectory: 
\begin{equation}
\lim\limits_{t\rightarrow \infty}\| \bm \sigma_i\lay\alpha - \bm \sigma_j\lay\alpha \|=0,~~ \forall i, j=1,\ldots,N
\end{equation}
The corresponding synchronization manifold is defined by the condition
$\bm \sigma_1\lay\alpha(t)=\bm \sigma_2\lay\alpha(t) = \ldots = \bm
\sigma_N\lay\alpha(t)$. Notice that intra-layer synchronization can be
observed in a single layer $\alpha$, or in more than one layer,
as in [Fig.~\ref{fig:schemisync}(a)].

In \emph{inter-layer synchronization} [Fig.~\ref{fig:schemisync}(b)], all
replicas asymptotically follow the same trajectory:
\begin{equation}
\lim\limits_{t\rightarrow \infty}\| \bm \sigma_i\lay\alpha - \bm \sigma_i\lay\beta \|=0, ~~\forall \alpha, \beta=1,\ldots,M, ~~\forall i=1,\ldots,N
\end{equation}

The corresponding synchronization manifold is defined by considering $\bm \sigma_i\lay 1(t)=\bm \sigma_i\lay 2(t) = \ldots = \bm \sigma_i\lay M(t)$, $\forall i=1,\ldots,N$.

Finally, \emph{complete synchronization} [Fig.~\ref{fig:schemisync}(c)] is defined by the condition that all the nodes of the system asymptotically follow the same trajectory:
\begin{equation}
\lim\limits_{t\rightarrow \infty}\| \bm \sigma_i\lay\alpha - \bm \sigma_j\lay\beta \|=0, ~~ \forall \alpha, \beta=1,\ldots,M, ~~\forall i,j=1,\ldots,N
\end{equation}

The corresponding synchronization manifold is $\bm \sigma_1\lay1(t)=
\ldots = \bm \sigma_N\lay1(t)=\ldots=\bm \sigma_1\lay M(t)= \ldots = \bm
\sigma_N\lay M(t)=\tilde{\bm \sigma}(t)$, or, equivalently,
$\tilde{\mathcal{S}}(t)=\bm 1_{M N} \otimes \tilde{\bm \sigma}(t)$,
where $\tilde{\bm \sigma}(t)$ is the common, synchronous trajectory.

Since inter-layer and intra-layer couplings are expressed in terms of 
Laplacian matrices, which are zero-row sum, the three
synchronization patterns are all invariant solutions of
Eqs.~(\ref{eq:syncmodel}). To observe synchronization in a system,
however, the invariance of the synchronized state
is not sufficient, but its stability, at least in a local
sense (that is, for initial conditions in a neighborhood of the
solution) is also required. Deriving the 
stability conditions for the three types of synchronized patterns 
is not straightforward and often requires additional assumptions 
on the structure of the system.


We start with \mattia{the stability analysis for} complete synchronization, focusing on a specific, but
instructive case study~\cite{sole2013spectral}, where
Eqs.~(\ref{eq:syncmodelKron}) can be significantly simplified.
Under the assumptions that the intra-layer and inter-layer
coupling functions are linear, layer- and node-independent and equal
each other (i.e., $H=V$),
Eqs.~(\ref{eq:syncmodelKron}) become:
\begin{equation}
\label{eq:syncmodelKronLINArenas}
\frac{d{\mathcal S}}{dt}  =   \bm F (\mathcal S)   -g(\mathcal{L}^s \otimes H)\,\mathcal S
\end{equation}
\noindent where $\mathcal{L}^s=\mathcal{L}^L+\frac{w}{g}\mathcal{L}^I$
is the supra-Laplacian matrix.  Notice that the system in 
Eq.~(\ref{eq:syncmodelKronLINArenas}) is equivalent to
$MN$ dynamical units of order $d$ coupled through a  
single-layer network whose structure is described by the 
$MN \times MN$ supra-Laplacian matrix.
This crucial observation makes possible the
application of the so-called Master Stability Function
(MSF), a standard approach to assess the local
stability of the complete synchronization manifold in
complex networks \cite{pecora1998master}.

Following this approach \cite{boccaletti2006complex,huang2009generic}, 
we consider a small perturbation of the synchronous trajectory,
${\mathcal S}(t) = \tilde{\mathcal S}(t) + \delta {\mathcal
  S}(t)$  with $\delta {\mathcal S} = \{ \delta \bm \sigma_1, \ldots,
\delta \bm \sigma_{MN} \}$, and linearize
Eqs.~(\ref{eq:syncmodelKronLINArenas})
around $\tilde{\mathcal{S}}(t)$ to obtain:
\begin{equation}
\frac{d(\delta {\cal S})}{dt}
  =
  \left[ I_{N}\otimes {\text J} {\bm f}( \tilde{\bm \sigma}) -
  g
   \mathcal{L}^s \otimes {\text J} {\bm h}( \tilde{\bm \sigma})\right] \delta\cal{S},
  \label{dyn_oscill2_compact}
\end{equation}
\noindent where ${\text J} \bm f$ and ${\text J} \bm h$ denote the
Jacobian matrices respectively of functions $\bm f$ and $\bm h$
evaluated at the synchronous solution $\tilde{\bm \sigma}$. 
The system in Eq.(\ref{dyn_oscill2_compact}) 
can be decoupled using a proper transformation based on
the eigenvectors of the supra-Laplacian matrix
$\mathcal{L}^s$. If all interactions in the multiplex network
are symmetric, $\mathcal{L}^s$ has the same properties of
the Laplacian of an undirected graph.
In particular, it is symmetric and positive semi-definite;
its eigenvectors form an
orthonormal base; and its eigenvalues are nonnegative quantities and
can, therefore, be ordered as: $0 = \lambda_{1} \le \lambda_{2} \le
\ldots \lambda_{MN}$.  If we further assume that 
the structure is connected, then, there is a single eigenvalue
equal to zero, i.e., $0 =
\lambda_{1} < \lambda_{2} \le \ldots \lambda_{MN}$. 
Let us now indicate with $T$ the matrix containing the left
eigenvectors of $\mathcal{L}_s$, and with $\Gamma$ the
diagonal matrix of its eigenvalues such that
$T^{-1} \mathcal{L}_s   T=\Gamma$. We can then
define a new set of variables by the transformation $\bm \zeta=\left (
T^{-1}\otimes I_d\right)\delta {\mathcal S}$, whose
dynamics is governed by:
\begin{equation}
  \frac{d {\bm \zeta}}{dt} = \left( I_{N}\otimes {\text J} {\bm f}
  -
  g
  \Gamma \otimes {\text J} {\bm h}
  \right)
         {\bm \zeta}
\label{dyn_oscill3_compact}
\end{equation}
Since $\Gamma$ is diagonal, Eq. (\ref{dyn_oscill3_compact}) gives $MN$ decoupled blocks:
\begin{equation}
  \frac{d\bm \zeta_{\nu}}{dt} = \bigl [ {{\text J}\bm f
    } - g \lambda_{\nu} {
      {\text J}      \bm h
    } \bigr ] \bm \zeta_{\nu},
\label{eq_eigmodes}
\end{equation}
\noindent where $\nu = 1, \ldots, MN$. The various blocks only differ
because of the eigenvalue $\lambda_{\nu}$. In more detail, the
eigenmode $\bm \zeta_1$ associated to $\lambda_{1}$ describes the
perturbation in the direction parallel to the synchronization
manifold, whereas for $\nu = 2, \ldots ,MN$ these equations correspond
to the transverse perturbations, which affect the stability of the
synchronized state.
Eq. (\ref{eq_eigmodes}) can be rewritten as a function of a single parameter $\gamma$ as follows:
\begin{equation}
  \frac{d\bm \xi}{dt} = \bigl [ {{\text J}\bm f
    } - \gamma {
      {\text J}      \bm h
    } \bigr ] \bm \xi,
\label{eq_eigmodes2}
\end{equation}
providing a variational equation from which the maximum Lyapunov
exponent \mattia{$\Lambda$} can be evaluated as a function of the parameter
$\gamma$. This defines the function $\Lambda(\gamma)$, the so-called MSF of
the system. For the synchronized state to be stable 
all the transverse modes must damp out, a condition that is fulfilled when 
$\Lambda(\gamma)<0$ for
$\gamma=\{g\lambda_2,\ldots,g\lambda_MN\}$. Notice that, 
the MSF $\Lambda(\gamma)$
depends on the node dynamics, the coupling function and the
synchronous solution, but not on the interaction structure. For
this reason, the MSF approach 
separates the role of the dynamics from that of the structure in
determining the synchronization stability of a system.

In full analogy with the single-layer case, for $\gamma>0$, the MSF can
exhibit three different behaviors: 
type I, where $\Lambda(\gamma)$ is always positive; type
II, where $\Lambda(\gamma)$ becomes negative in an open interval
$\gamma\in (\gamma_c,\infty)$; type III, where $\Lambda(\gamma)$
is negative in a finite range of values, i.e., for $\gamma \in
(\gamma_{c1}, \gamma_{c2})$. For systems in the first class
synchronization never occurs. Conversely,
synchronization can always be achieved by systems in the second class,
provided that the coupling strength $g$ is sufficiently strong, i.e. when
$g>\gamma_c/\lambda_2$. Finally, in the third class, stability is
ensured when, simultaneously, $g\lambda_2>\gamma_{c1}$ and
$g\lambda_{MN}<\gamma_{c2}$, a condition impossibile to meet if
$\lambda_{MN}/\lambda_2 > \gamma_{c2} / \gamma_{c1}$. Consequently,
for class II systems, the larger is the first non-zero
eigenvalue, the easier is to reach synchronization, and the 
\emph{synchronizability} can be measured by the value
of $\lambda_2$. Instead, for class III systems, the eigenratio
$r=\lambda_N/\lambda_2$ is usually adopted to study
synchronizability, with smaller values of $r$ indicating
higher levels of synchronizability. 

To study the stability of the synchronized state of
Eq.~(\ref{eq:syncmodelKronLINArenas}), we therefore need to evaluate
$\lambda_2$ and $r=\lambda_N/\lambda_2$ of the supra-Laplacian matrix
$\mathcal{L}^s$. For class II systems, the analysis of 
the first non-zero eigenvalue $\lambda_2$ of the supra-Laplacian 
$\mathcal{L}^s$ as a function of ${w}/{g}$ shows the onset of
different regimes \cite{sole2013spectral}.
For small ${w}/{g}$, i.e.~when the
inter-layer is weak compared to the intra-layer coupling,
$\lambda_2$ of the supra-Laplacian $\mathcal{L}^s$ is typically dominated
by the second largest eigenvalue of $\mathcal{L}^I$,  
whereas increasing ${w}/{g}$ the emergence of super-diffusion
discussed in Section \ref {diff1} and in Fig.~\ref{fig:diffusion_diff}
also leads to an enhanced syncronizability of the multiplex structure
compared to the syncronizability of the individual layers.
On the contrary, for class III systems, \textcite{sole2013spectral} have
shown that there is an optimal value of the parameter ${w}/{g}$ that yields
the minimum of $r=\lambda_N/\lambda_2$, and have derived such optimal
value in the case of a multiplex networks whose layers are
Erd\H{o}s-R\'enyi (ER) random graphs.

\mattia{A second analytically tractable case is studied in~\cite{tang2019master}. There, the strong assumption that the inter- and intra-layer coupling functions are equal to each other is relaxed, and the more general case of different (possibly nonlinear) functions is considered. This leads to the following linearization of Eq.~(\ref{eq:syncmodelKron}) around the synchronous state $\tilde{\mathcal S}$}: 
\begin{equation}
\frac{d(\delta \mathcal S)}{dt} = \left ( I_{M \times N} \otimes {\text J} \bm f- g  (\mathcal L^{L} \otimes {\text J} \bm h ) - w (\mathcal L^{I} \otimes {\text J} \bm v )\right ) \delta \mathcal S
\label{eqpertinter}
\end{equation}
\noindent
\mattia{Remarkably, Eq.~(\ref{eqpertinter}) can be simultaneously
diagonalised and decoupled under the further assumption that $\mathcal L^{L}$ and $\mathcal L^{I}$ commute, i.e. $\mathcal L^{L}\mathcal L^{I} = \mathcal L^{I} \mathcal L^{L}$:}
\begin{equation}
\frac{d\bm \zeta_{h}}{dt} = \bigl [ { {\text J} \bm f} - g
  \lambda_{h}^{L} { {\text J} \bm h - w \lambda_{h}^{I} {\text J} \bm
    v} \bigr ] \bm \zeta_{h},
\label{eq_eigmodesbis2}
\end{equation}
\mattia{The commutativity of $\mathcal{L}^L$ and $\mathcal{L}^I$ is a strong assumption, but it is strictly required to decouple the equations. However, \textcite{tang2019master} present numerical examples of multiplex structures where $\mathcal{L}^L$ and $\mathcal{L}^I$ do not commute, yet network synchronization can still be predicted using this method.}

In Eqs.~(\ref{eq_eigmodesbis2}) $\lambda_{h}^{L}$ and $\lambda_{h}^{I}$, with $h=1,\ldots,MN$, 
are the eigenvalues of $\mathcal L^{L}$, and $\mathcal L^{I}$.  
Given the definition of $\mathcal L^{L}$ and $\mathcal L^{I}$ in
Eq.~\eqref{eq:LLLI}, both these matrices have multiple zero
eigenvalues. Similarly to Eq.~\eqref{eq_eigmodes2}, also 
Eq.~\eqref{eq_eigmodesbis2} can be rewritten 
in a parametric form: 
\begin{equation}
  \frac{d\bm \xi}{dt} = \bigl [ { {\text J} \bm f} - \gamma { {\text J} \bm h
      - \chi {\text J} \bm
    v} \bigr ] \bm \xi,
\label{eq_eigmodesbis3}
\end{equation}
\noindent where we now have two parameters, $\gamma$ and $\chi$. 
For $\gamma=\chi=0$ we have the mode along the
synchronization manifold, while in all the other cases the modes are
transverse to it. There are, however, two other important cases that
arise when either $\chi=0$ or $\gamma=0$. In the first case,
Eq.~(\ref{eq_eigmodesbis3}) reduces to 
$d\bm \xi /dt = \bigl [ {
    {\text J} \bm f} - \gamma {\text J} \bm h \bigr ] \bm
\xi$, which accounts for when there is no inter-layer
coupling. This equation allows to calculate the MSF for each independent
intra-layer network, provided that the synchronous
state around which each master stability equation is calculated is the
same for all layers. Similarly, for $\gamma=0$, Eq.~(\ref{eq_eigmodesbis3})
reduces to $d\bm \xi /dt = \bigl [ { {\text J} \bm f} - \chi {\text J} \bm v \bigr ] \bm \xi$, which represents the situation where there is no intra-layer coupling and, hence, allows one to calculate the MSF
for each independent inter-layer network, provided that the 
synchronous state is the same for all networks.

From Eq.~(\ref{eq_eigmodesbis3}) and its reductions, three regions 
where the maximum Lyapunov exponent is negative are obtained:
$R_{\gamma,\chi}=\{(\gamma,\chi)|\Lambda(\gamma,\chi)<0\}$,
$R^{intra}_{\gamma,\beta}=\{(\gamma,\chi)|\Lambda(\gamma)<0\}$, and
$R^{inter}_{\gamma,\chi}=\{(\gamma,\chi)|\Lambda(\chi)<0\}$. For any given 
connectivity, these three regions can be parametrized in
terms of the coupling coefficients $g$ and $w$, i.e., $R_{g,w}$,
$R^{intra}_{g,w}$, $R^{inter}_{g,w}$, allowing one to compute the
values of the coupling strengths leading to
synchronization. Complete synchronization requires that all the
transverse modes damp out, and, hence, that $g$ and $w$ lie in the
intersection among the three regions: $(g,w) \in R_{g,w} \cap
R^{intra}_{g,w} \cap R^{inter}_{g,w}$. 


Let us now move to discuss the \mattia{stability analysis for} inter-layer and
intra-layer synchronization. Inter-layer synchronization takes places
when each unit is synchronized with all its replicas, regardless of
whether or not it is synchronized with the other units of its
layer. In a dual manner, intra-layer synchronization occurs when each
unit is synchronized with the other nodes of its layer, regardless of
whether or not it is synchronized with its replicas. As we have
already mentioned, in the presence of coupling functions that are
diffusive, the synchronous manifolds associated to these forms of
synchronization are guaranteed to exist, but their stability in
general requires further conditions, \mattia{which have been investigated using the MSF approach in the case of identical nodes \cite{sevilla2016inter,tang2019master}, as well as in the case of nodes mismatched across layers \cite{anwar2023synchronization}.}


For simplicity, we \mattia{restrict the analysis to the case of identical nodes and} illustrate the method by referring to a multiplex
structure of two layers with the same topology and with linear
coupling. Under these assumptions, Eq. (\ref{eq:syncmodel}) becomes:
\begin{equation}
\label{eq:syncmodelLINDuplex}
\begin{array}{c}
d{\bm \sigma_i\lay{1}}/dt
  =   \bm f (\bm \sigma_i\lay{1})  - g  \sum\limits_{j=1}^N l_{ij} H\, \bm \sigma_j\lay{1} + w  V (\bm \sigma_i\lay{2}-\bm \sigma_i\lay{1})\\
d{\bm \sigma_i\lay{2}}/dt
  =   \bm f (\bm \sigma_i\lay{2})  - g  \sum\limits_{j=1}^N l_{ij} H\, \bm \sigma_j\lay{2} + w  V (\bm \sigma_i\lay{1}-\bm \sigma_i\lay{2})
\end{array}
\end{equation}

We begin our discussion illustrating the intra-layer synchronization
case. Let us denote with $\bm{\tilde{\sigma}}\lay 1$ and
$\bm{\tilde{\sigma}}\lay 2$ the synchronized trajectories in the two
layers. These states obey the following equations:
\begin{equation}
\label{eq:syncmodelLINDuplex2}
\begin{array}{c}
d\bm{\tilde{\sigma}}\lay 1/dt
  =   \bm f (\bm{\tilde{\sigma}}\lay 1)  + w  V (\bm{\tilde{\sigma}}\lay 2-\bm{\tilde{\sigma}}\lay 1)\\
d\bm{\tilde{\sigma}}\lay 2/dt
  =   \bm f (\bm{\tilde{\sigma}}\lay 2)  + w  V (\bm{\tilde{\sigma}}\lay 1-\bm{\tilde{\sigma}}\lay 2)
\end{array}
\end{equation}
which are obtained from Eq.~(\ref{eq:syncmodelLINDuplex}), since in the synchronization manifold we have 
$\bm \sigma_1\lay 1=\ldots=\bm
\sigma_N\lay 1=\bm{\tilde{\sigma}}\lay 1$ and $\bm \sigma_1\lay
2=\ldots=\bm \sigma_N\lay 2=\bm{\tilde{\sigma}}\lay 2$.
Considering now small perturbations 
and linearizing Eq. (\ref{eq:syncmodelLINDuplex}) around 
$\bm{\tilde{\sigma}}\lay 1$ and $\bm{\tilde{\sigma}}\lay 2$, we get:
\begin{equation}
\label{eq:syncmodelLINDuplexKRON}
\frac{d(\tilde{\bm{\delta \sigma}})}{dt}
  = \left [ I_N\otimes \left ( \tilde{J\bm f} -w\,L^I\right) \right] \tilde{\bm{\delta \sigma}}-g\left (L\,\otimes I_2 \otimes H \right)\tilde{\bm{\delta \sigma}}
\end{equation}
with {\small$\tilde{\bm{\delta \sigma}}_i=\left[(\tilde{\bm{\delta \sigma}}_i\lay{1})^T,(\tilde{\bm{\delta \sigma}}_i\lay{2})^T\right]^T$,
$\tilde{\bm{\delta \sigma}}=\left[\tilde{\bm{\delta
      \sigma}}_1^T,\ldots,\tilde{\bm{\delta \sigma}}_N^T\right]^T$,
$\tilde{J\bm f}=\left ( \begin{array}{cc} J\bm f (
\bm{\tilde{\sigma}}\lay 1 ) &
0 \\ 0 & J\bm f ( \bm{\tilde{\sigma}}\lay 2
) \end{array}\right)$,
$L^I=\left ( \begin{array}{cc} 1 & -1 \\ -1 &
  1 \end{array}\right)$}, and the Laplacian matrix of the intra-layer
connectivity, shared by the two layers, is denoted by $L$. If
the intra-layer network is undirected and connected, then new
transformed variables $\bm \zeta=(T^{-1}\otimes
I_{2d})\tilde{\bm{\delta \sigma}}$ can be defined, where
$T$ is the matrix containing the orthonormal
eigenvectors of $L$. In this new reference system,
we have $N$ decoupled equations:
\begin{equation}
\label{eq:syncmodelLINDuplexKRONdecoupled}
\frac{d\bm \zeta_h}{dt}
  = \left ( \tilde{J\bm f} -w\,L^ I\right) {\bm \zeta_h}-g\lambda_h\left (I_2 \otimes H \right) \bm \zeta_h
\end{equation}
where the transverse modes correspond to $h=2,\ldots,N$.
Eq. (\ref{eq:syncmodelLINDuplexKRONdecoupled}) can be parametrized as a function of $\alpha=g\lambda_h$:
\begin{equation}
\label{eq:syncmodelLINDuplexKRONdecoupledGeneric}
\frac{d\bm \xi}{dt}
  = \left ( \tilde{J\bm f} -w\,L^I\right) {\bm \xi}-\alpha\left(I_2 \otimes H \right) \bm \xi
\end{equation}
from which the MSF can be calculated. Notice that, at variance with the
MSF of the previous cases, here $\xi \in \mathbb{R}^{2d}$.

We now move to discuss the stability of inter-layer synchronization.
The synchronous trajectory  $\bm{\tilde{\sigma}_i}$ at node
$i$ obeys the equation:
\begin{equation}
\label{eq:manifoldinterlayer}
\frac{d\bm{\tilde{\sigma}_i}}{dt}
  =   \bm f (\bm{\tilde{\sigma}_i})  -g\sum\limits_{j=1}^Nl_{ij} H\,\bm{\tilde{\sigma}_j}
\end{equation}
which represents the dual equation of Eq.~\eqref{eq:syncmodelLINDuplex}.
In fact, here it is the coupling term
between the layers that vanishes as, on the synchronization manifold,
we have that $\bm \sigma_i\lay1=\bm \sigma_i\lay2$, $\forall
i$. Linearization around the synchronous solution yields:

\begin{equation}
\label{eq:syncmodel2LINDuplexLIN2compact}
\frac{d(\bar{\bm{\delta \sigma}})}{dt}
  =   J\bm F (\bm{\tilde{\sigma}_1},\bm{\tilde{\sigma}_2},\ldots,\bm{\tilde{\sigma}_N})  - g  \left( L \otimes H \right) \bar{\bm{\delta \sigma}} -2 w  V \bar{\bm{\delta \sigma}}
\end{equation}

\noindent where $\bar{\bm{\delta \sigma}}=[\bar{\bm{\delta
      \sigma}_1}^{T},\ldots,\bar{\bm{\delta \sigma}_N}^{T}]^T$ with
$\bar{\bm{\delta \sigma_i}}=\bm{\delta \sigma_i}\lay{1}-\bm{\delta
  \sigma_i}\lay{2}$, and $J\bm F
(\bm{\tilde{\sigma}_1},\bm{\tilde{\sigma}_2},\ldots,\bm{\tilde{\sigma}_N})=\mathrm{diag}\{J\bm
f (\bm{\tilde{\sigma}_1}),\ldots,J\bm f
(\bm{\tilde{\sigma}_N})\}$. Eq. (\ref{eq:syncmodel2LINDuplexLIN2compact})
describes the dynamics of the modes transverse to the interlayer synchronization
manifold, such that the computation of its maximum Lyapunov exponent
allows to assess whether the synchronous solution is stable or not.

\begin{figure}
\begin{center}
\includegraphics[width=0.6\textwidth]{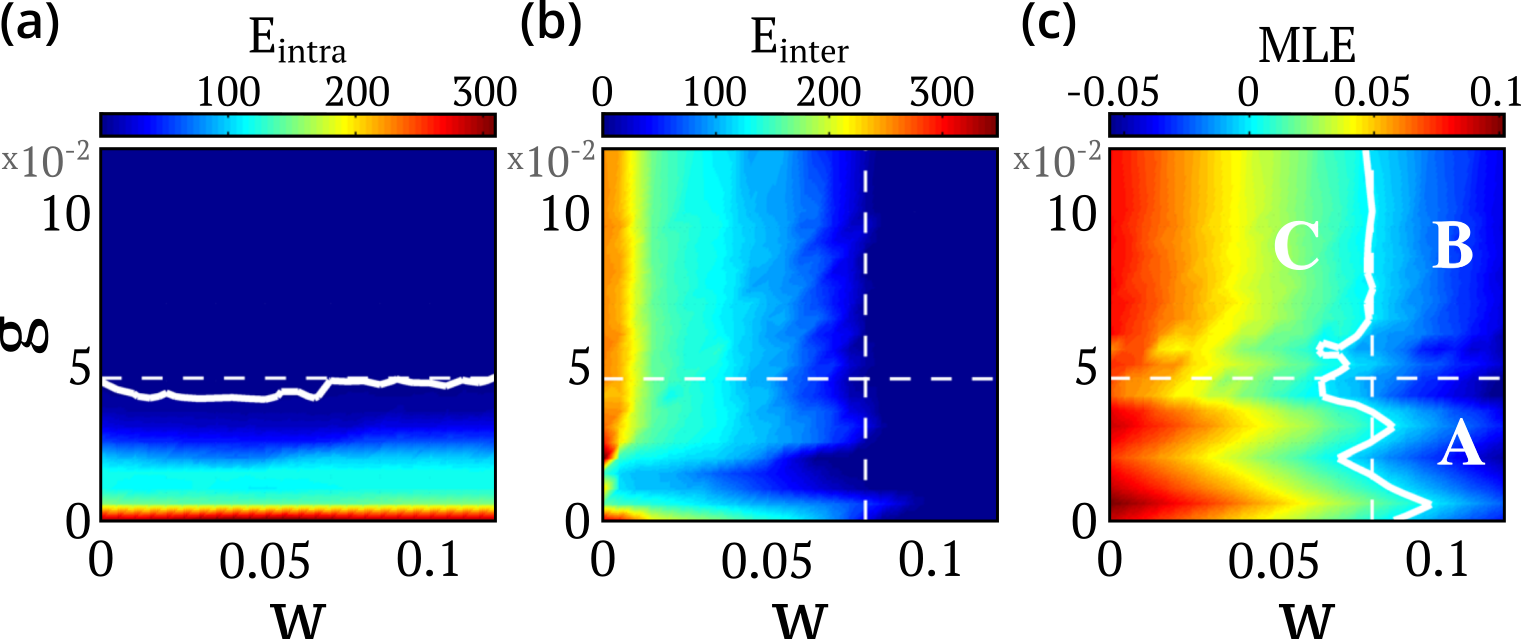}
\caption[]{Synchronization diagram of a duplex network of $N=500$ R\"ossler
  oscillators. (a) Intra-layer synchronization error $E_{intra}=\lim\limits_{T\rightarrow \infty}\frac{1}{T}\int\limits_{0}^T\sum\limits_{j=2}^N\|\bm \sigma_j(t) -\bm \sigma_1(t)\|dt$. (b) Inter-layer synchronization error $E_{inter}=\lim\limits_{T\rightarrow \infty}\frac{1}{T}\int\limits_{0}^T\|\bar{\bm{\delta \sigma}}\|dt$. (c) Maximum Lyapunov exponent (MLE) of Eq. (\ref{eq:syncmodel2LINDuplexLIN2compact}). The horizontal dashed lines represent the synchronization threshold of the isolated layers (i.e., $w=0$), whereas the vertical ones the synchronization threshold for a pair of nodes (i.e., $g=0$). Finally, the white bold curve in panel (c) marks the transition from negative to positive sign of the MLE.
Figures adapted from \textcite{sevilla2016inter}.\label{fig:synchro_boccchaos}}
\end{center}
\end{figure}

As an example of inter- and intra-layer synchronization,
we consider $2N$ chaotic R\"ossler
oscillators placed on the nodes of the two layers of a duplex,
as in~\textcite{sevilla2016inter}. Here, since R\"ossler
oscillators are 3-dimensional and $M=2$, then $d=3$ and $D=6$. An interesting scenario arises when
the coupling functions are selected such that the intra-layer coupling yields 
a type III MSF, while the inter-layer coupling yields a type II MSF. In
this situation inter-layer and intra-layer synchronization can
coexist. The phase diagram of a duplex network,  
with connectivity being 
an ER random graph with $\langle k \rangle=16$ at each
layer, is shown in
Fig.~\ref{fig:synchro_boccchaos}. Here, three different regions
(marked as A, B, and C) are identified in the parameter space $w-g$. 
Region A denotes the occurrence of inter-layer synchronization without
intra-layer synchronization; in region B both inter-layer and
intra-layer synchronizations exist, such that the system undegoes
complete synchronization; finally in region C intra-layer synchronization
without inter-layer synchronization occurs. Notice that, for
intermediate values of $g$, inter-layer synchronization is achieved
for values of $w$ below the synchronization threshold for a pair of
nodes (represented by the dashed vertical line in 
figure). The remarkable effect here is that inter-layer and
intra-layer synchronization may enhance each
other in a multiplex network. 

\mattia{So far, we have considered identical node dynamics. Although, in principle, the more general case of non-identical dynamics could be addressed by extending MSF-based approaches in analogy with single-layer networks, this scenario has instead been investigated using other techniques. Specifically, methods based on Lyapunov functions \cite{han2025intralayer} or spectral graph theory \cite{liu2023intralayer} have been used for the purpose of determining the conditions for synchronization stability of heterogeneous multiplex networks.}

\subsection{Relay synchronization}

In single-layer networks synchronization can also occur through
dynamical relaying
\cite{banerjee2012enhancing,fischer2006zero,gutierrez2013generalized}. In
its simplest form, this type of synchronization appears in a set of
three nodes, where oscillator A is connected to oscillators B and C,
but B and C are not connected each other.  In such configuration, node
A can act as a dynamical relay, enabling synchronization between B and
C without synchronizing with them. Quite often, B and C display a
stronger form of synchronization (e.g., zero-lag synchronization),
compared to the synchronization between A and B or A and C
(e.g., lag synchronization). 
Relay (also known as remote)
synchronization has been found in networks with symmetries 
\cite{nicosia2013remote} or in the presence of heterogeneties 
\cite{gambuzza2013analysis,gambuzza2016inhomogeneity}, and is of high
relevance in brain networks
\cite{vuksanovic2014functional,vlasov2017hub} where the transfer of
information between distant cortical areas \mattia{may be} mediated by the
thalamus, which acts as a relay through the thalamo-cortical pathways
\cite{guillery2002thalamic}.

\mattia{Relay synchronization has also been observed in multiplex networks, where the phenomenon takes peculiar forms. In particular, an entire layer (or more than one) can act as a relay for the nodes in other layers. This results in two distinct scenarios:  
{\em ii)} inter-layer synchronization is induced between layers that are not directly connected by inter-layer links; {\em ii)} intra-layer synchronization emerges among all nodes of a sparse or even disconnected layer.}

\begin{figure}[t!]
\begin{center}
\includegraphics[width=0.6\textwidth]{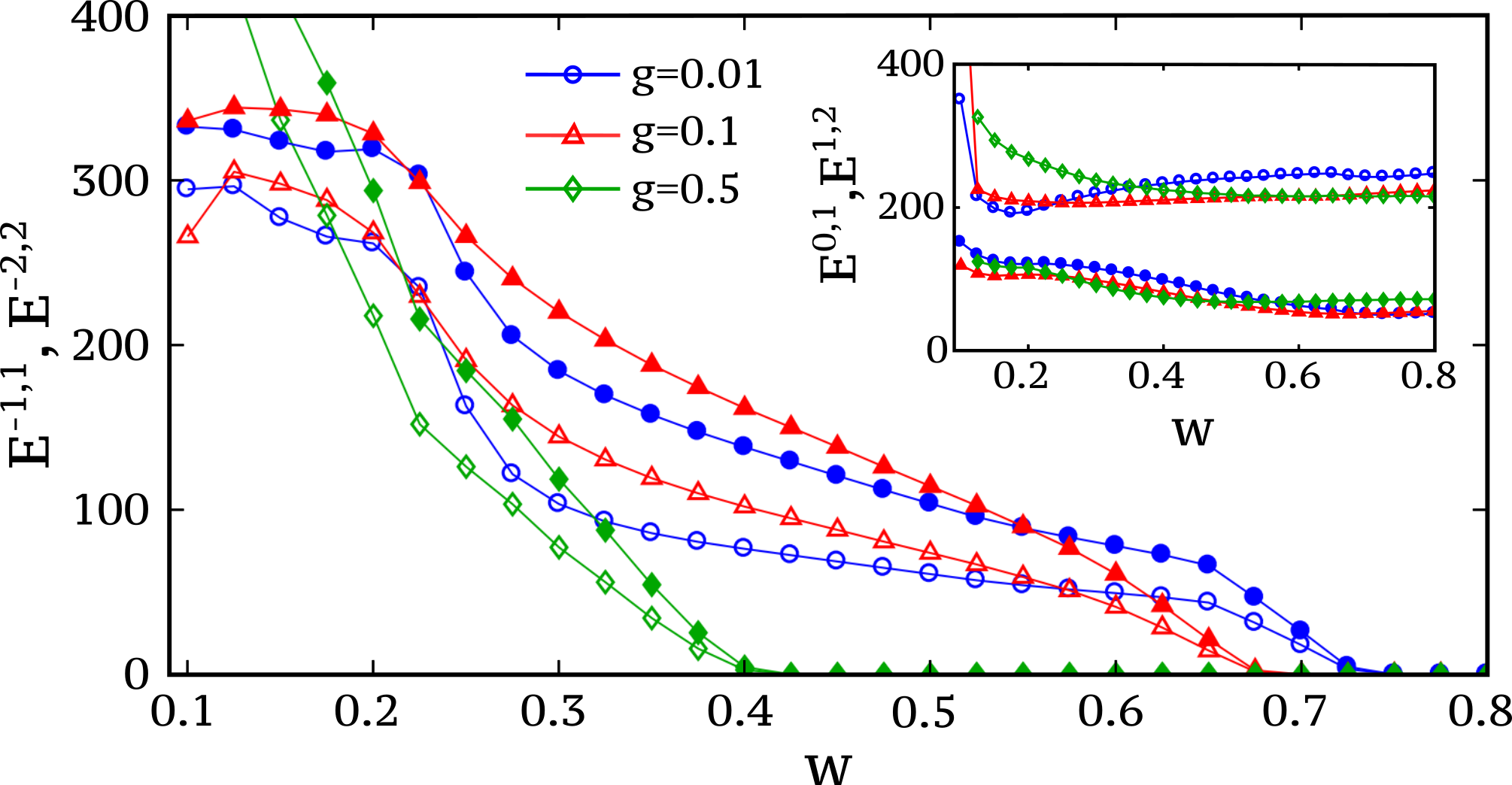}
\caption[]{Inter-layer synchronization mediated by dynamical
  relaying. Synchronization errors, defined as
  $E\laysup{\alpha,\beta}=\lim\limits_{T\rightarrow
    \infty}\frac{1}{T}\int\limits_{0}^{T}\sum\limits_{i=1}^N\|
  \sigma_i\lay\alpha(t)-\sigma_i\lay\beta(t)\|dt$, between paired
  layers, $E\laysup{-1,1}$ and $E\laysup{-2,2}$, for different values of the
  intra-layer coupling $g$ (main panel) and between one of the outer
  layers and the relay layer, $E\laysup{0,1}$ and $E\laysup{0,2}$, (inset). In all
  layers the topology is given by an ER network. Figure adapted from 
  \textcite{leyva2018relay}.
  \label{fig:5interlayersync}}
\end{center}
\end{figure}

As an example of the first form of synchronization,
\textcite{leyva2018relay} have considered a multiplex network
composed by an odd number of
layers, $M=2M_R+1$, indexed with $\alpha=-M_R,\ldots,0,1,\ldots,M_R$,
and such that layer $-\alpha$ and $\alpha$ have the same structure, i.e.  
$L\lay {-\alpha} = L\lay \alpha$, and node dynamics,
 i.e., $\bm f\lay\alpha=\bm
 f\lay{-\alpha}$.
 When 
%
$\bm f\lay\alpha \neq \bm f\lay{\alpha'}$ for
$\alpha \neq \pm \alpha'$, the multiplex network 
admits an inter-layer synchronous solution where 
layers $-\alpha$ and $\alpha$ 
evolve synchronously,
irrespectively of the presence or not of intra-layer synchronization.
The stability of this state can be studied by characterizing the maximum
Lyapunov exponent of a properly defined set of linearized equations
representing the modes transverse to the inter-layer synchronous
manifold. One finds that it is indeed 
possible to observe this type of relay synchronization, e.g. 
in a multiplex with $M=5$ layers of
$N=500$ R\"ossler oscillators, MSF class I  
intra-layer coupling functions, and MSF class II 
inter-layer coupling functions.
With this setting, intra-layer synchronization is not stable when   
the layers are considered in isolation or are weakly coupled.   
A sufficiently large inter-layer coupling
can instead induce synchronization in replica nodes of layers
$-\alpha$ and $\alpha$. 
%
%
Inter-layer synchronization errors are reported in
Fig.~\ref{fig:5interlayersync} for several values of the intra-layer
coupling $g$. We observe a critical value of
the inter-layer coupling $w$ 
beyond which inter-layer synchronization is observed
($E\laysup{-1,1}$ and $E\laysup{-2,2}$ approach zero)  
without intra-layer coherence ($E\laysup{0,1}$ and $E\laysup{1,2}$, reported in the inset, 
are significantly larger than zero).  
Moreover, the critical value of $w$ decreases as the intra-layer
coupling $g$ increases. 
Quite interestingly, the two pairs
of layers (layers -2 and 2, and layers -1 and 1) reach synchronization
simultaneously, denoting that inter-layer synchronization mediated by
dynamical relays is a global state for the multiplex network.

%

\begin{figure}[tb!]
\begin{center}
\includegraphics[width=0.46\textwidth]{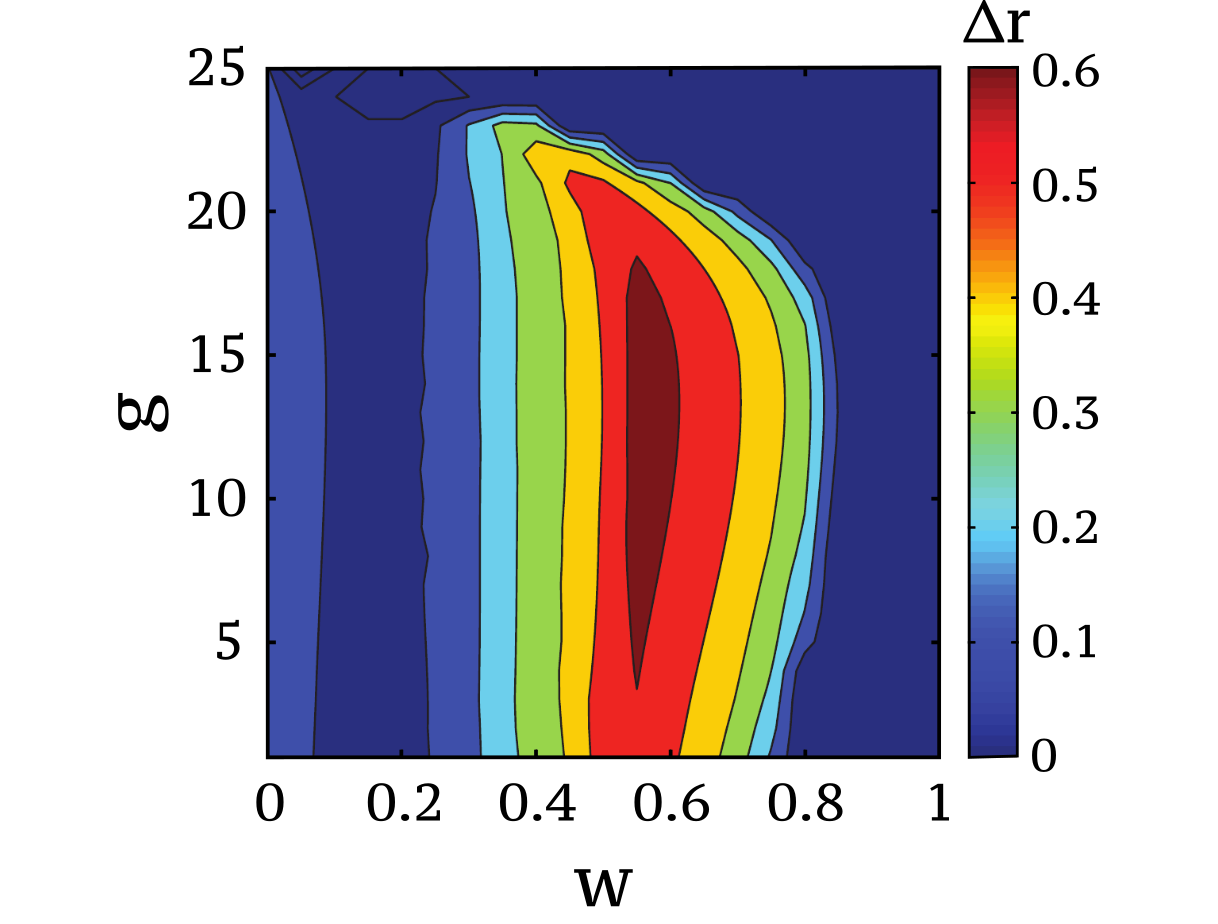}
\caption[]{Intra-layer synchronization mediated by dynamical relaying.
  The value of $\Delta r =  r\lay{\alpha} - r^{I}$ for a multiplex
  network with $N=100$ nodes, one empty and one fully-connected
  layer, is shown as a function of $g$ and $w$.  
  Figure adapted from \textcite{gambuzza2015intra}.}
\label{fig:synchro_gambuzza}
\end{center}
\end{figure}

%
Multiplexity is also able to produce intra-layer synchronization
of layers that would otherwise be asynchronous.  
\textcite{gambuzza2015intra} have considered the extreme case  
of a multiple network whose first layer is a complete graph and
the second layer is an empty graph. Despite the total absence of links
in the second layer, intra-layer dynamical synchronization 
can be achieved due to the inter-layer coupling. 
Remarkably, a complete intra-layer
synchronization in the empty layer can even be obtained in the absence
of inter-layer synchronization. At variance with the previous case, 
\textcite{gambuzza2015intra} have adopted 
Stuart-Landau oscillators with different 
natural frequencies in the two layers, so to implement a parametric
mismatch.  
The system behavior has been characterized by introducing 
%
a phase coherence between two oscillators, $i$ in
layer $\alpha$ and $j$ in layer $\beta$, as $r_{ij}\lay{\alpha
  \beta}=| \langle e^{{\mathrm i} (\theta_i\lay\alpha-\theta_j\lay\beta)}
\rangle_t|$, and by computing the phase coherence within a layer as
$r\lay{\alpha}=\frac{1}{N(N-1)}\sum\limits_{i,j=1}^N r_{ij}\lay{\alpha
  \alpha}$, and the inter-layer coherence as
$r^{I}=\frac{1}{N}\sum\limits_{i=1}^N r_{ii}\lay{\alpha
  \beta}$.
The values of $\Delta r = r\lay{\alpha} - r^{I}$ reported
in Fig.~\ref{fig:synchro_gambuzza} indicate that there exists a wide
range of inter-layer coupling strength $w$, intermediate
between desynchronization and global synchronization, where nodes
are synchronized inside each layer, despite the lack of inter-layer
coordination.

\subsection{Cluster synchronization and chimera states}
\label{subsec:ClusterSynchronization}
\label{subsec:ChimeraStates}

In cluster synchronization, the nodes split into different groups
where the units within each cluster converge to the same trajectory,
that is however distinct from that of the other groups. In
single-layer networks this form of synchronization is associated to
the presence of symmetries
\cite{pecora2014cluster,sorrentino2016complete} or equitable
partitions \cite{schaub2016graph,gambuzza2019criterion} \mattia{in the network structure}.
%
The multiplex intra- and inter-layer synchronization discussed in the
previous section are two examples of this form of synchronization,
where groups correspond to layers or to ensembles of replicas,
respectively.


\mattia{An experiment showing cluster synchronization in a multiplex structure has been
carried out using a configuration of Colpitts circuits with 
resistive and magnetic couplings \cite{blaha2019cluster}.} Despite the small number of units considered (four
periodic oscillators), the system exhibits a very rich dynamical
behavior including bistability, hysteresis, quasiperiodicity, and a
clustered quasiperiodic state that should be ascribed to the presence
of symmetries in the highly regular structure. \mattia{The theoretical conditions for the emergence of cluster synchronization in multiplex networks can be derived as a special case of the more general framework developed multilayer networks in \cite{della2020symmetries}.}

The term cluster synchronization is sometimes used with a different
meaning, namely to refer to groups of phase-synchronized nodes in a
network of non-identical weakly coupled oscillators
\cite{jalan2005synchronized,amritkar2005synchronized}.  These
phase-synchronized clusters are also found in multiplex networks. For
instance, \textcite{jalan2016cluster,singh2017interplay} have considered the
case of coupled maps, showing that
the cluster synchronizability of a layer can be either enhanced or
degraded, with a non-trivial dependence of the phenomenon on the
strength of the coupling, the density of connectivity in the two
layers, and their architecture.

Chimera states represent another dynamical regime where not all the
nodes of the network converge to the same synchronous solution. In
particular, they are characterized by the coexistence, in a
symmetrical structure, of a coherent domain, formed by synchronous
oscillators, and an incoherent one, where the units are not
synchronized
\cite{kuramoto2002coexistence,abrams2004chimera,panaggio2015chimera}. These
states are of great theoretical interest, especially in the context of
neuroscience, where they are related to unihemispheric sleep
\cite{rattenborg2000behavioral,abrams2008solvable} and spatial
patterns involved in the cognitive organization of the brain
\cite{bansal2019cognitive}, and in experimental settings of light
modulators \cite{hagerstrom2012experimental}, mechanical
\cite{martens2013chimera,matheny2019exotic} and electronic systems
\cite{gambuzza2014experimental,gambuzza2020experimental}.

As an example of chimera states in multiplex networks 
\textcite{majhi2017chimera} have considered two layers of
Hindmarsh-Rose oscillators. Aiming at 
studying neuronal activity in populations of uncoupled neurons, they
assume that the first layer is composed of isolated neurons,
while the second of coupled neurons. The intra-layer
coupling function in the second layer models electrical synapses 
accounting for a form of non-local interaction, while  
the inter-layer coupling function represents instead
chemical synapses. Varying the strength of the inter-layer
coupling, the system dynamical behavior can be
modulated from incoherent to coherent, through chimera states and cluster
synchronization. In particular, due to the onset of inter-layer
synchronization, in the chimera states the regions of coherence and
incoherence correspond to the same neurons in the two layers. 
These chimera states can not be observed in the absence of chemical synapses,
and are therefore a phenomenon emerging from the interplay between the
two types of coupling in the multiplex network.
This result can also be seen as an 
example of relay synchronization, given that the units in the second layer
partially or fully synchronize despite the lack of direct
interactions.

\textcite{sawicki2018synchronization,sawicki2019chimeras} have studied
a multiplex of three layers of FitzHugh-Nagumo oscillators, where each
layer implements a non-locally coupled topology. In this system, relay
synchronization of chimera states is observed, with coexisting
coherent and incoherent domains appearing in the outer layers. In
addition to this, the coherent domains in the two outer layers can
synchronize each other, while the incoherent domains remain
desynchronized. 
These results show that is possible to elicit a desired
state in a certain layer without a direct manipulation of
its parameters (that can be not accessible or difficult to maneuver),
but acting on it through another layer.
This is further demonstrated in other works on chimera states in multiplex networks \cite{ghosh2016emergence,ghosh2018non,mikhaylenko2019weak,rybalova2022multiplexing}. In particular, \textcite{mikhaylenko2019weak} consider a 
two-layer of FitzHugh-Nagumo oscillators, where: {\em i)} by tuning the
coupling strength in one layer it is possible to induce chimera states
with desired mean phase velocity profiles; {\em ii)} by tuning the
intra-layer coupling strength, \mattia{chimera states with a single incoherent domain may be
suppressed and in-phase synchronization and chimera states with two incoherent domains may be
induced.}

\subsection{Explosive synchronization}

\mattia{Phase oscillators are possibly the simplest model to study synchronization of interacting units; they reproduce the case when the coupling strength is weak compared to the attraction to the limit cycle, such that the dynamics of each oscillator can be represented by a single \emph{phase} variable \cite{acebron2005kuramoto,rodrigues2016kuramoto}.} In single-layer networks of phase oscillators, such as in the Kuramoto model, the transition to synchronization is usually second-order 
\cite{acebron2005kuramoto}. 
There are however cases where an abrupt, first-order
transition, known as explosive synchronization, is observed
\cite{gomez2011explosive,leyva2013explosive,zhang2013explosive,zhang2015explosive}. \mattia{Explosive synchronization may arise under different conditions, including degree–frequency correlations, dynamical feedback between coupling strength and local coherence, the presence of delays, and structural modularity. A common feature of these mechanisms is that they hinder the formation of large synchronized groups, while promoting the persistence of small clusters until a sudden global transition occurs.}

\begin{figure}
\begin{center}
\includegraphics[width=0.55\textwidth]{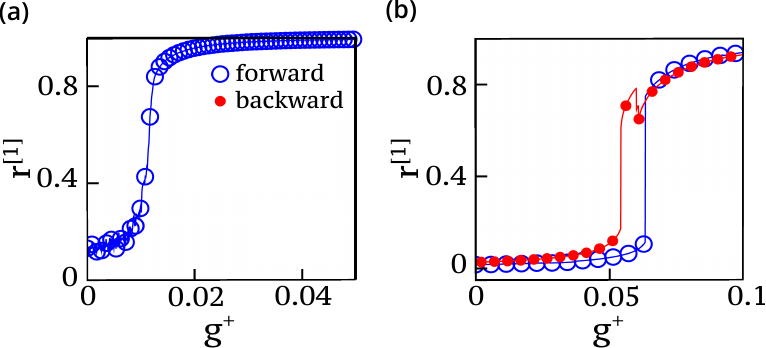}
\caption[]{Transition to synchronization in a multiplex network with an excitatory layer and an
  inhibitory one. The order parameter $
  r\lay \alpha=\lim\limits_{T\rightarrow +\infty}{\frac{a}{T}\int\limits_{t_r}^{t_r+T}\left| \frac{1}{N}\sum\limits_{i=1}^N e^{ \mathrm{i} 
      \sigma_i\lay\alpha}\right|dt}$ with $\alpha=1$ is reported as a function of the coupling strength  
  of the excitatory layer. (a) When $w=0$, the transition in the excitatory layer is second-order. (b)
  When $w=2$, the transition becomes first order, while the inhibitory layer still remains incoherent. Figures adapted from \textcite{jalan2019inhibition}.\label{fig:ESmultiplex}}
\end{center}
\end{figure}

Explosive synchronization also occurs in multiplex structures, where it features unique
characteristics. 
A first relevant result is that multiplexing one layer, which in
isolation is not supporting an explosive transition, with one that, on the contrary,
is exhibiting this type of transition, triggers off explosive transition 
also in the first layer~\cite{kachhvah2017multiplexing}. Even more interestingly, explosive transition
can be induced in a multiplex in cases where no layer in isolations 
would exhibit it~\cite{jalan2019inhibition}.
To show this, let us consider a multiplex
of one excitatory and one inhibitory layer with  
$d=1$, so that the system state variables 
$\sigma_i\lay{1}$ and $\sigma_i\lay{2}$, with $i=1,\ldots,N$,
are nodes phases as in the Kuramoto model. The dynamical equations read:
\begin{equation}
\label{eq:SarikaExplosive}
{\small
\begin{array}{cc}
d \sigma_i\lay{1}/dt & =   \omega_i\lay{1} + g^+ \sum\limits_{j=1}^N a_{ij}^1 \sin(\sigma_j\lay{1}-\sigma_i\lay{1})+w\sin(\sigma_i\lay{2}-\sigma_i\lay{1}) \\[0.2 em]
d \sigma_i\lay{2}/dt & =   \omega_i\lay{2} + g^- \sum\limits_{j=1}^N a_{ij}^2 \sin(\sigma_j\lay{2}-\sigma_i\lay{2})+w\sin(\sigma_i\lay{1}-\sigma_i\lay{2})
\end{array}
}
\end{equation}

\noindent where $\omega_i\lay{1}$ and $\omega_i\lay{2}$ are the
natural frequencies in the two layers. The parameter $g^+>0$ ($g^-<0$)
represents the positive (negative) coupling 
in the excitatory (inhibitory) layer 1 (layer 2).
An example of inhibition-induced explosive synchronization 
is obtained in a duplex with $N=50$, a fully connected excitatory layer,
and a regular ring as the inhibitory layer.
Fig.~\ref{fig:ESmultiplex}(a) shows that, 
when the two layers do not interact, namely $w=0$, the transition to
synchronization in the excitatory layer is second-order.
Conversely, when $w > 0$ \mattia{(Fig.~\ref{fig:ESmultiplex}(b))}, a first
order transition with an abrupt change of the order parameter $r\lay 1$ and the presence
of an hysteresis loop can be observed.
This result is robust to the topology of both layers, provided that
the inhibitory coupling strength is significantly larger than the
excitatory, i.e., $g^-\gg g^+$. A strong inhibitory coupling 
hinders the onset of global synchronization in 
the inhibitory layer, favoring on the contrary the generation of local
clusters. This effect propagates via the inter-layer coupling to the
excitatory layer, where the formation of large synchronized clusters is
also inhibited. The onset of synchronization is therefore hampered until 
$g^+$ reaches a critical value at which an abrupt
transition occurs. \mattia{Analogously to single-layer networks, explosive synchronization is driven by a mechanism that prevents the emergence of large synchronous clusters. In contrast, in multiplex networks this inhibition arises inherently from the multiplex structure itself.}

\subsection{Oscillators coupled \mattia{via edge-colored graphs}}
\label{subsec:multilayersameobject}

At variance with the previous sections, here, we consider the case
of a single set of oscillators coupled through different
types of interactions. This situation can be described by a multiplex
network where the nodes across the layers are exactly the same physical
elements, \mattia{namely an edge-colored graph.}
%
%
%
%
%
More in details, the multiplex network is composed by $N$ nodes
interacting through $M$ layers, each one generally having its own
topology and representing a different type of interaction. 
Accordingly, the coupling function, indicated as $\bm
h\lay{\alpha}$, is generally layer-specific.
In the case of diffusive
coupling, the dynamics of each oscillator is described
by~\cite{delgenio2016synchronization}:  
\begin{equation}
\frac{d{\bm \sigma_i}}{dt} = \bm f (\bm \sigma_i) - \sum_{\alpha=1}^M
g\lay{\alpha} \sum_{j=1}^N l_{ij}\lay{\alpha} \bm h\lay{\alpha}(\bm
\sigma_j),
\label{dyn_oscill_mult}
\end{equation}
\noindent where the dynamical state $\bm \sigma_i$ 
of node $i$ is a vector of    
$\mathbb{R}^d$ (since in this case $D=d$), and 
$l_{ij}\lay {\alpha}$ are the entries of the Laplacian at layer
$\alpha$.
Once again, the existence of an invariant solution
${\mathcal{S}^*}(t)=\bm 1_{N} \otimes {\bm \sigma}^*(t)$,
where ${\bm \sigma}^*(t)$ is the common, synchronous trajectory,
namely  $\bm \sigma_1(t)=
\ldots = \bm \sigma_N (t)={\bm \sigma}^*(t)$, is guaranteed by the property
that the Laplacians are zero row-sum matrices.  
To study stability of the synchronization manifold,
Eqs.~(\ref{dyn_oscill_mult}) are linearized around $\mathcal{S}^*$ and the
dynamics of the synchronization error $\delta \mathcal{S} = \{ \delta \bm
\sigma_1, \ldots, \delta \bm \sigma_N \}$, with $\delta \bm \sigma_i
\equiv \bm \sigma_i - \bm \sigma^*$, is derived:
\begin{equation}
\frac{d(\delta \mathcal{S})}{dt} = \biggl ( I_N \otimes {\text J} \bm f({\bm \sigma}^*) -  \sum_{\alpha=1}^M g\lay {\alpha} L\lay {\alpha} \otimes {\text J} \bm h\lay {\alpha}  \biggr ) \delta \mathcal{S},
\label{dyn_oscill2_mult}
\end{equation}

Next, new variables defined via the transformation
${\bm \zeta}=(T^{-1} \otimes  I_D)\delta \mathcal{S}$, where
$T$ is the matrix containing the orthonormal eigenvectors of
$L\lay {1}$, are considered. In this way, we obtain the dynamics of the
transverse modes:
\begin{equation}
\begin{array}{lll}
\frac{d {\bm \zeta}_h}{dt} & = & \biggl(J \bm f ({\bm \sigma}^*) - g\lay{1}
\lambda_h\lay{1} J \bm h\lay{1} ({\bm \sigma}^*) \biggr ) \bm \zeta_h \\ & &
- \sum\limits_{\alpha=2}^M g\lay{\alpha} \sum\limits_{j=2}^N
{\tilde l}_{hj}\lay{\alpha} J \bm h\lay{\alpha} (\bm \sigma^*) \bm \zeta_j,
\end{array}
\label{dyn_oscill3}
\end{equation}
\noindent where $\tilde{L}\lay{\alpha} =  T^{-1} L\lay{\alpha} T$ and 
$h=2,\ldots, N$. Eq.~(\ref{dyn_oscill3}) provides the
condition for synchronization stability. Consider the norm of $ \bm
\Omega $, where $\bm \Omega \equiv (\bm \zeta_2, \ldots, \bm
\zeta_N)$. Since $ \| \bm \Omega \| (t) \approx \exp (\Lambda t)$, with
$\Lambda$ being the maximum Lyapunov exponent, stability of the
synchronization manifold requires that $\Lambda < 0$.
%
%
%
In the special case when the Laplacians commute, such that they are
diagonalized by the same matrix $T$, Eq.~\eqref{dyn_oscill3} simplifies into:
\begin{equation}
\frac{d {\bm \zeta}_h}{dt} = \biggl(J \bm f (\bm \sigma^*)  - \sum\limits_{\alpha=1}^M g\lay{\alpha} \lambda_h\lay{\alpha} J \bm h\lay{\alpha} ({\bm \sigma}^*) \biggr ) \bm \zeta_h,
\label{dyn_oscill4}
\end{equation}
\noindent where $h=2,\ldots,N$ and the transverse modes are also decoupled each other.

\begin{figure}
\begin{center}
\includegraphics[width=0.425\textwidth]{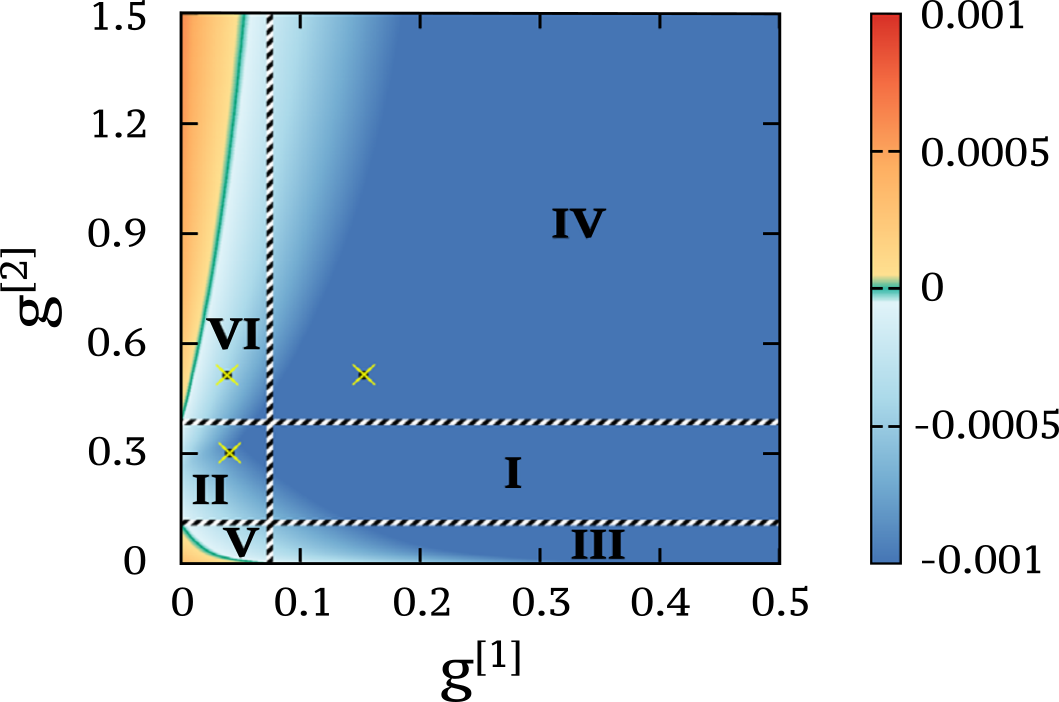}
\caption[]{Maximum Lyapunov exponent for \mattia{a set of R\"ossler oscillators coupled via an edge-colored graph with
$M=2$ layers (each constituted by an ER network)} as a function of the coupling strengths $g\lay 1$ and $g\lay 2$. The
  dashed lines are the boundaries of the stability regions in the
  isolated layers, i.e., when either $g\lay 1=0$ or $g\lay
  2=0$. Figure adapted from \textcite{delgenio2016synchronization}.}
\label{fig:synchro_delgenio}
\end{center}
\end{figure}

As an example of the rich dynamics that can be obtained, let us consider 
\mattia{a set of R\"ossler oscillators coupled via an edge-colored graph including
$M=2$ layers of interactions. Each layer is} characterized by an ER topology, with coupling
functions selected such that the oscillators interacting exclusively
through layer 1 (i.e., when $g\lay 2=0$) have class II MSF, \mattia{where synchronization can be obtained for a large enough coupling coefficient,} and, when
interacting exclusively through layer 2 (i.e., when $g\lay 1=0$), have
class III MSF, \mattia{where the values of the coupling coefficient leading to synchronization are bounded (see Sec. \ref{sec:secVA})}. Six different regions appear when the two coupling
strength $g\lay{1}$ and $g\lay{2}$ are varied (Fig.~\ref{fig:synchro_delgenio}).
In region I both layers are stable if considered in isolation. Regions
II, III and IV demonstrate how synchronization can be obtained thanks
to the stabilizing effect of one of the two layers on the instability
of the other. Finally, regions V and VI show that even  
two unstable layers can stabilize synchronization, thus offering another remarkable example of emergent
multiplex behavior.

\mattia{In the more general case, when the Laplacians do not commute, analyzing the stability of synchronization in a set of oscillators coupled via an edge-colored graph requires computing the maximum Lyapunov exponent from Eq.~\eqref{dyn_oscill3}, which consists of a set of coupled linear differential equations. As a result, this analysis can become computationally demanding for edge-colored graphs with a large number of nodes. A possible approach to mitigate this complexity is based on a mean-field approximation, as developed in \cite{del2022mean}, which yields the following equation for the transverse modes:}

\begin{equation}
\begin{array}{lll}
\frac{d {\bm \zeta}_h}{dt} & = & \biggl(J \bm f (\bm \sigma^*)  - \sum\limits_{\alpha=1}^M g\lay{\alpha} \lambda_h\lay{\alpha} J \bm h\lay{\alpha} ({\bm \sigma}^*) \biggr ) \bm \zeta_h -\\
& & \sum\limits_{\alpha=2}^M \varepsilon\lay{\alpha}g\lay{\alpha} \left [ \sum\limits_{k=2}^N |\lambda_h\lay{\alpha}-\lambda_k\lay{\alpha}| J \bm h\lay{\alpha} ({\bm \sigma}^*) \bm \zeta_k
\right],
\end{array}
\label{dyn_oscill5}
\end{equation}

\noindent \mattia{where $\varepsilon\lay{\alpha}$ represents the rotation angle by which the eigenvectors of the first layer are rotated (in every direction) in the mean-field rotation matrix associated with layer $\alpha$ (for more details on the meaning and computational aspects of the rotation angle, see \cite{del2022mean}).  Eq.~\eqref{dyn_oscill5} offers a significant decrease in the complexity with respect to Eq.~\eqref{dyn_oscill3}, and allows the computation of the stability diagrams for multiplexes of much larger size. Notice also that the second term in the right hand part vanishes if the Laplacians commute, so that Eq.~\eqref{dyn_oscill5} reduces to Eq.~\eqref{dyn_oscill4}. This term can, thus, be viewed as a first-order correction based on a mean-field perturbative approximation of the dynamics. This correction strictly holds under the assumption of quasi-identical layers. However, \cite{del2022mean} show that it remains accurate even for networks with highly dissimilar layers, suggesting an underlying mean-field nature of synchronization stability in multilayer networks.}

\mattia{Lastly, we note that within the framework of oscillators interacting through edge-colored graphs, it is also possible to study the case of oscillators coupled via both continuous (modeled in one layer) and impulsive (modeled in a second layer) interactions \cite{jin2021stochastic}.}

\subsection{Control via multiplexing}

\mattia{Synchronization in multiplex networks can be controlled through several techniques that build upon methods originally developed for single-layer structures. Examples include adaptive strategies~\cite{jin2021synchronization,han2025distributed}, intermittent control~\cite{liang2024synchronization}, and sliding mode control~\cite{wu2023fixed}—the latter being able to achieve all three types of synchronization discussed above: global, intra-layer, and inter-layer synchronization. These techniques aim to induce synchronization in systems that, without control, would either fail to synchronize or fail to follow a prescribed trajectory. This section focuses on a distinctive feature of multiplex networks, namely the possibility of actively exploiting multiplexing for control.} In Section~\ref{subsec:ChimeraStates} we have seen examples in which
one layer induces specific dynamics in other layers of a multiplex
network, effectively acting as a \emph{controller} for the system. In
a similar fashion, global synchronization may be induced in a layer
only targeting (that is, multiplexing) a selection of nodes~\cite{gutierrez2012targeting}). Here, we expand this concept
illustrating other ways to perform a distributed control acting
on one or more layers of a multiplex network.
In particular, we consider the same setup as in
Section~\ref{subsec:multilayersameobject}. We assume 
that the first layer is the network we want to control, namely the physical layer, 
%
%
while the other layers represent additional
forms of interactions through which the control actions can be
implemented.
%
%
\begin{figure}
\begin{center}
\includegraphics[width=0.47\textwidth]{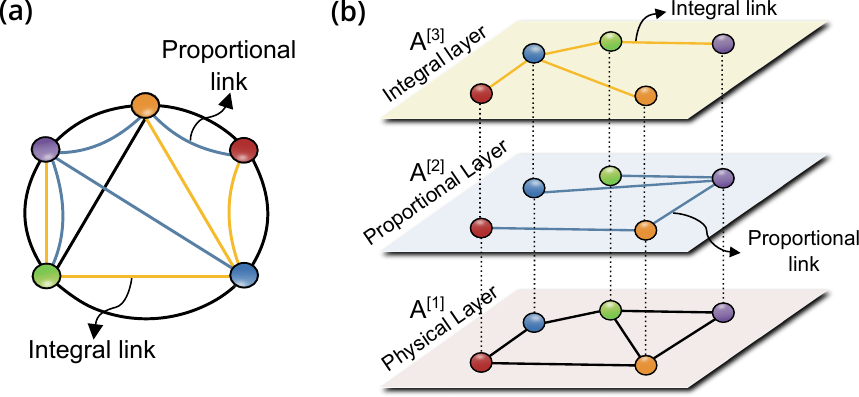}
\caption[]{Control via multiplexing. (a) Network representation: the links of the system to control (the physical layer) are shown as black lines,
  while blue and yellow lines stand for the proportional and integral
  links used for control. (b) Multiplex representation of a network
  under proportional and integral distributed controllers.
  Figures adapted from \textcite{lombana2016multiplex}.}
\label{fig:multiplexingcontrol}
\end{center}
\end{figure}
Although this control scheme is general, here we restrict to its
application to synchronization. To this aim, let us rewrite
Eqs.~(\ref{dyn_oscill_mult}) in the case of linear coupling as follows: 
\begin{equation}
    \small{
\begin{array}{lll}
\frac{d{\bm \sigma_i}}{dt} & = & \bm f (\bm \sigma_i) - g\lay{1} \sum_{j=1}^N a_{ij}\lay{1} H \lay{1}\bm \sigma_j- g\lay{2} \sum_{j=1}^N a_{ij}\lay{2} H \lay{2}\bm \sigma_j \\
& & -g\lay{3} \sum_{j=1}^N a_{ij}\lay{3} H\lay{3}\int\limits_0^t \bm \sigma_j d \tau
\end{array}
}
\label{dyn_oscill_mult_control}
\end{equation}
to better distinguish $A\lay{1}=\{a_{ij}\lay{1}\}$, accounting for the
connections in the physical layer from the control layers
$A\lay{2}$ and $A\lay{3}$.
Here, a link $(i,j)$ in layer 2 implements a proportional action
between the nodes $i$ and $j$, i.e., a feedback control law
proportional to the difference of the states of these units, whereas a
link $(i,j)$ in layer 3 implements an integral action, that is, a
feedback control term proportional to the integral of the difference
of the states (Fig.~\ref{fig:multiplexingcontrol}). Eqs.~(\ref{dyn_oscill_mult_control}) can be further
generalized to incorporate the derivative action by considering a
fourth layer, in this way realizing the most widely used control law
in industrial processes, namely the Proportional-Integral-Derivative 
(PID) controller~\cite{marlin1995process}. \mattia{This control scheme may be, for instance, applied to power grids, where the units represent power generators or consumers.}

Here, for simplicity, we only consider PI control in a multiplex
network whose layers have the same topology,
i.e. $A\lay{1}=A\lay{2}=A\lay{3}$, and
$H\lay {1}=H\lay {2}$, as in
\cite{burbano2016synchronization}. Under these assumptions, following
the usual steps of the MSF approach, the equations of the (decoupled)
transverse modes can be derived:
\begin{equation}
\frac{d{\bm \zeta_h}}{dt} = \left [D\bm f (\bm s) - (g\lay{1}+g\lay 2)\lambda_i H\lay {1}\right ]\bm \zeta_h-g\lay{3} \lambda_i H\lay{3}\int\limits_0^t \bm \zeta_h d\tau
\label{dyn_oscill_mult_controlblockdiag}
\end{equation}
%
%

Letting $\alpha= (g\lay{1}+g\lay2)\lambda_i$ and
$\beta=g\lay3\lambda_i$, from
Eqs. (\ref{dyn_oscill_mult_controlblockdiag}) one obtains a MSF
$\Lambda_{max}=\Lambda_{max}(\alpha,\beta)$ characterizing the
stability of synchronization for the controlled network. This MSF may
be used to \emph{tune} the parameters of the control, i.e., $g\lay 2$
and $g\lay 3$, so that to achieve synchronization. Remarkably, the
presence of dynamical coupling, in the form of an integral term, is
shown to significantly expand the region of synchronization stability.
%

This approach can be further extended to incorporate other
functionalities of the control into additional layers. For instance,
in \cite{kempton2017self} strategies for link weight adaptation
operating in a fully decentralized way have been embedded in the
network. In this case the additional layers have been designed to estimate  
the first non-zero eigenvalue of the Laplacian matrix, and to perform a
weight optimization aimed at maximizing it, 
so that to improve the synchronizability
of a class II MSF system.

\section{Spreading processes} 
\label{sec:spreading} 

  Spreading processes are at the core of many collective phenomena at the societal level, ranging from epidemics to the emergence of social movements. These processes been extensively studied in the field of network dynamics \cite{pastor2015epidemic,kiss2017mathematics,arruda2018fundamentals} as it provides the natural theoretical framework to accommodate the microscopic mechanisms underlying the spread of information, rumors and pathogens. 

The basic building blocks of spreading models are the so-called compartmental models, originally introduced in the context of epidemiology in the first half of the 20th century~\cite{kermack1927contribution,ross1911prevention}.
In these models, the state of each agent in a population can take a discrete set of values called compartments. For instance, the \emph{Susceptible-Infected-Susceptible} (SIS) model only has two compartments, susceptible (S) and infected (I), while the \emph{Susceptible-Infected-Recovered} 
(SIR) model adds a third compartment for recovered individuals (R). The transitions between compartments are determined by the infection, $\lambda$, and recovery, $\mu$, parameters, which represent either transition probabilities in discrete-time models or transition rates in continuous-time models. In the SIS model, a susceptible agent can be infected (and become infectious) with probability $\lambda$ by an infected neighbor, while infected agents return to a susceptible state with probability $\mu$. In contrast, in the SIR model, infected agents move to a recovered compartment with probability $\mu$ and cannot return to the susceptible state. These models provide the basic mechanisms for studying endemic regimes (SIS) and time-limited epidemics outbreaks (SIR) and, when applied to real epidemics, can be enriched with additional epidemiological, clinical and socioeconomic information to design effective containment interventions \cite{anderson1992infectious,keeling2007modeling}.

The structure of the underlying network of contacts also plays an important role. This means that the same pathogen  can cause a large epidemic in one network and, at the same time, be harmless in another network with different structural properties. \textcite{pastor2001epidemic} revealed that the degree heterogeneity of real complex networks leads to epidemic vulnerability. The epidemic threshold $\lambda_c$, i.e., the minimum infectivity  that a pathogen must have to produce an epidemic in a network, depends, in fact, on the precise structure of the underlying graph:
\begin{equation}
 \lambda_c= \frac{\mu}{\Lambda ({A})}\;,
\label{eq:thresholdA}
\end{equation} 
where $\Lambda (A)$ is the maximum eigenvalue of the adjacency matrix 
$A$.  The value of $\Lambda (A)$ can be approximated by considering the ensemble of networks with a given
expected degree sequence as $\Lambda (A)\simeq \langle k^2\rangle/\langle k\rangle$ \cite{chung2003spectra},   which yields to a vanishing epidemic threshold in
scale-free networks~\cite{pastor2001epidemic}.
These results opened the path to  the design of efficient containment strategies leveraging the heterogeneous nature of human interactions~\cite{pastor2002immunization,cohen2003efficient}.   

Beyond epidemiology, compartmental models have also been widely used for the study of the spread of ideas, news and rumours. In these cases common epidemic models, such as the SIS or SIR, are used to mimic the transfer of information between pairs of agents. Focusing on the SIR model we can map the epidemiological compartments into socially-inspired ones as $S$: ignorant, $I$: spreader, $R$: stifler (without any interest in transmitting the information). Apart from this map, socially-inspired compartmental models include subtle variations to the transition rules between compartments. In particular, we can distinguish those variations in the contagion process, $S\rightarrow I$, (adding e.g. threshold-mechanisms for the study of complex contagions of ideas~\cite{centola2010spread}) or in the recovery step, $I\rightarrow R$ such as the Daley-Kendall~\cite{daley1964epidemics} and Maki-Thompson \cite{maki1973mathematical} SIR-like models. These socially-inspired have been extensively studied in networks during the last decade in the same fashion as epidemic models [see \cite{castellano2009statistical} for a comprehensive review].

In this section, we review the main applications of the multiplex formalism to spreading processes~\cite{7093190,de2016physics,DEARRUDA20181} in the context of both the spread of information/ideas and pathogens. Although a variety of dynamical approaches have been applied to analyze spreading problems in multiplexes, such as the heterogeneous mean field (HMF) or the Generalized Epidemic mean field (GEMF)~\cite{sahneh2013generalized}, in what follows we will illustrate the different spreading models by means of the so-called Microscopic Markov Chain Approach (MMCA)~\cite{gomez2010discrete,gomez2011nonperturbative}.  This formalism allows to include the specific structure of each layer in the equations  (thus going beyond the averaging over networks with identical degree distribution implicit in the HMF) and to cast the specific models covered in this section under the general formulation given in sec.~\ref{subsec:dynamics}. Thus, although relevant results will be highlighted regardless of the dynamical framework at work, for the sake of coherence illustrate the different dynamical setups by means of the MMCA.

\subsection{Multimodal contagion processes} 
\label{sec:epid1}

We begin by exploring the case of social contagions and the spread of information. In this context, the multiplex formulation allows to easily analyse the case when several, $M$, information dissemination platforms coexist, each one represented as the layers of a multiplex. To this aim we can define the state of each node $i$ as a vector $\bm \sigma_i = [ \sigma_i \lay 1, \ldots, \sigma_i \lay M]^T$, where each component denotes the state of agent $i$ at each layer, taking values among the set of compartments at work, {\em e.g.} $\sigma_i\lay{\alpha}=\{S,I\}$ for the SIS model or $\sigma_i\lay{\alpha}=\{S,I,R\}$ for the SIR one. This way the multiplex formulation allows that an agent can be spreader in some of the transmission layers while remaining silent in the rest of them. Importantly, apart from a different structure of connections, each layer has its own contagion and recovery probabilities ($\lambda\lay{\alpha}$, $\mu\lay{\alpha}$). Moreover, the multimodal contagion framework in multiplexes includes an interlayer coupling represented by a set of contagion probabilities ${\lambda\lay{\alpha\beta}}$ where $\alpha$, $\beta=1,...,M$ and $\alpha\neq\beta$. Each interlayer coupling $\lambda\lay{\alpha\beta}$ signifies the additional effort a given user $i$ encounters when switching to another transmission channel ($\beta$) to disseminate the information that $i$ is currently spreading through a different source ($\alpha$).

Different works \cite{cozzo2013contact,min2016layer,xiang2016cooperative} have tackled the central question about the behavior of the contagion threshold and its connection with the thresholds of each network layer in isolation. In~\cite{cozzo2013contact}  Cozzo {\em et al.} have extended the MMCA for the SIS dynamics of single-layer networks to address multimodal contagions. In this way, the state of each node $\bm \sigma_i$ is monitored as a vector ${\bm \rho}_{i}$ whose component $\rho_{i}\lay{\alpha}$ account of the probability that node $i$ in layer $\alpha$ is infected, $\sigma_i\lay{\alpha}=I$. Thus, by assuming the statistical independence of these probabilities across nodes and layers
  (a situation that usually holds in sparse and lowly clustered networks)
  one can write the evolution equations for each of the $M$ components of the state vector, ${\bm \rho}_{i}(t)$, of node $i$ as:
\begin{equation}
\begin{split}
\rho_{i}\lay{\alpha} (t+1) = & [1 - \rho_{i}\lay{\alpha} (t)] [1 - q_{i}\lay{\alpha}(t)] + (1 - \mu\lay{\alpha}) \rho_{i}\lay{\alpha}(t)
\end{split}
\label{SIScozzo1}
\end{equation}
where the first term accounts for the probability that node $i$ at layer $\alpha$ is healthy at time $t$ but becomes infected at time $t+1$. On the other hand, the second term is the probability that, when infected at time $t$, node $i$ at layer $\alpha$ remains infected for the next time step. The term $q_{i}\lay{\alpha}(t)$ is the probability that node $i$ at layer $\alpha$ is not infected by any other node. This probability reads:
{\small
\begin{equation}
q_{i}\lay{\alpha}(t)=\prod_{\beta=1,\;\beta\neq\alpha}^{M}\left(1-\lambda\lay{\beta\alpha}\rho_{i}\lay{\beta}(t)\right)\prod_{j=1}^{N}\left(1-\lambda\lay{\alpha}a\lay{\alpha}_{ij} \rho_{j}\lay{\alpha}(t)\right)\;.
\label{SIScozzo2}
\end{equation}}

To analyze the stability of the disease-free solution, $\rho_{i}\lay{\alpha}=0$ one can linearize Eq.~(\ref{SIScozzo1}) by considering $\rho_{i}\lay{\alpha}=\epsilon_{i}\lay{\alpha}\ll1$. In addition it is useful to consider a simplified scenario where {\it (i)} all the intra-layer infection and recovery probabilities are identical, i.e., $\lambda\lay{\alpha}=\lambda$ $\forall\alpha$ and $\mu\lay{\alpha}=\mu$ $\forall\alpha$,  and {\it (ii)} for the inter-layer contagion probabilities we set: $\lambda\lay{\alpha\beta}=\eta\lambda$.  This way, by imposing the stationary condition Eq.~(\ref{SIScozzo1}) transforms into:
\begin{equation}
\frac{\mu}{\lambda}\epsilon_{i}\lay{\alpha}= \sum_{\beta=1}^{M}\sum_{j=1}^{N}\left( \delta_{\alpha,\beta}a\lay{\alpha}_{ij}\epsilon_{j}\lay{\alpha} 
+\eta(1-\delta_{\alpha\beta})\epsilon_{i}\lay{\beta}\right)\;.
\label{eq:linsimple}
\end{equation}

\begin{figure}[t!]
\begin{center}
\includegraphics[width=0.47\textwidth]{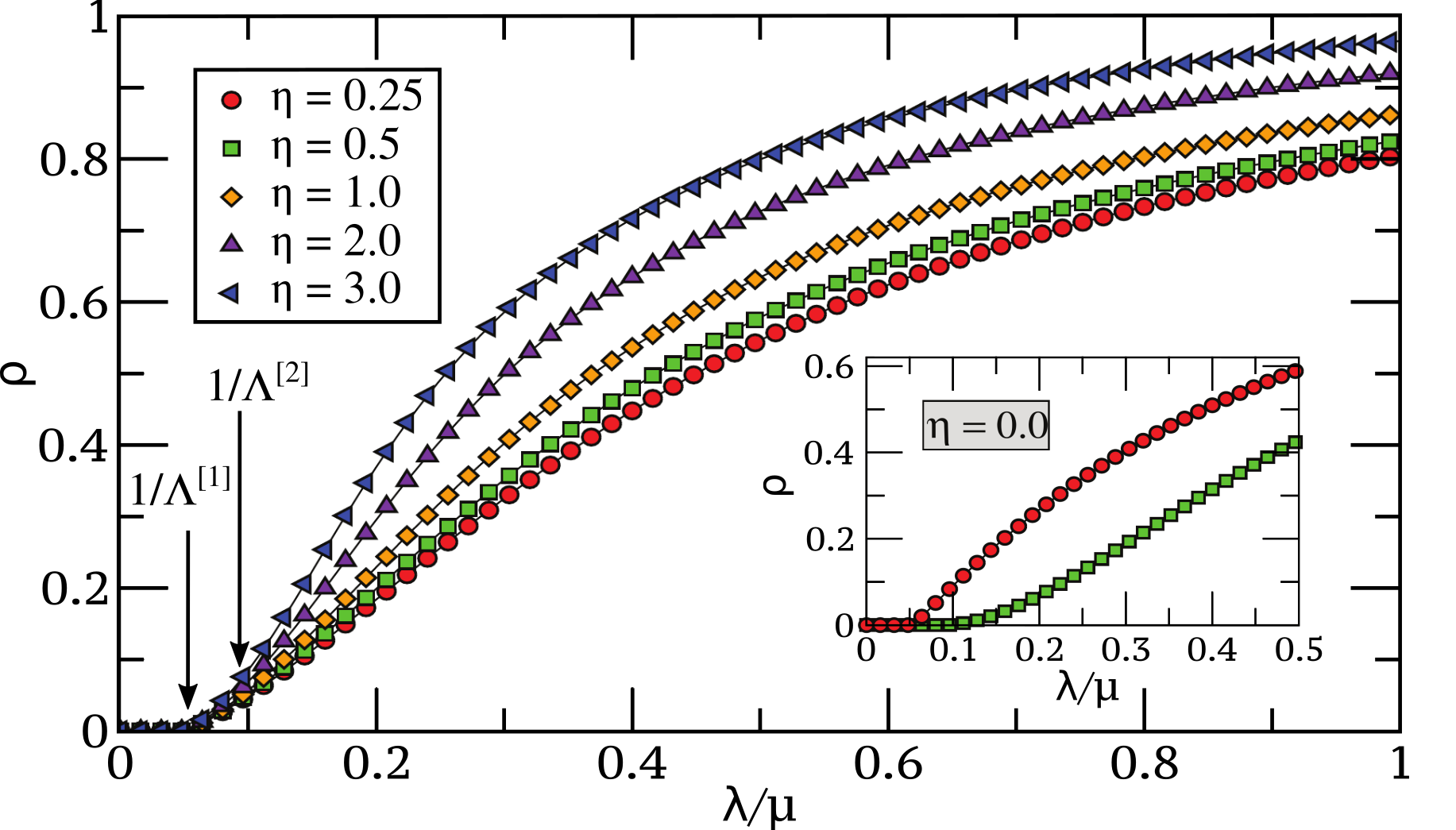}
\caption[]{Fraction of infected nodes ($\rho$) versus the re-scaled intralayer contagion probability $\frac{\lambda}{\mu}$ for a multiplex system composed of two layers with $N=10^4$ nodes each for different values of the ratio $\eta=\frac{\lambda^{[\alpha\beta]}}{\lambda}$. The arrows indicate the value of the inverse of the largest eigenvalues of the two layers, $\Lambda(A\lay{\alpha})\equiv \Lambda_{\alpha}$, whereas the inset shows the case in which the layers are independent. Figure adapted from \textcite{cozzo2013contact}.}
\label{fig:epid_cozzo}
\end{center}
\end{figure}

The former set of equations can be easily casted in a compact form by considering the following supra-adjacency matrix of the multiplex ${\cal A}(\eta)$:
\begin{equation}
     \cal{A}(\eta)  =\left( \begin{array}{cccc}
       A\lay{1}  &  \eta I  & \ldots & \eta I \\
       \eta I & A\lay{2} & \ldots &\eta I\\
       \ldots & \ldots &  \ldots & \ldots \\
       \eta I & \ldots & \eta I & A\lay{M}\\
    \end{array}
    \right)
\label{eq:matrixcozzo}
\end{equation}
where the diagonal blocks correspond to the set of $M$ adjacency matrices ${\bf A}=\{A\lay{1},A\lay{2},...,A\lay{M}\}$. This way, considering the $M\cdot N$ dimensional state vector ${\cal S}=\{ \sigma\lay{1}_1, \ldots, \sigma\lay{1}_N, \ldots , \sigma\lay{M}_1, \ldots, \sigma\lay{M}_N\}$ introduced in Eq.~(\ref{eq:difvect}) for the case of diffusion processes, and substituting its components by the infection probabilities, ${\cal S}=\{ \epsilon\lay{1}_1, \ldots, \epsilon\lay{1}_N, \ldots , \epsilon\lay{M}_1, \ldots, \epsilon\lay{M}_N\}$, Eq.~(\ref{eq:linsimple}) can be written as:
\begin{equation}
    \frac{\mu}{\lambda} {\mathcal S}  =   \cal{A}(\eta) S \;,      
\label{eq:matrix}
\end{equation}
so that  the epidemic threshold $\lambda_c$, i.e., the minimum value of $\lambda$ that satisfies the former equation, reads: 
\begin{equation}
\lambda_c = \frac{\mu}{\Lambda[{\cal{A}(\eta)}]}\;,
\label{eq:autovec}
\end{equation}
where $\Lambda[{\cal{A}(\eta)}]$ is the largest eigenvalue of ${\cal A}(\eta)$.

Although one can easily numerically compute $\lambda_c$ for a given multiplex, in \cite{cozzo2013contact} the analytical solution of Eq.~(\ref{eq:autovec}) is further studied by means of perturbation theory in the limit of weak inter-layer coupling (or alternatively large layer switching cost), $\eta\ll 1$. 
For the case of $M=2$ layers 
two main scenarios are obtained. When $\Lambda(A\lay{1}) \gg \Lambda(A\lay{2})$, so that $\Lambda[{\cal{A}}(0)]= \Lambda(A\lay{1})$, the effect of $\eta$ is negligible and the dynamics is dominated by layer $1$, $\Lambda[{\cal{A}(\eta)}]= \Lambda(A^1)$ (see Fig.~\ref{fig:epid_cozzo}).
Conversely, when the layers share the same structural properties and  $\Lambda(A\lay{1}) \gtrsim \Lambda(A\lay{2})$ , $\Lambda[{\cal{A}(\eta)}]= \Lambda(A\lay{1}) + \Delta \Lambda$ with  $\Delta \Lambda>0$ so that the critical point $\lambda_c$ decreases with respect to $\eta=0$, being $\Delta \Lambda$ dependent on the relation between the eigenvector centralities of the nodes at both layers. 

Another important result of this multimodal framework is presented in \cite{ferraz2017disease,ferraz2020universality} 
where the authors explore the structure
of the eigenvector ${\bm v}$ of the maximum eigenvalue of the corresponding supra-adjacency matrix
by looking the inverse partipation ratio (IPR):
\begin{equation}
{\text{IPR}}({\bm v})=\sum_{\alpha=1}^{M}\sum_{i=1}^{N}\left(v\lay{\alpha}_{i}\right)^4\;.
\end{equation}
The IPR allows to identify whether certain localization patterns appear near the epidemic onset $\lambda_c$, i.e., when the disease prevalence associated to each node is proportional to its contribution to the leading eigenvalue. At variance with monoplex networks \cite{goltsevPRL2012,Van_MieghemEPL2012} for which the disease is mainly localized on a subset of vertices, in multiplexes, just after the epidemic threshold, disease localization takes place on the layers when $\eta\ll 1$ while it becomes delocalized as $\eta$ increases. In addition, they show that multiple peaks for the susceptibiliy appear for $\lambda>\lambda_c$ when $\eta\ll 1$, pinpointing that, in this weak inter-layer regime, the different critical points corresponding to each layer show up. Again, when $\eta$ increases the secondary susceptibility peaks decay and only the main one located at $\lambda_c$ remains.

  The former results refer to undirected networks. However, in most online social platforms attention is not reciprocal and an agent $i$ is only able to pass information to those agents that follow $i$. The role of directionality in the dependence of the epidemic threshold $\lambda_c$ with the inter-layer contagion probability $\eta\lambda$ was explored in~\cite{wang2019directionality} observing that, when $\eta$ was large enough,  the directionality of inter-layer links produce a much less pronounced decrease of the epidemic threshold with the increase of $\eta$ than the case when directionality is placed in the intra-layer links.

  Finally, it is worth recalling that multimodal contagions have also been studied in epidemiological contexts. However, unlike in social contagions when an agent $i$ contracts the pathogen from an infectious contact in layer $\alpha$, $i$ becomes automatically infectious in all the layers. Thus, in this context inter-layer couplings are absent and the dynamical state of each node in all the layers is identical, $\sigma_i \lay {\alpha}=\sigma_i\lay{\beta}=\sigma_i$, which implies that multimodal epidemic spreading takes place in edge-colored graphs. Although theoretical works~\cite{buono2014epidemics,zhao2014multiple} on multimodal epidemic spreading yields no remarkable effect with respect to the case of single layer networks (thus highlighting the importance of inter-layer wiring in multiplexes), the use of edge-colored graphs has provided a better modelization of real epidemiological problems problems such as the possibility of a correct measuring of basic and effective reproduction numbers in population with multiple and simultanous interaction contexts (represented as layers)~\cite{liu2018measurability} or the design of control strategies of parasite spreading across different transmission routes~\cite{stella2017parasite,stella2018ecological}.

\subsection{Spreading processes in competition/cooperation} 
\label{sec:epid2}

Now we focus on the case of the simultaneous spread of competing and cooperative communicable diseases. At variance with multimodal contagions,
  the mathematical formulation of multiple disease interplaying~\cite{8955870} implies that there is no explicit contagion probability $\lambda\lay{\alpha\beta}$ between layers in which the different spreading processes take place, but an indirect influence that turn the epidemiological parameters of the disease spreading in one layer, say $\alpha$, dependent on the epidemiological state of the other layers $\beta\neq\alpha$. 

For illustrating the interplay between two pathogens, let us focus, for the moment, on two identical SIR processes that spread in a sequential way (one after the other) through the same network, as proposed by~\citet{newman2005threshold}, or, alternatively, in a duplex~\cite{funk2010interacting}, being each layer the transmission backbone for each pathogen. In these models pathogens are mutually exclusive, 
so the spread of the first pathogen provides immunity from the subsequent disease. Thus, while in the first layer all nodes have identical SIR parameters ($\lambda\lay{1}=\lambda$, $\mu\lay{1}=\mu$), given a pair of nodes, $i$ and $j$, linked in layer $2$, the probability of node $i$ infecting node $j$ and vice versa is $\lambda_{ij}=\lambda_{ji}=\lambda$ when $\sigma_i\lay{1}=\sigma_j\lay{1}=S$ and $\lambda_{ij}=\lambda_{ji}=0$ otherwise. 

The more complicated case of simultaneous spreading pathogens across the same network was subsequently studied by a number of works \cite{karrer2011competing,grassberger2013outbreaks,grassberger2015avalanche,grassberger2016phase,chen2017fundamental}. From the case of sequential spreading, it is clear that tackling the simultaneous spread of interacting diseases demands the formulation of link- and node-dependent epidemiological parameters for each disease $\alpha$. In particular the transmission probability of pathogen $\alpha$, $\lambda\lay{\alpha}$, is no longer constant but depends on the overall (across all the layers) epidemiological state of each pair of connected nodes in layer $\alpha$:
\begin{equation}
\lambda\lay{\alpha}_{ij}(t)=f\lay{\alpha}(\bm{\sigma}_i(t),\bm{\sigma}_j(t))\;.
\label{interact_CONT}
\end{equation} 
The recovery probability is also affected by the epidemiological state of each node in the other layers, i.e.,
\begin{equation}
\mu_{i}\lay{\alpha}(t)=g\lay{\alpha}(\bm{\sigma}_i(t))\;.
\label{interact_REC}
\end{equation}
To assign the specific form of the functions $f({\bm x}, {\bm y})$ and $g({\bm x})$ one should consider the specific interplay between the studied diseases. 
For competitive (cooperative) diseases and the SIS/SIR models one typically considers that when agent is infected with pathogen $\alpha$, then this infection can:
\begin{itemize}
\item[I.] decrease (increase) the probability of being infected with pathogen $\beta$. 
\item[II.] decrease (increase) the likelihood of passing the infection with pathogen $\beta$.
\item[III.] increase (decrease) the recovery probability when infected with pathogen $\beta$.
\end{itemize} 

The competitive case constitutes an ideal benchmark for studying multi-strains diseases, such as Influenza~\cite{influenza} or Dengue~\cite{Dengue}, for which there is total or partial cross-immunity. The first study in this line~\cite{karrer2011competing} was tackled for two SIR diseases spreading in a single network, i.e., identical layers, and assumed the existence of perfect cross-immunity for the concurrent disease
The main result is the existence of a phase diagram displaying: a disease-free phase, an endemic phase for either disease, and a phase where both diseases coexist. Interestingly, coexistence appears despite the fact that the diseases are mutually exclusive and its existence depends on the respective values of infection and recovery probabilities of the two SIR diseases in isolation. The opposite case of cooperative pathogens spreading across identical network layers, is addressed in~\cite{grassberger2013outbreaks,grassberger2015avalanche,dufresne2015complex,grassberger2016phase,chen2017fundamental,cui2019effect} and aims at capturing the synergistic effects between pathogens, as observed in the case Tuberculosis and HIV \cite{lawn2009epidemiology} or Influenza and Pneumonia \cite{acuna2011influenza}. Under cooperative conditions, the usual second-order epidemic transition turned into an abrupt (explosive) one, as a direct consequence of the cooperativity of infections. These two frameworks are analyzed in a unified way in \cite{soriano2019markovian} showing
that degree heterogeneity
enlarges the coexistence phase in the case of competition while it smooths the epidemic onset for cooperative diseases. 



Interacting diseases in multiplexes have been studied for two competitive \cite{marceau2011modeling,sahneh2014competitive} and two cooperative \cite{azimi-tafreshi2017cooperative}  scenarios, while different frameworks for treating  the spread of interacting diseases across multiplexes in a unified way have been also introduced \cite{sanz2014dynamics,zhao2014unified,wu2020spreading}. 
In these works, each infection probability
$\lambda\lay{\alpha}$ ($\alpha=1,2$) accounts for the probability of a fully susceptible individual contracting the disease $\alpha$ after coming into contact with a neighbour in layer $\alpha$ infected exclusively with the pathogen $\alpha$, i.e., being susceptible to all other diseases $\beta\neq\alpha$. Similarly, $\mu\lay{\alpha}$  ($\alpha=1,2$) describes the probability that an individual infected only with pathogen $\alpha$ will recover from that disease. The interaction between the two diseases comes into play when individuals involved in a potential transmission event of the pathogen $\alpha$ are infected by other pathogens $\beta\neq\alpha$. In \citet{sanz2014dynamics,wu2020spreading}, the particular form of Eq.~(\ref{interact_CONT}) to accommodate rules I and II for the interaction between pathogens reads:
\begin{equation}
\lambda\lay{\alpha}_{ij}(t)= \left\{ \begin{array}{lcc}
             \lambda\lay{\alpha} &  {\text{if}}  & \sigma_{i}\lay{\beta}=\sigma_{j}\lay{\beta}=S \\
             \\ \gamma\lay{\alpha}_{\alpha}\lambda\lay{\alpha} &  {\text{if}}  & \sigma_{i}\lay{\beta}=I\; \&\; \sigma_{j}\lay{\beta}=S \\
             \\ \gamma\lay{\alpha}_{\alpha}\gamma\lay{\alpha}_{\beta}\lambda\lay{\alpha} &  {\text{if}}  & \sigma_{i}\lay{\beta}=\sigma_{j}\lay{\beta}=I
             \end{array}
   \right.\;,
\end{equation} 
with $\alpha\neq\beta$. The set of parameters $\bm \gamma=\{\gamma\lay{1}_{1}, \gamma\lay{1}_{2},\gamma\lay{2}_{1},\gamma\lay{2}_{2}\}$ take positive values, while they have values $\gamma\lay{\alpha}_{\beta}>1$ for cooperative diseases and $\gamma\lay{\alpha}_{\beta}<1$ for competing ones. Analogously, the recovery probabilities $\bm \mu$ are modified by a set of parameters $\bm \eta=\{\eta\lay{1},\eta\lay{2}\}$ that capture the effect of the interaction between diseases on their respective  recovery probabilities. Specifically, to address rule III Eq.~(\ref{interact_REC}) reads: 
\begin{equation}
\mu_{i}\lay{\alpha}(t)= \left\{ \begin{array}{lcc}
            \mu\lay{\alpha} &  {\text{if}}  & \sigma_{i}\lay{\beta}=S \\
             \\ \eta\lay{\alpha}\mu\lay{\alpha}&  {\text{if}}  & \sigma_{i}\lay{\beta}=I
             \end{array}
   \right.\;,
\end{equation}
with $\beta\neq\alpha$. The interaction parameter for the recovery probability is $\eta\lay{\alpha}<1$ when pathogen $\beta$ competes with $\alpha$ and $\eta\lay{\alpha}>1$ for the cooperative case.

\begin{figure}[t!]
\begin{center}
\includegraphics[width=0.55\textwidth]{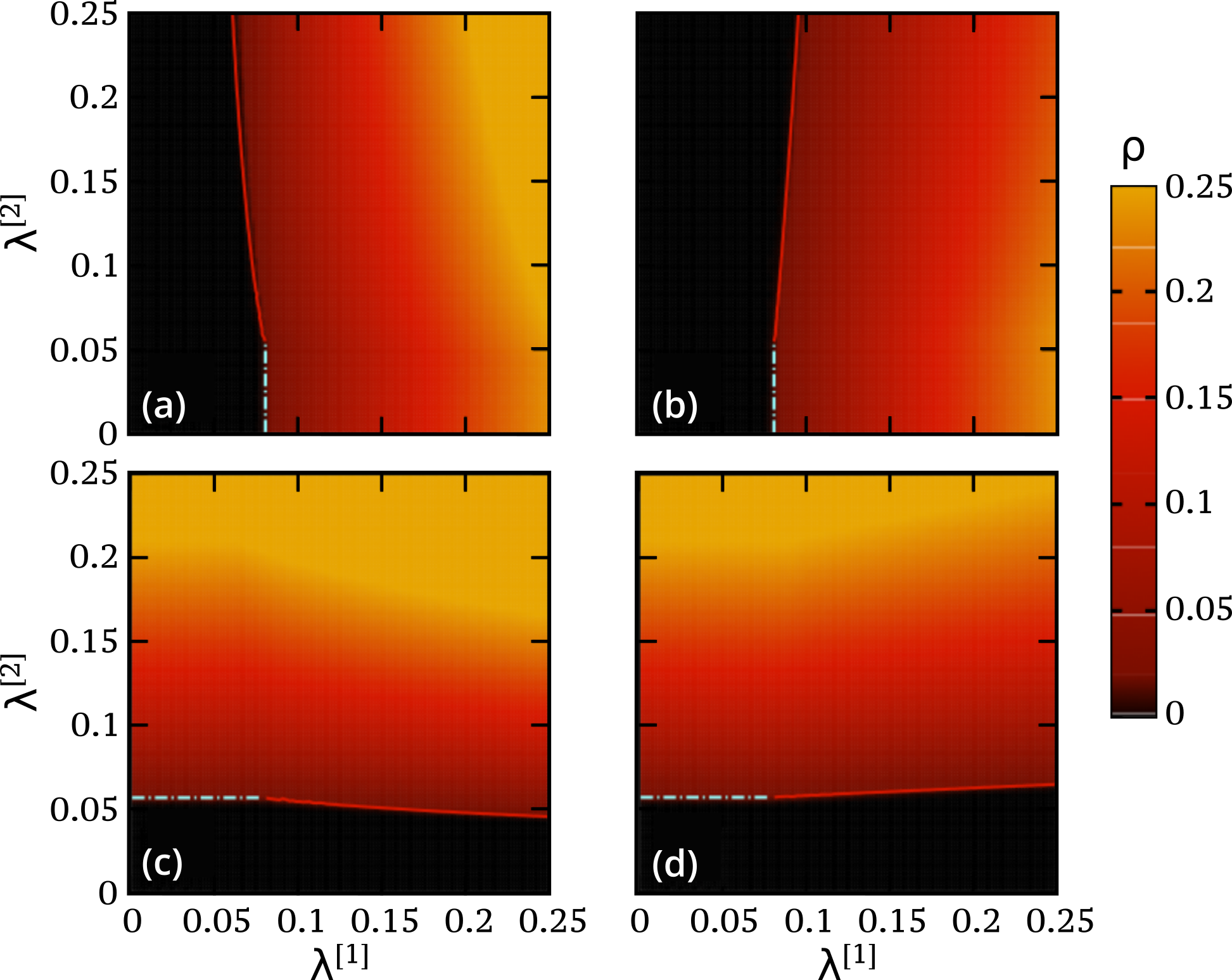}
\caption[]{Epidemic diagrams $\rho(\lambda\lay{1},\lambda\lay{2})$ for a multiplex composed of two uncorrelated scale-free networks of $N=5000$ nodes. Panels (a) and (c) show the cooperative case ($\gamma=1.3$, $\eta=0.8$) while (b) and (d) are for the competitive one ($\gamma=0.8$, $\eta=1.3$). Figure adapted from \textcite{sanz2014dynamics}.}
\label{fig:coinfection2}
\end{center}
\end{figure}

The former formulation is analyzed in~\citet{sanz2014dynamics}, by means of a HMF, and  in~\cite{wu2020spreading}, through a generalization of the MMCA presented in \cite{soriano2019markovian} to multiplexes. These works derive the epidemic thresholds 
  both in the absence of interaction (primary thresholds) and when it exists (secondary thresholds)
  highlighting the effects of the cooperation/competition between pathogens.
    In Fig.~\ref{fig:coinfection2} the epidemic prevalence in the two layers of a multiplex is plotted for cooperative [left panels (a) and (c)] and competitive [right panels (b) and (d)] cases when the interplay parameters are simplified by setting $\gamma\lay{\alpha}_{\beta}=\gamma$ and $\eta\lay{\alpha}=\eta$. In the cooperative case the epidemic threshold of one disease is anticipated only when the incidence of the other one is nonzero. However, when diseases are competitive their respective thresholds are delayed as soon as the competitor disease is present in the multiplex. Interestingly,  in the case of positive correlations between the degree sequences of the two layers, a non-vanishing epidemic threshold can be found even for scale-free networks. This is a qualitatively different behavior emerging from the multiplex dynamics which can not be obtained in single-layer heterogeneous networks. 

\subsection{Interplay of disease spreading and human response}
\label{sec:awareness}

Competitive and cooperative diseases are not the unique interaction mechanisms that two spreading processes can share~\cite{8955870}. Imagine that the contagion disease $\alpha$ confers some immunity to $\beta$ while contracting $\beta$ enhances the susceptibility to $\alpha$. This case was treated in \cite{ahn2006epidemic} and illustrated as the interplay between preventive information propagation and contagion dynamics in~\cite{funk2009spread,granell2013dynamical,granell2014competing,wang2014asymetrically,Massaro2014epidemic,wang2017epidemic}. On one hand, preventive information (awareness of the risk of contracting a disease) can spread across one layer promoting protection to the infection in the other one. On the other hand, the contagion by the pathogen induces the awareness about contagion risk, initiating a cascade of awareness propagation in the other layer. 

In \cite{granell2013dynamical,granell2014competing}, the former problem is analyzed as two coupled SIS processes, although they call the SIS associated to information spreading UAU (for Unaware-Aware-Unaware). 
Similarly to competitive and cooperative SIS dynamics primary infections: $(S,U)+(I,U)\rightarrow 2(I,U)$ and $(S,U)+(S,A)\rightarrow 2(S,A)$ have probabilities $\lambda\lay{1}$ and $\lambda\lay{2}$ respectively, while the corresponding recovery processes, $(I,U)\rightarrow (S,U)$ and $(S,A)\rightarrow(S,U)$, are characterized by $\mu\lay{1}$ and $\mu\lay{2}$.
The two spreading processes become coupled in the following way. Once a healthy agent becomes aware of the disease, $(S,A)$, the probability of being infected is automatically reduced by a factor $\gamma<1$, becoming  $\gamma \, \lambda\lay{1}$. Note that the case $\gamma=0$ implies complete immunization against the disease. Additionally, an infected agent becomes immediately aware of the disease, i.e., the state $(I,U)$ is not allowed. Thus, individuals can be in three different states, namely $(S,U)$, $(S,A)$ and $(I,A)$. 

\begin{figure}[t!]
\begin{center}
\includegraphics[width=0.46\textwidth]{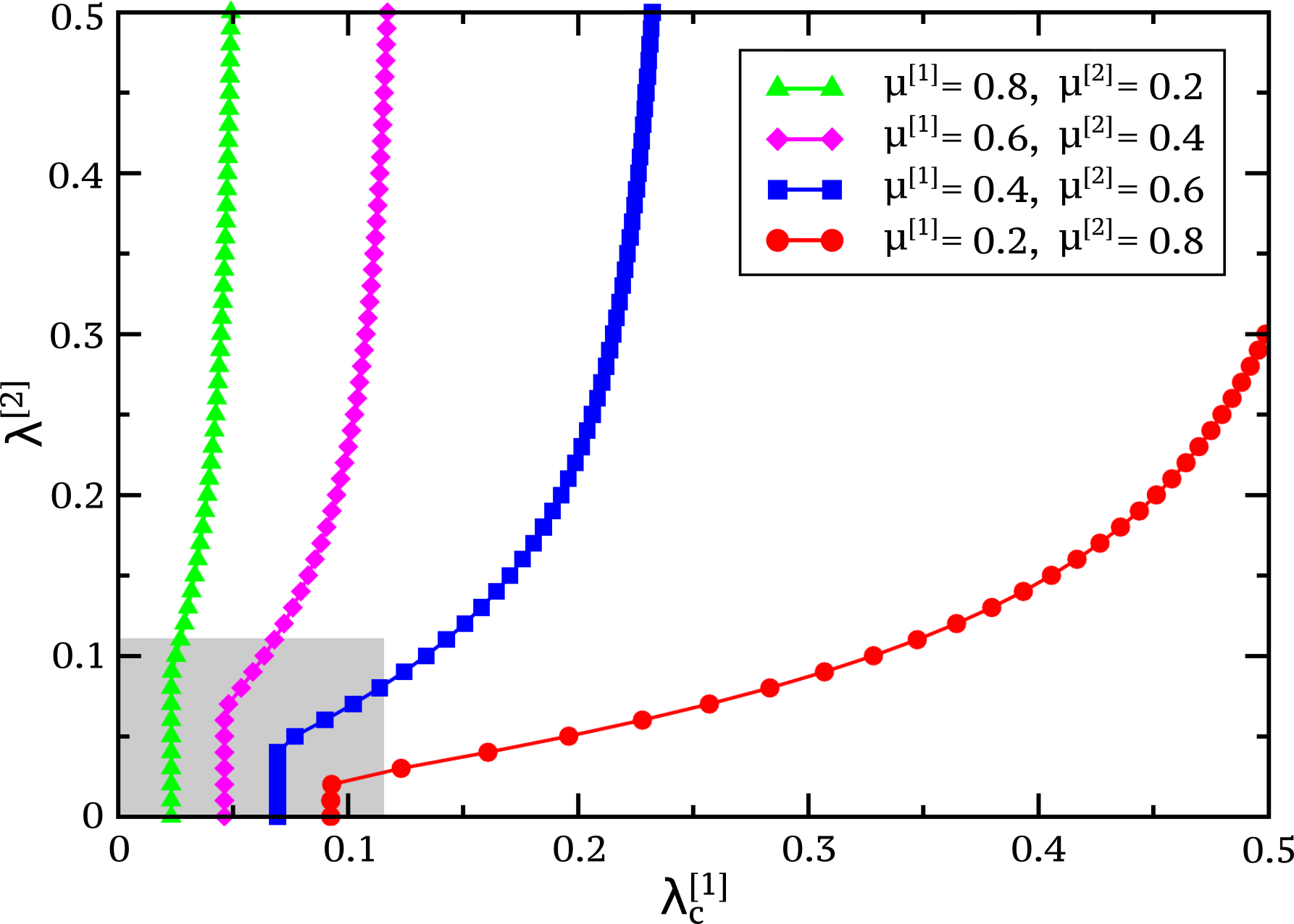}
\caption[]{Dependence of the onset of the epidemics  $\lambda\lay{1}_c$ as a function of $\lambda\lay{2}$ for different values of the recovery probabilities $\mu\lay{1}$ and $\mu\lay{2}$.  The shaded rectangle corresponds to the area where the metacritical points may be. Figure adapted from \textcite{granell2013dynamical}.}
\label{fig:UAUSIS}
\end{center}
\end{figure}

By applying the MMCA one associates to each node $i$ the probabilities of being in each of the former states at time $t$: $\rho_i^{SU}(t)$, $\rho_i^{SA}(t)$ and $\rho_i^{IA}(t)$. By solving the evolution equations in the stationary regime and linearizing around the epidemic-free state
one obtains an eigenvalue problem for a matrix ${\bm H}$ with elements
\begin{equation} 
h_{ij}= \left[1-(1-\gamma)\left(\rho_j^{SA})^*+(\rho_j^{IA})^*\right)\right]a_{ji}\lay{1}\;,
\label{eq:cond}
\end{equation} 
and eigenvalues $\mu\lay{1}/\lambda\lay{1}$. Thus, the epidemic threshold reads $\lambda\lay{1}_c =\mu\lay{1}/\Lambda ({\bm H})$,
where $\Lambda ({\bm H})$ is the maximum eigenvalue of matrix ${\bm H}$. Note that if $\gamma=1$ 
we recover ${\bm H}={\bm A}\lay{1}$ and the threshold becomes identical to Eq.~(\ref{eq:thresholdA}).  However, when awareness generates protection, i.e., $\gamma<1$, the epidemic onset depends on the awareness level of the system, being this threshold larger as awareness increases. Moreover, if $\lambda\lay{2}$ is small enough so that awareness cannot spread macroscopically in the second layer, i.e., $(\rho_j^{SA})^*=(\rho_j^{IA})^*=0$ the epidemic threshold takes the same value as if the two contagion processes were decoupled. By increasing $\lambda\lay{2}$ up to its critical value, the epidemic threshold $\lambda\lay{1}_c$ starts to depend on awareness, being the point where $\lambda\lay{1}$ stops being independent of $\lambda^2$ a metacritical point, i.e., the point in which the two critical onsets get intertwined and, in particular, the onset of the epidemic $\lambda_c\lay{1}$ starts depending on the prevalence of aware individuals. This behavior is shown in Fig.~\ref{fig:UAUSIS}.  

The former results were obtained by assuming that infected individuals become immediately aware. However, this assumption can be relaxed to allow agents in the IU state while providing a transition $U\rightarrow A$ with probability $\kappa$ that accounts of the self-awareness \cite{granell2014competing,gao2016competing,kan2017effects}.  Although it has no direct effect on the expression of the epidemic threshold, in the supercritical phase the former works show that the mitigation action of awareness is significantly more efficient when the contact and information layers are highly correlated.

%

The UAU-SIS model is a minimal benchmark capturing the influence that the diffusion of information has on the spread of a disease. To better capture this interplay other features such as the enhancing mass media effect on awareness \cite{granell2014competing} or the adoption of
more refined information spreading models \cite{velasquez2017interacting,pan2018impact} have been also addressed. Interestingly, in \cite{ventura2019epidemic} a Maki-Thomson rumour model is implemented for the information layer, so that the aware state does not transition to unaware but to a stifler one in which aware individuals cease to spread the voice but do not forget the risk. In this setting the authors explore how the speed of information propagation relative to that of infections affects the epidemic mitigation. Counterintuitively, the authors find that when awareness spread too fast a large fraction of stiflers are formed thus causing an increase of the prevalence which decrease the mitigation effect. This counter-intuitive effect was also observed in \cite{wang2017epidemic,velasquez2020disease} using the original UAU-SIS formulation.

Awareness transmission models focus on the negative impact of social response on the propagation capacity of the disease. However, human response can also positively interfere with the spread of the pathogen by enhancing the recovery probability of infected agents. This type of human response to epidemics in the form of social support is modeled under the multiplex lens in \cite{chen2018suppressing,chen2018optimal} by coupling a SIS dynamics in the first (contact) layer with a resource diffusion dynamics running the second (social) one. This way, the recovery probability of the SIS dynamics becomes: 
\begin{equation}
\mu\lay{1}_{i}(t)= \mu\lay{1} h(\sigma\lay{2}_i(t))\;,
\end{equation}
where $\sigma\lay{2}_i(t)$ represents the resource located at node $i$. In this case disease propagation in the contact layer is affected by a random walk process in the social one that, in turn, is also affected by the epidemiological state in such a way that only susceptible nodes in the contact network can participate in the diffusion. This interplay provides an interesting tension when layers are heterogeneous, as hubs tend to be both the most affected by the pathogen and also they tend to allocate most of the circulating resources, and produce the emergence of hybrid epidemic transitions \cite{chen2018suppressing} while optimal resource diffusion strategies depend crucially on the correlation between contact and social layers \cite{chen2018optimal}.


The former approaches to model human behavior during epidemics typically rely on simple mechanisms of awareness or resource transmission between agents. However, the adoption of protective measures like vaccines comes at a cost, and agents face a dilemma of whether to adopt these measures based on perceived risk. Evolutionary game theory provides a way of analyzing this dilemma \cite{funk2010modelling, fu2011imitation, wang2015coupled} by adding a new compartment for vaccinated individuals. Each node is assigned a payoff that takes into account the cost of protection ($C_V$) and the cost of contracting the disease ($C_I$). The choice of whether to vaccinate or not is made by comparing an agent's own payoff with her acquaintances' performance, which is driven by the disease's spread. This creates a feedback loop between the epidemic and game dynamics.


The transmission and strategy update processes often take place in different networks, and the multiplex formulation have recently open the door to disentangle the spreading and imitation networks \cite{jentsch2018spatial,kabir2020impact}. In \textcite{fukuda2015influence} the authors studied a sequence of (SIR) epidemic outbreaks, where a fraction of the population becomes immunized before the outbreak by paying $C_{V}$. After the epidemic, agents compare their payoffs with their neighbors and update their strategy, followed by another epidemic outbreak. When the transmission and imitation networks coincide, \textcite{cardillo2013evolutionary} showed that degree heterogeneity enhances vaccination behavior after a number of SIR outbreaks has taken place. However, interesting effects arise when the symmetry between the networks is broken. When the transmission network is heterogeneous and the imitation graph is homogeneous, vaccination coverage decreases significantly and the outbreak size increases due to the inability to extend vaccination to low-degree nodes connected to the hubs. On the other hand, when the transmission graph is homogeneous and the imitation network is heterogeneous, the vaccination coverage is slightly the same as in the case of homogeneous monoplexes since the multiplex structure favors the creation of small and uniformly distributed vaccination clusters that act as containment barriers to the spread of the disease.


\subsection{\vito{Spreading in multiplex metapopulations}}

  To conclude this section, we briefly mention a different type of spreading dynamics that have been tackled under the lens of multiplex networks: reaction-diffusion processes. These processes are usually studied using a networked metapopulation, which consists of nodes representing geographical patches where agents interact (reaction), and links representing mobility flows that agents can take when changing locations (diffusion). Reaction-diffusion processes are frequently employed to investigate the spatiotemporal trajectory of epidemics \cite{colizza2007reaction,balcan2009multiscale,belik2011natural}.

The use of multiplex metapopulations has allowed considering the interplay between different reaction-diffusion processes taking place in the same set of geographical areas. In particular, the most paradigmatic situations deal with those scenarios previously explained for contact multiplex networks. Examples of the studies carried out on multiplex metapopulations include the analysis of multimodal transmission \cite{apolloni2014metapopulation,Soriano2018spreading}, the spread of competing pathogens  \cite{poletto2013host}, or the simultaneous spread of information and diseases, extending the use of the UAU information spread model coupled to an SIR epidemic \cite{lima2015disease}. The interest in multiplex metapopulations has also been applied in real contexts where two different kinds of populations (with different mobility patterns) interact in the same collection of patches. In~\cite{bosetti2020heterogeneity}, this framework allows calculating the increase of measles re-emergence in a country (Turkey) with a large vaccine coverage when an incoming flow of refugees from Syria. Interestingly, this formalism allowed calculating the risk of reemergence of measles as a function of the interaction between the two populations at patches, proposing a maximal dispersal of refugees to avoid such outbreaks from happening.


\section{Social dynamics}
\label{sec:SD} 

{\color{black}
Statistical physics has long provided powerful approaches to the study
of social dynamics. Although humans are far more complex than physical
particles, the emergence of regular patterns from the interactions of
social agents provides grounds for employing the perspective of 
statistical physics. Within this framework, several models inspired by statistical
physics have been used to study a variety of social phenomena such as opinion 
formation, polarization, and fragmentation, as well as cultural dissemination
and the spreading of social behaviors \cite{galam1982sociophysics,castellano2009statistical,starnini2025opinion}.

Social networks are one of the most natural and representative examples 
of multiplex networked systems, since individuals are simultaneously 
embedded in multiple different layers of 
interactions \cite{wasserman1994social,szell2010multirelational,battiston2014structural}. 
Consequently, the study of social dynamics requires a multiplex perspective, 
and many studies have addressed how multiplexity alters social phenomena. 
The influence of multiplex structure on social dynamics manifests in 
various forms. For instane, nodes can display different states (opinions) 
across layers, reflecting the multifaceted nature of individuals. 
Links also may act on different types of relations, as in edge-colored 
networks, where each layer represents a distinct mode of interaction. 
Moreover, inter-layer connections can generate non-trivial effects, 
since activity or consensus in one layer may promote or hinder dynamics in another.

In the following, we review current advances in social dynamics on multiplex 
networks, with a particular focus on how multiplexity alters the behavior 
of classical models. We begin with the voter model as a paradigmatic model
of opinion dynamics, and its variants. We also discuss other opinion 
dynamics models including Hamiltonian-based formulations and rule-based 
approaches of opinion formation. Finally, we examine cultural 
dissemination and social contagion processes in multiplex settings. 
}

\subsection{Voter model and its variants} 

\begin{figure}[t]
	\includegraphics[width=0.55\textwidth]{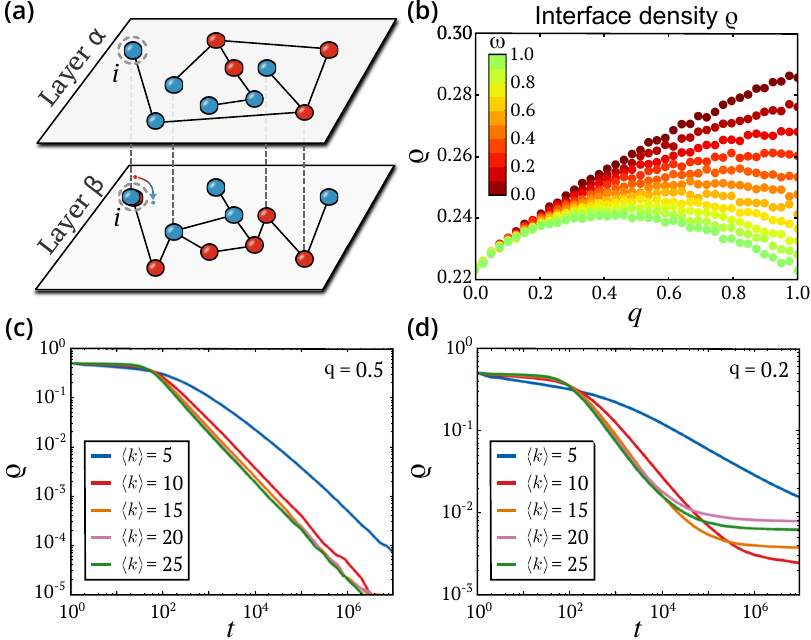}
	\caption[]{
		\newtext{
			Voter models on multiplex networks.	
			(a) Any change in the state of node $i$ at one layer will enforce a change 
			of its state connected by an inter-layer link on the other layer.}
		(b) The interface density is a non-monotonous function of the multiplexity
		$q$ when $ \omega \neq 0$, i.e., when the dynamics on the two layers are
		not independent. (c) In a voter model with ageing, the system
		converges to consensus for large values of $q$. 
		\vito{(d) Dynamically trapped configuration with coexisting opinions
			are instead possible, when $q$ is smaller that a
			critical value.} 
		Figures adapted
		from~\textcite{diakonova2016irreducibility} and~\textcite{artime2017ageing}.}
	\label{fig:votermodel}
\end{figure}

{\color{black}
The voter model, where a population of interacting individuals endowed
with a binary variable evolves based on imitation dynamics, is one of
the simplest and most studied models of opinion 
dynamics \cite{liggett1985interacting,castellano2003incomplete,sood2005voter}. 
Specifically, each node $i$ in a network is assigned a binary variable 
$\sigma_i \in \{ -1,1\}$, representing its opinion.  
At each update step, a node $i$ is selected at random and this node then chooses 
one of its neighbors uniformly at random and adopts that neighbor’s state. 
The dynamics can be characterized by the interface density $\rho$, 
i.e. the fraction of links connecting nodes in different states. 
In uncorrelated networks with average degree $\mu$, $\rho$ in surviving 
runs becomes a plateau
\begin{equation}
  \rho = \frac{\mu - 2}{3(\mu - 1)},
  \label{eq:rho_plateau}
\end{equation}
before the system reaches full consensus~\cite{vazquez2008analytical}.
On any finite networks, the voter model always reaches consensus due to 
finite-size fluctuations. The consensus time $T_N$, which is the 
characteristic time to reach consensus for network size $N$, is highly
sensitive to underlying network topology, and
for uncorrelated networks it is given by
$T_N \sim N \mu_1^2 /\mu_2$,
where $\mu_k$ is the $k$-th moment of the degree distribution \cite{sood2005voter}.

\textcite{diakonova2016irreducibility}
have considered a voter model on a two-layer multiplex network, where 
a fraction $q$ of the nodes has replicas on both layers, while the
remaining nodes are present only in one layer. The fraction $q$ is
called ``multiplexity'', so that, for $q=1$ we have that all the nodes
of each layer are also present on the other layer, while for $q=0$ the
two layers are effectively independent. At each discrete time-step,
one of the two layers is selected and one step of the single-layer
voter model is run on it. That is, a node is chosen uniformly at 
random on the chosen layer, and it adopts the opinion of one of its 
neighbors, also chosen at random.} 
The additional ingredient of the model is that if a node is
present on both layers, its opinions at the two layers will be always
identical. This means that if node $i$ changes its opinion $s_i^{\alpha}$ 
on one layer $\alpha$ due to a voter model interaction, its 
opinion $s_i^{\beta}$ on the other layer is immediately set to the 
same value as well, as shown in Fig.~\ref{fig:votermodel}(a).

The dynamics of consensus in this system can be studied by using 
the interface density $\rho$ as a function
of the multiplexity $q$ and of the fraction $\omega$
of overlap edges present in both layers. 
The interface density in multiplex networks is defined as 
$\rho = 1/M \sum_{\alpha}  \rho\lay{\alpha}$, where
\begin{equation}
  \rho\lay{\alpha} = \frac{1}{  \sum_{i}\sum_{j<i}a\lay{\alpha}_{ij}  }
  \sum_{i}\sum_{j<i}a\lay{\alpha}_{ij}|\sigma\lay{\alpha}_i -
  \sigma\lay{\alpha}_j| 
\end{equation}
is the fraction of edges connecting nodes with different 
opinions at layer $\alpha$. 
\newtext{
The interface density of the multiplex voter model depends heavily 
on $q$ and $\omega$ as shown in Fig.~\ref{fig:votermodel}(b). 
In a single-layer network the interface density depends only on 
the mean degree $\mu$ as in Eq.~\ref{eq:rho_plateau}, while in 
multiplex networks $\rho(q,\omega)$ depends in a non-trivial way 
on both multiplexity $q$ and overlap $\omega$. It means that
the multiplex voter model cannot be reduced into a voter model on 
a single-layer graph obtained by simply aggregating the two original layers. 
In addition, with increasing $\omega$, the dependence of 
$\rho(q,\omega)$ on $q$ is no longer monotonic and a maximum of 
$\rho(q,\omega)$ appears. Interestingly the range of multiplexity 
observed in many real-world social networks  
is similar to the typical values of $q$ at which $\rho(q,\omega)$ is 
maximal. This implies that social systems might tend to self-organise
in order to guarantee the survival of a variety of different opinions. 
}


\textcite{artime2017ageing} have studied the effect of multiplexity
$q$ in a voter model where aging is also considered. In this
model, the propensity of an agent to change opinion depends on the
amount of time elapsed $\tau_{ij}$ since the last change occurred, so called 
aging, effectively causing the freezing of opinions which have 
persisted for longer. \newtext{Here aging is implemented through a persistence-time
dependent update probability $p_i(t)=b/\tau_i$, so that older opinions 
become increasingly resistant to change.}
The main result is that there exists a value $q^*$ of multiplexity above
which the multiplex voter model always converges towards consensus, 
as shown in Fig.~\ref{fig:votermodel}(c). 
Conversely, as shown in Fig.~\ref{fig:votermodel}(d), when $q<q^*$ the 
convergence to the absorbing state, i.e., consensus is indefinitely delayed, 
thus allowing for the persistence of coexisting opinions for long times. 


\textcite{gastner2019hypocrisy} studied the effects of hypocrisy in
the voter model, using a two-layer multiplex network. One of the two
layers represents the ``external'' opinions, i.e., those declared by
each agent, while the other layer contains only the ``internal'' or
true opinions actually held by the agents, which might differ from
their external ones. In this model the agents interact only on the
external layer, through the usual single-layer voter model
dynamics. The relation between the external and the internal opinion
of an agent is determined by a combination of imitation of other
external opinions, externalisation of the internal opinion, or
internalisation of the external opinion. The multiplex formulation
allows to conclude that the presence of hypocrisy effectively slows
down the attainment of consensus, and the effect increases with system
size, thus allowing again for long-lived states where both opinions
coexist.

Finally, \textcite{gradowski2020pair} have proposed a variant of 
$q$-voter model on multiplex networks. 
\newtext{The $q$-voter 
model \cite{castellano2009nonlinear} describes 
opinion updates in which an individual samples $q$ neighbors at random 
and, if all share the same state, adopts that state, while with probability 
$p$ the individual instead changes state independently. They extended 
the dynamics to multiplex networks by introducing two natural rules. 
In the first case, the voter considers $2q$ neighbors across both layers 
as a single group and adopts their state if all of them share the same 
opinion. In the second case, the voter updates only when the $q$ neighbors 
chosen within each layer separately are unanimous, requiring agreement 
in both layers simultaneously. Using pair approximation, they showed 
that multiplexity can alter both the critical point and the order 
of the transition from continuous to discontinuous, 
depending on whether opinions are aggregated across 
the two layers or require agreement within each layer separately.
}

\subsection{Hamiltonian-based opinion formation}

\newtext{
Beyond the voter model, another major class of opinion dynamics 
consists of Hamiltonian-based approaches directly inspired by spin 
models in statistical physics. Such Hamiltonian-based approaches 
provide a natural framework to incorporate both social interactions 
and external influences, and have been widely applied to study opinion 
formation ~\cite{galam1982sociophysics,galam1991towards}.
For instance, Ising-like formulations describe opinion formation as 
a symmetry-breaking phase transition, and account for polarization
or fragmentation under external fields and competing interactions.
}

\textcite{battiston2016interacting} endowed the agents of a multiplex
system with a set of opinions on different topics, whose dynamics,
modelled as coupled Ising models, depends both on external factors,
such as peer-pressure, and internal ones, such as the tendency of
agents to choose opinions on different topics which are considered to
be in agreement. In particular, they considered a generic multiplex
network where each node on layer $\alpha$ is associated to the
functional:
\begin{equation}
F_i\lay \alpha = J_i \sum_{j=1}^N a_{ij}\lay \alpha \sigma_j\lay{\alpha}
+\gamma\frac{\chi_i}{J_i} \sum_{\mathclap{\substack{\beta = 1 \\ \beta
      \neq \alpha}}}^M \sigma_i\lay{\beta} + h\lay{\alpha}
\label{eq:ising_node}
\end{equation}
where $\sigma_i\lay{\alpha}$ is the spin of node $i$ at layer
$\alpha$, $J_i$ is the strength of coupling with neighbours of $i$
nodes at the same layer, representing the permeability of $i$ to
direct social pressure, and $\chi_i$ is the level of internal
coherence of node $i$, representing its tendency to prefer a certain
arrangement of its spins on all the layers. The parameter $\gamma$
sets the relative importance of internal coherence and social
pressure. Finally, $h\lay{\alpha}$ models an external force that
drives the spins of a layer in a specific direction. The underlying
assumption is that the system tends to optimize the Hamiltonian
$ H=-\sum_{\alpha = 1}^{M} \sum_{i=1}^{N} F_i\lay{\alpha}
  s_i\lay{\alpha} $.
The struggle between coherence and social pressure at node level
produces new critical behaviours and allows for the emergence of
different phases, depending on the intensity and homogeneity of the
opinions coupling across the population. 
\newtext{For instance, the system may display global consensus, consensus with 
internal incoherence among a fraction of agents, or states dominated 
by internal coherence without external influence.}


Along similar lines, \textcite{chmiel2017tricriticality} studied a $k$-neighbour 
Ising model in a two-layer clique, where the decision to flip a spin depends
only on $k$ of the neighbouring spins. They show that, at difference
with the single-layer case, the transition to consensus is always
continuous if the two layers consist of cliques. Conversely, if the
multiplexity $q$ of the system is tuned, and one of the two layers has
an incomplete clique of nodes which is also active on the other layer,
the transition become discontinuous at a tricritical value $q^*(k)$.

\newtext{
The Ashkin–Teller model \cite{ashkin1943statistics}, a classical spin 
system with four-body interactions, was adapted by \textcite{jang2015ashkin} 
to multiplex networks in order to study opinion formation. They considered 
a two-layer multiplex network where each node has two Ising variables 
(one per layer), coupled through four–spin interactions. By varying the
interlayer coupling strength and the degree distribution, they found 
a rich phase structure, including not only continuous and discontinuous 
transitions but also successive and mixed-order ones. These 
behaviors were explained in terms of Landau 
theory, showing that multiplexity can qualitatively change collective 
states compared to single-layer spin models.
In addition, \textcite{kim2021ashkin} have analyzed the effect of 
link overlap on the same Ashkin-Teller model. Using a generalized 
Ashkin-Teller Hamiltonian defined on multiplex networks with distinct 
overlapping and non-overlapping links, they provided a full mean-field 
solution and revealed complex phase diagrams. They show that overlap 
enhances interlayer correlations and thereby increases the coherence 
of opinions across layers.}

\subsection{Other models of opinion formation}

\begin{figure}[t]
    \includegraphics[width=0.48\textwidth]{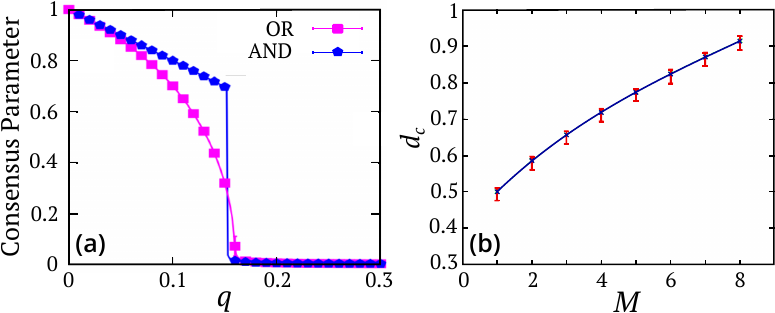}
    \caption[]{Multiplex majority-vote and Deffuant models. (a) Consensus parameter (magnetization) \vito{as a function of the noise parameter $q$}  
      in the multiplex majority-vote model where voters follow the majority opinion 
among the layers (OR model, continuous transition), and when the opinion of all
layers are considered (AND model, abrupt transition). (b) Critical confidence bound $d_c$ to achieve consensus in a multiplex Deffuant model with continuous opinions as a function of the number of layers $M$. Figures adapted from \cite{choi2019majority} and \cite{shang2015deffuant}.}
    \label{fig:isingmodel}
\end{figure}

{\color{black} 
In this section, we review opinion dynamics models beyond the voter and 
Ising-type frameworks that have been extended to multiplex networks. 
A wide spectrum of such models has been proposed in the literature of
social physics \cite{castellano2009statistical,starnini2025opinion}, 
and their generalizations to multiplex networks have also been explored. 
Here we focus on a few paradigmatic cases for which multiplexity has 
revealed qualitatively new phenomena, i.e., majority-vote models, 
bounded-confidence models, and opinion competition models. 

The majority-vote model \cite{de1992isotropic} is a nonequilibrium 
spin system in which each node adopts the majority state of its neighbors 
with probability $1-q$ and the opposite state with probability $q$, where 
$q$ is a noise parameter. On both regular lattices 
and complex networks, the MV model exhibits a continuous order–disorder 
phase transition as noise $q$ is varied \cite{de1992isotropic,pereira2005majority}.
\textcite{choi2019majority} extended the majority-vote model to multiplex networks, 
where every node has a replica on each layer and the replicas share the same state. 
They proposed two different majority rules depending on the type of inter-layer interactions: 
the OR rule and the AND rule. 
Under the OR rule, a node updates its state when the majority condition is 
satisfied in at least one layer, so a single layer is sufficient to drive 
a change. Under the AND rule, by contrast, the majority condition must hold 
in all layers simultaneously, making updates more restrictive. 
Figure \ref{fig:isingmodel}(a) shows that the the magnetization, i.e., 
the average opinion state of the system shows a continuous transition 
in the OR model, while it turns discontinuous for the AND model. 

\textcite{amato2017opinion} extended the Abrams–Strogatz model of language 
competition to multiplex networks in order to study opinion competition dynamics. 
In the original Abrams–Strogatz model \cite{abrams2003modelling}, individuals can have 
one of two opinions and the probability of switching opinions depends on two factors: 
the prevalence of the opinion, defined as the fraction of the entire population currently 
holding it, and its prestige, a fixed parameter that represents the inherent attractiveness 
of the opinion. In a single-layer network setting \cite{castello2006ordering}, this 
dynamics drives the system toward consensus, leading to the extinction of the competing 
alternative (language death). 
\textcite{amato2017opinion} generalized this framework to multiplex 
networks by assigning different prestige values to each layer, while 
inter-layer coupling enforces consistency across replicas of the same individual. 
Under these conditions, if different layers favor different opinions, multiplexity 
generates a stable coexistence phase in which both opinions persist, 
which is unstable in a single layer network.
}

Beyond the case of discrete opinions, \textcite{shang2015deffuant}
considered a multiplex version of the Deffuant model
\cite{deffuant2000mixing}, where agent opinions are continuous, and
the probability for two agents to interact and modify their opinions
depends on how different those opinions are, with smaller differences
yielding higher interaction probability.
\newtext{In the multiplex Deffuant model, each agent is represented by replicas 
on multiplex networks that always share the same continuous opinion value. 
At each update step, one layer is chosen at random, and two neighboring 
agents on that layer may interact if their opinions differ by less than the confidence bound $d$. 
When this condition is satisfied, both agents adjust their opinions  
toward each other.} They provided numerical
evidence that the multiplex version of the model does not allow any
ordering. In particular, Fig.~\ref{fig:isingmodel}(b) shows that the
critical confidence bound $d_c$ to obtain consensus grows with the
number of layers $M$ of the multiplex network.
\newtext{
A subsequent study by \textcite{antonopoulos2018general} generalized 
this setting by allowing different layers to have distinct confidence 
bounds by considering more general initial opinion distributions.}
They confirmed that multiplexity generally hinders the attainment of consensus for
a large range of confidence levels.

\subsection{\change{Social contagion and cultural dissemination}} 

Not limited to opinion formation, multiplex networks provide a powerful 
framework to describe a variety of other social dynamics based on interactions 
that naturally happen and develop across several social spheres. 
Two notable examples include models of cultural dynamics and behaviour
adoption models. The former ones are often considered an extension of
opinion dynamics model, since they consider the evolution of a
certain number of ``cultural traits'' which evolve according to a
specific set of imitation rules. The latter ones, instead, are similar
to the models used to reproduce disease spreading dynamics.
\newtext{
Unlike epidemic spreading where spreading occurs through a single contact, 
social contagion such as behavior adoption or information spreading typically 
requires that the fraction of adopting neighbors exceeds an individual threshold.
In social contagion models, multiplexity may alter cascade conditions 
and can either facilitate or hinder large-scale adoption, depending 
on the interplay between layers. 
In the following, we review representative models of social contagion
and cultural dynamics on multiplex networks
}

{\color{black}
Several works have focused on describing the diffusion of social 
behaviours \cite{schelling1973hockey} and the adoption of 
novelties \cite{bass1969new} across different social contexts, 
following the framework of complex contagion~\cite{guilbeault2018complex} 
and threshold models \cite{watts2002simple}. Social contagion 
is often modeled through the threshold mechanism, and the prototypical model 
that implements this idea is the Watts threshold model.
The threshold model shows that a small initial fraction of adopted nodes can trigger
global cascades.
In this framework, nodes adopt when a sufficient fraction of their 
neighbors have already adopted. \textcite{brummitt2012multiplexity} have
studied a multiplex generalization of the Watts threshold model, where 
adoption requires that a node’s threshold be exceeded in at least one 
of the layers. They show that multiplexity can facilitate cascading dynamics
that would not occur in isolated layers.
\textcite{lee2014threshold} introduced heterogeneous response 
rules across layers, distinguishing between OR-type nodes (adopt if the 
threshold is satisfied in any layer) and AND-type nodes (adopt only if 
thresholds are satisfied simultaneously in all layers). They found that 
OR rules tend to promote cascades, while AND rules suppress them and 
can lead to abrupt transitions. In addition, heterogeneous mixing of 
OR- and AND-type nodes yields nontrivial cascading behavior on multiplex
networks.

Another important type of social contagion is reinforcement models 
in which individuals adopt only after accumulating multiple pieces of 
behavioral information from their neighbors \cite{wang2015dynamics}. 
In these models, each adopted neighbor transmits information with some 
probability, and a susceptible individual adopts once the cumulative number
of exposures exceeds a given threshold.
Extending this framework to multiplex structures, 
\textcite{wang2018social} study the reinforcement model on 
correlated multiplex networks and show that interlayer degree correlations 
strongly affect the final adoption size: negative degree correlations 
facilitate spreading at low transmission probabilities but 
suppress it when transmission is high, whereas positive correlations 
show the opposite effect.} Another study on behaviour adoption is present
in~\textcite{zhu2018heterogeneous}, where each layer has a
different adoption threshold and nodes are divided into activists and
conservatives depending on their individual willingness to adopt the
new behaviour. In particular, the activists will adopt as soon as the
total accumulated pieces of information on any layer exceeds the
corresponding threshold, while conservatives wait until the thresholds
on both layers have been exceeded. The most interesting result is that
the fraction of activists determines the character of the transition
to global adoption, which is a hybrid transition (a discontinuous jump 
coexisting with critical behavior) for lower fractions of
activists and becomes continuous when that fraction increases.




\newtext{
Understanding how culture spreads and organizes is a key topic in
social dynamics and the Axelrod model provides a simple yet widely 
used framework for exploring these dynamics.}
The Axelrod model describes the diffusion of cultural traits into a
population based on the mechanisms of
social influence (individuals tend to imitate each other) and
homophily (similar individuals tend to stick together) \cite{axelrod1997dissemination}.  
In such a model the cultural profile of each individual is described by
a feature vector of $F$ integer variables $(\sigma \lay 1 , \ldots,
\sigma \lay F )$. Each feature $f$, with $f=1,\ldots,F$, takes one of
$\kappa$ possible traits, $\sigma \lay f=0,1,\ldots, \kappa-1$.  
Given two individuals $i$ and $j$, their cultural similarity is measured 
through the so-called {\em cultural overlap} $\psi_{ij}$, defined as:
\begin{equation}
\psi_{ij}=\frac{1}{F}\sum_{f=1}^{F}\delta({\sigma_i \lay f, \sigma_j \lay f})
\label{eq:eq1}
\end{equation}
where $\delta({\sigma_i \lay f, \sigma_j \lay f})$ is Kronecker's
delta and $0 \le \psi_{ij} \le 1$. According to homophily, each pair
of individuals $i,j$ interacts with probability $\psi_{ij}$. The
largest cultural component of the system $S$ includes all
individuals with $\psi=1$ and belonging to the same connected
component. For a fixed number of features $F>2$, at a critical
number of cultural traits $\kappa_c$ the model undertakes
a non-equilibrium phase transition in the size of the largest cultural 
component $S$, from a globalised ($S=1$) to a multicultural 
phase ($S=0$)~\cite{castellano2000nonequilibrium, vilone2002ordering}.  

\begin{figure}[bt!]
	\begin{center}
		\includegraphics[width=0.55\textwidth]{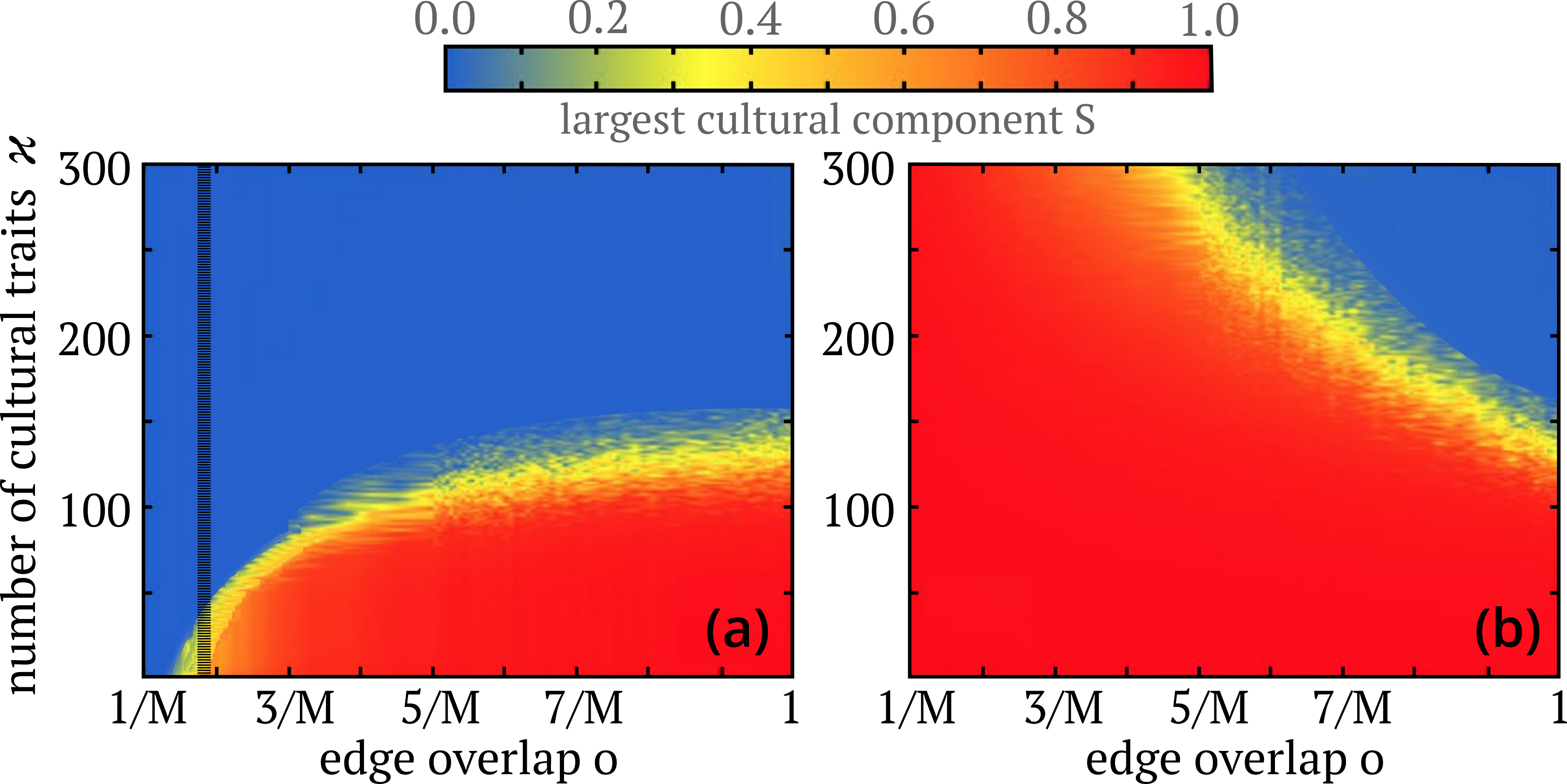}
		\caption[]{Multiculturality in the multiplex Axelrod model. 
			The size of the largest     cultural component as a function of
			the number of cultural traits $\kappa$ in the
			the Axelrod model in multiplex networks with tuneable edge overlap $o$ (a) and their corresponding aggregated networks (b). 
			Figure adapted from~\textcite{battiston2017layered}.}
		\label{fig:social_axelrod}
	\end{center}
\end{figure}

In real systems, individuals tend to
diversify their links according to the topic of the
interactions.  
However, under layered social influence on multiplex networks, individuals may copy a neighbor’s trait only on layers where they are connected.
\textcite{battiston2017layered} showed that the presence of such layered
interactions can explain empirical observations on the presence of multiculturality 
in reality. In order to study the impact of layered social influence on
culture diffusion, the structure of a social network can be controlled
by tuning its level of edge overlap $o$, while the interaction probability is set to be proportional to
the number of shared traits on the connected features $\sum_{f=1}^{F}
a_{ij} \lay f \delta({\sigma_i \lay f, \sigma_j \lay f})$. 
In Fig.~\ref{fig:social_axelrod} the size of the largest cultural
component $S$ is shown as a function of the number of cultural traits $\kappa$ 
and for different values of the structural overlap $o$ in the multiplex
network (a) and the corresponding aggregated network (b).  
For $o=1$ the single and multiplex networks models are undistinguishable, and
multiculturality can only be achieved for large values of $\kappa$.
\newtext{As $o$ decreases, the critical value $\kappa_c$ separating multiculturality 
from globalization becomes smaller in the multiplex networks and 
at a critical value $o_c$, it vanishes. This implies that 
multiplexity promotes multiculturality, with the dynamics 
converging to a multicultural state regardless of the number 
of cultural traits.}



\section{Evolutionary games}
\label{sec:games} 

Interactions mechanisms such as peer pressure and imitation are fundamental to model the emergence of collective phenomena in a population. \change{Yet, in many cases interactions involve strategic decision-making, where the choices of the different agents influence one another, and the outcome for each agent depends on the choices of all others involved}. Such situations are investigated by game theory, a branch of mathematics which has found applications ranging from economics to psychology, ecology and biology. \change{A typical case is that of social dilemmas, where individual self-interest conflicts with the collective good, often leading to outcomes that are worse for everyone involved compared to what could be achieved by cooperating.}

\change{Moving beyond single games and focusing on the dynamics of strategy change,} in the early 1970s Maynard Smith started considering repeated strategic interactions among agents, setting the basis for the emerging field of evolutionary game theory~\cite{smith1972game,smith1982evolution}. In well-mixed populations, the dynamics of the abundances of different strategies is captured by a set of non-linear differential equations known as the replicator equations~\cite{diederich1989replicators,hofbauer1998evolutionary,opper1999replicator,chawanya2002large}. Let us consider a population where each individual is associated to a state $\sigma$, representing its strategy or behavior. At each time, the relative abundance of the different groups is described by $\bm x =\{ x_1, \ldots, x_k \}$, where $x_i$ accounts for the fraction of individuals whose state is $\sigma_i$, and hence $\sum_{1=1}^K x_i =1$.  We have that
\begin{equation}
\dot x_i = x_i [ f_i(\bm x) - \sum_{j=1}^K x_j f_j(\bm x)] \qquad i=1,\ldots,K
\label{games:eq_replicator1}
\end{equation}
where $f_i$ the average payoff obtained by players with strategy $i$. \change{Essentially, the dynamics is such that strategies with higher-than-average fitness increase in frequency because of their success, while those with lower-than-average fitness decrease. Analyzing the the solutions of the replicator equation and their stability helps determine which strategies will eventually survive or be displaced in the long run.}

In real-world systems interactions among agents do not occur at random, but can be conveniently described by networks. The complexity of real-world network structures often limits the insights available from analytical treatments, and the game dynamics is often investigated by agent-based models. ~\textcite{nowak1992evolutionary} first placed agents on the nodes of a square lattice, and ran numerical simulations by letting them repeatedly play with their neighbours, updating their strategy according to a best-response mechanism. \change{They discovered that clusters of cooperators could survive even in adverse conditions, i.e. for systems where full defection would be the expected outcome among rational agents (the so-called Nash equilibrium).} The same result is achieved by considering more flexible update rules, widely used in more modern literature on games on networks, where the probability that an individual $i$ copies the strategy of individual $j$ depends on the difference in earnings of the two agents, $P(\sigma_j \to \sigma_i) \propto f_j - f_i$. At the heart of this phenomenon, known as \textit{network reciprocity}, is that repeated games between neighbors allow for the creation of robust mutual interactions based on trust, more rewarding over time despite the temptation to defect in a single round. This is not possible in well-mixed populations, where the continuous mixing of the agents prevents the formation of special bonds among interacting individuals. 

\change{While the effectiveness of network reciprocity has been challenged by experiments with humans in the lab~\cite{roca2009evolutionary,sanchez2018physics}, its discovery has also given rise to a rich stream of literature investigating analytically and computationally how network structures impact the emergence of prosocial behavior~\cite{nowak2006five, szabo2007evolutionary, perc2013evolutionary, wang2015evolutionary,perc2017statistical}. For instance, small-worldness~\cite{abramson2001social, kim2002dynamic}, clustering~\cite{assenza2008enhancement} and heavy-tailed degree distributions~\cite{santos2005scale, gomez2007dynamical} have all been shown to further promote the emergence of prosocial behavior. Beyond simple graphs, the layers of a multiplex network allow to describe a richer scenario where agents can be involved in distinct contexts and games, for which different strategies might work the best. Hence, the state of each agent $i$ can be described as a vector $\sigma_i=(\sigma_i\lay 1, \ldots, \sigma_i\lay M)$, where at each layer the player has the right to choose to cooperate or defect. Here we provide an  overview of evolutionary games on multiplex networks, focusing on patterns and behaviors which can not be obtained on single networks, and discussing how multiplex structures can help support cooperation beyond the limit of single-layer networks.}

\subsection{Pairwise games} 
\label{sec:pairwisegames}

A lot of attention has been devoted to dyadic settings, where individuals or species interact on pairs defined by the links of a network. 
We consider two-strategy games, where the state of the agents is described by a binary variable, associated for instance to perform a cooperative ($\sigma=C$) or defective ($\sigma=D$) action. 
These dilemmas can be described by the following payoff matrix,
\begin{equation}
\bordermatrix{
	& C & D \cr
	C & R & S \cr
	D & T & P \cr}
\label{games:payoffmatrix}
\end{equation}
where the four payoffs represent the reward $R$ or the sucker $S$ earned by a cooperator against another cooperator or a defector (first line), and the temptation $T$ or the punishment $P$ obtained by a defector against another cooperator or a defector (second line). Depending on the relative values of $R$, $S$, $T$ and $P$, Eq.(\ref{games:payoffmatrix}) defines qualitatively different classes of games. 
When $T > R > P > S$, the payoff matrix describes the prisoner's dilemma~\cite{axelrod1984evolution, rapoport1966taxonomy}. This dilemma presents the most adverse conditions for cooperation, and has $P$ as the Nash equilibrium of the system. When the sucker is higher than the punishment, i.e. $T > R > S > P$, agents play the snowdrift game, sometimes known as chicken's game~\cite{gui2005economics}. This game has two pure Nash equilibria associated to the agents playing opposite strategies, and is hence known as an anti-coordination game. In ecology, this payoff ordering also describes the hawk-dove game~\cite{smith1982evolution, smith1976logic, cressman1995evolutionary}. If instead the order of the other two payoffs is inverted, i.e. $R > T > P > S$, individuals play the stag-hunt game~\cite{rousseau1997discourses, luce1957games}. Once again the game has two pure Nash equilibria, one of and full defection (risk-dominant) and one of full cooperation (payoff-dominant), and coordination is necessary to synchronize on the latter. Finally, when $R$ is the highest payoff and $P$ the lowest, independent on the relative values of $S$ and $T$, the absence of conflict defines the harmony game, which has full cooperation as the dominant strategy and the only Nash equilibrium.

\begin{figure}[t!]
\begin{center}
\includegraphics[width=0.48\textwidth]{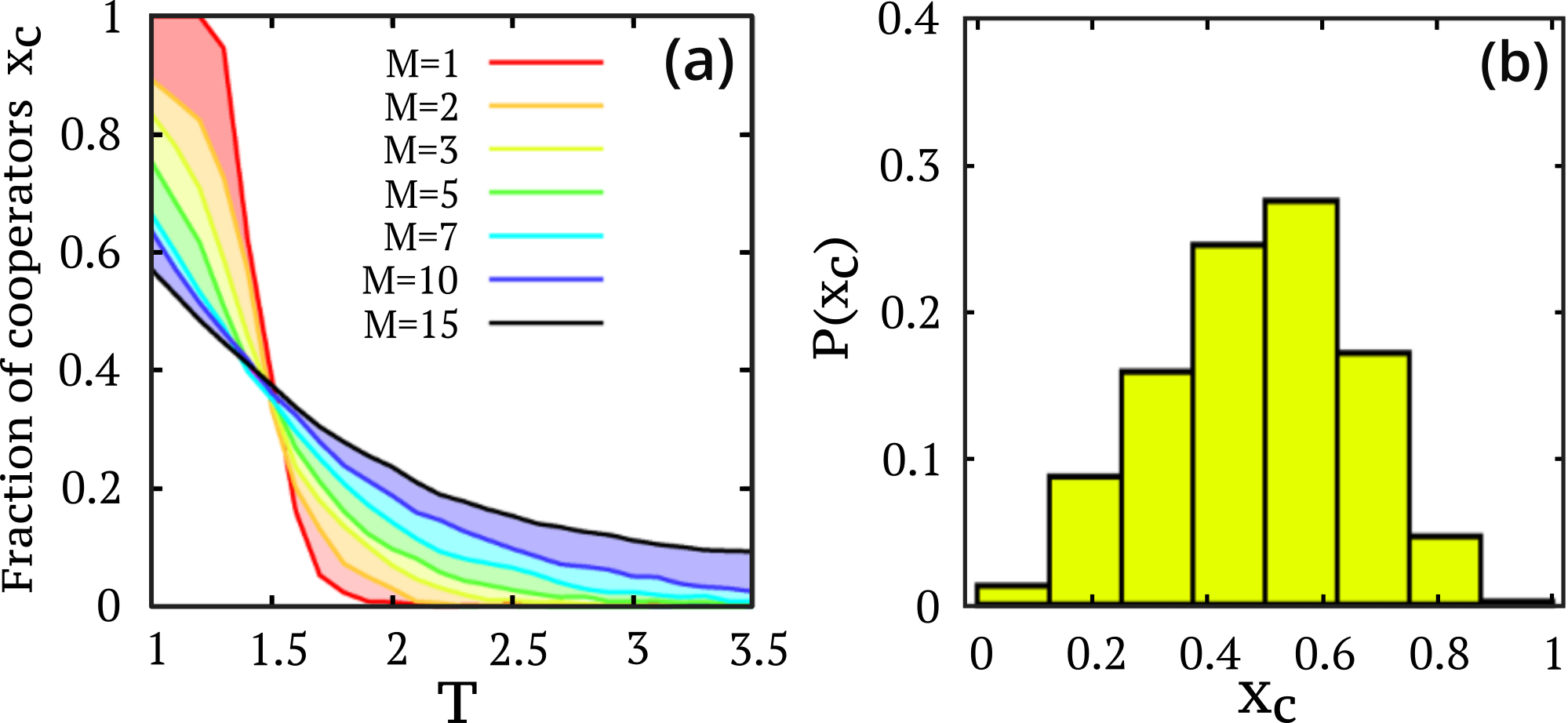}
\caption[]{Multiplexity enhances cooperation in the prisoner's dilemma. (a) Average fraction of cooperators $x_c$ as a function of the temptation parameter $T$ in a multiplex network with uncorrelated ER layers with $\langle k \rangle = 3$. Individuals play a weak prisoner's dilemma with $R=1$, $P=0$ and $S=0$. (b) Heterogenous distribution of the fractions of cooperators $x_c\lay \alpha$ across layers for $T=1.4$. Figures adapted from~\textcite{gomez2012evolution}.
}
\label{fig:games1}
\end{center}
\end{figure}

The investigation of games on multiplex networks was kicked-off by
~\textcite{gomez2012evolution}, which first studied individuals
playing a weak version of the prisoner's dilemma ($P=S$) on
uncorrelated ER multilayer networks, allowing for different strategies
on different layers. In a time step of a Montecarlo simulation, two
connected agents $i$ and $j$ accumulate earnings $f_i^{\alpha}$ and
$f_j^{\alpha}$ by playing games with each of their neighbors at each
layer $\alpha$. Interdependence between layers is achieved through
coupling of the fitness functions: when an individual $i$ updates its
strategy $\sigma_i \lay \alpha$, the probability $P(\sigma_j\lay \alpha \to
\sigma_i\lay \alpha)$ does not depend on the fitness at the
corresponding layer $\alpha$, but on the aggregated fitness:
\begin{equation}
f_i=\sum_{\alpha=1}^M f_i\lay{\alpha},
\label{games:aggregation1}
\end{equation} 
mimicking the lack of fine-grained information on the earnings in different social contexts. 
This novelty introduces additional complexity for neighbors of a node $i$ in a $\alpha$ layer, since imitation based on the net benefit of $i$ may lead to the choice of strategies that do not perform well in the specific strategic context of layer $\alpha$. Interestingly, such condition strengthen the resilience of cooperation, which is attained even in adverse conditions, by suppressing feedback of individual success at the single layers, hence reducing the aggressive invasion of defectors into the population. This is shown in Fig.~\ref{fig:games1}(a), where cooperators across all layers survive even for extremely large values of the temptation $T$, and the higher the number of layers the stronger the enhancement of cooperation. This behavior is the opposite of what would be observed if players would play in the corresponding aggregated network, where the addition of uncorrelated layers makes the network denser and closer to the limit of well-mixed population, where cooperation extinguishes quickly. Besides, 
the distribution of cooperators $\bm x_c =\{x_c\lay 1, \ldots, x_c\lay M \}$ across the layers was found to be heterogenous, as shown in Fig.~\ref{fig:games1}(b). 

A different setting considers that the strategy update at one layer might depend for a fraction $\gamma$ of the earnings on the opposite layer. In general higher values of $\gamma$ tend to further promote prosocial behavior. Besides, spontaneous symmetry breaking in the number of cooperators of the two layers arises naturally only when the coupling in the payoff earnings is above a given critical threshold $\gamma_c$~\cite{jin2014spontaneous}. If only a fraction of nodes bias their fitness calculations on earnings on the opposite layer, an intermediate fraction of inter-layer couplings was found to maximise cooperation~\cite{wang2013optimal}.

\begin{figure}[b!]
	\begin{center}
		\includegraphics[width=0.55\textwidth]{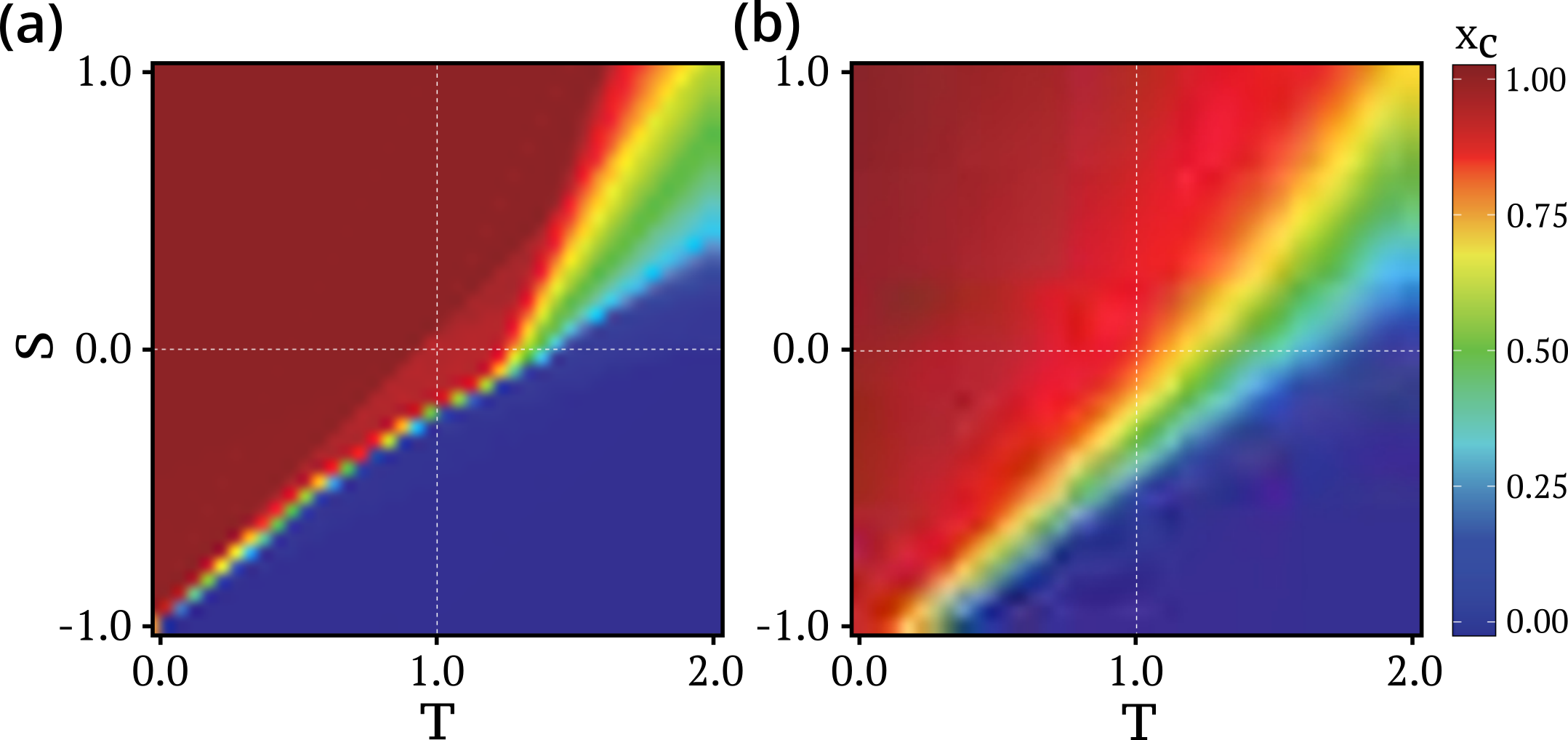}
		\caption[]{Fraction of cooperation in the $T-S$ plane, for (a) a single network and (b) a multiplex network with $M=10$ ER uncorrelated layers with $\langle k \rangle = 3$. Reward and punishment are set to $R=1$ and $P=0$.  The values of temptation $T$ and sucker $S$ payoffs determine four different games: harmony (upper-left quadrant), snowdrift (upper-right), the stug-hunt (lower-left) and the prisoner's dilemma (lower-right). Multiplexity enhance cooperation in the latter three games, but allows defectors to survive in a population playing the harmony game. 
			Figures adapted from~\textcite{matamalas2015strategical}
		}
		\label{fig:games2}
	\end{center}
\end{figure}

~\textcite{matamalas2015strategical} uses the same set-up of ~\textcite{gomez2012evolution} to investigate the emergence of cooperation in other social dilemmas. This can be easily done by setting $R=1$ and $P=0$ and letting vary $T$ and $S$. In Figs.~\ref{fig:games2}(a, b) the fraction of cooperators $x_c$ is shown in the full $T-S$ for a single network, as well as for a multiplex networks with $M=10$. Similarly to the prisoner's dilemma, multiplexity promotes cooperation also in the stag-hunt and the snowdrift games. However, interestingly, the multiplex structure can also hamper prosocial behavior, with the survival of defectors for a wide range of the parameters in the harmony game for which full cooperation is always achieved on single networks. For all games, the population was found to split into coherent agents -- which played the same strategy across different layers -- and highly incoherent ones, explaining the failure of mean-field approaches to describe the system dynamics.

\begin{figure}[t!]
\begin{center}
\includegraphics[width=0.55\textwidth]{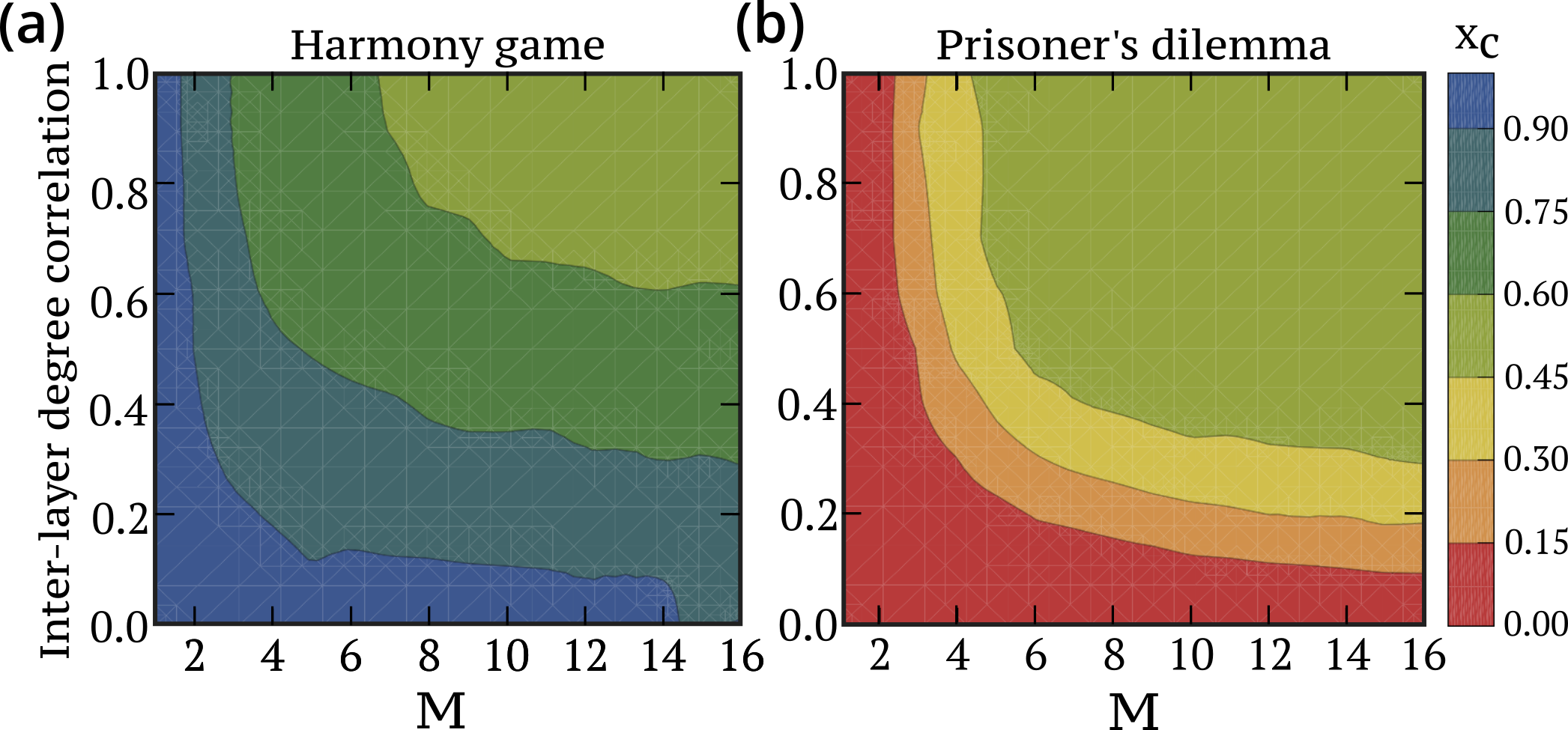}
\caption[]{Multiplex structure induces topological enslavement in heterogeneous structures.
Average fraction of cooperators as a function of inter-layer degree correlations and number of layers $M$ for (a) the harmony game ($T=0.5$, $S=0.5$) and (b) the prisoner's dilemma ($T=1.5$, $S=-0.5$). All layers are obtained by using the multiplex geometric model introduced in Ref.~\cite{kleineberg2016hidden}, have power-law distribution with exponent $\gamma=2.6$ and non-vanishing clustering coefficient $C=0.4$. Coupling the fitness at the different layers with the presence of hubs that can accumulate disproportionate earnings by playing more games makes the game dynamics and the emerging levels of cooperation depend on the topology rather than on the exact values of the game parameters for correlated systems with many layers. Figures adapted from~\textcite{kleineberg2018topological}.
 }
\label{fig:games3}
\end{center}
\end{figure}

~\textcite{kleineberg2018topological} first investigated evolutionary games \change{on scale-free multiplex networks, focusing on the prisoner's dilemma and harmony game. They considered a fixed-cost-per-game scenario~\cite{santos2008social}, where hubs play more game at no additional cost and hence have the potential to earn higher payoffs than poorly connected individuals. While on single networks the two games yield very different outcomes, on a multiplex networks their emerging dynamics become increasingly similar in heterogenous systems as the number of layers $M$ and inter-layer degree correlation increase (Fig.~\ref{fig:games3}). This phenomenon was dubbed topological enslavement, and occurs when the evolutionary dynamics becomes dominated by the hub nodes (i.e., the
network topology), such that the outcome of the game is determined by initial conditions rather than the game
parameters.}

The basic setting considered so far, where new behavior is induced by the coupling of the payoffs at different layers, can be further enriched. For instance,~\textcite{xia2015heterogeneous} modified the original implementation of the multiplex prisoner's dilemma by making the strength of the interdependent coupling player-dependent. 
Other options include introducing individual features such as influence~\cite{meng2015spatial, jia2019ability} and reputation~\cite{wang2017inferring}, adding the effect of strategies popularity~\cite{liu2018popularity}, introducing memory in the strategy update process~\cite{luo2016cooperation}, considering layers with weighted links~\cite{meng2016interdependency} or systems with limited resources~\cite{luo2018co,sun2020aspiration}.While all these scenarios contribute to a better description of the microscopic mechanisms behind the evolution of cooperation in real-world systems, yielding more complex and richer patterns, they do not alter qualitatively the findings obtained by the simplest scheme previously described.

A different scenario was considered in~\cite{wang2014degree}, where an interaction layer is used by the agents to play and accumulate payoff, coupled to a second layer which determines the neighbours chosen for the strategy update. in presence of intra-layer degree-degree correlation on both layers cooperation is hindered for the prisoner's dilemma, the snowdrift and the stag-hunt game, whereas prosocial behavior may be further promoted in case of disassortative mixing. Finally, it is worth to mention that the literature has not been limited to games described by the payoff matrix in Eq.~\ref{games:payoffmatrix}. Multiplex extensions of other pairwise dilemmas have been proposed, including the ultimatum game~\cite{deng2020effects} and the traveler's game~\cite{xia2014evolution}. 

\subsection{Multiplayer games}
\label{sec:multiplayergames}

In many cases social dilemmas involve more than two individuals at a time. The public goods game, first introduced in the context of experimental economics, is the most-well known multiplayer game~\cite{sigmund2010calculus, archetti2012review}. In its simplest implementation, each individual in a group of $G$ players can decide to cooperate and contribute to a common pool by mean of a token $t$, or defect. The amount in the pool is then multiplied by a synergy factor $R$, and shared equally among all agents, regardless of their strategy. Hence, cooperators earn a payoff $f_c = t (N_c R / G -1)$, whereas defectors receive $f_d = t (N_c R / G)$, where $N_c$ is the number of cooperators in the group. The game can be studied as a function of a single parameter $r=R/G$, known as the reduced synergy factor. For $r<1$ rational behavior would suggest agents to defect. However, if everybody follows this strategy no dividends will be available, no player will make an individual profit, and the population will follow into the so-called tragedy of the commons~\cite{hardin1968tragedy}. Hence, for $r<1$ the  game is widely regarded as the generalization of the prisoner's dilemma to groups of arbitrary size. By contrast, for $r>1$ it is assimilable to the harmony game previously introduced. 

In a network, groups can be inferred by considering an agent and all its linked neighbours. First observed in pairwise games, computer simulations showed that network reciprocity also holds for group interactions in homogeneous systems~\cite{szabo2002phase}. Clustering~\cite{rong2010feedback}, and in particular a heavy-tailed degree distribution~\cite{santos2005scale, santos2006evolutionary} sustain prosocial behavior, though assortative mixing can reduce cooperation by limiting the evolutionary advantage of hubs~\cite{rong2009effect}. In traditional graph implementations multiplayer games are intrinsically different from their pairwise counterpart, due to the formation of indirect links among players which are not directly connected in the network but play within the same group~\cite{szolnoki2009topology, szolnoki2013correlation}. For instance, increasing group size does not necessarily lead to mean-field behavior, as observed for pairwise games~\cite{szolnoki2011group}. However, such behavior is obtained when group evolutionary dynamics are modeled through higher-order interactions~\cite{alvarez2021evolutionary}, where group associations are described by different hyperlinks~\cite{battiston2020networks, battiston2021physics}, or through a bipartite representation between agents and groups~\cite{gomez2011evolutionary}. For a survey on the public goods game on structured populations, we refer the reader to the recent reviews~\cite{perc2013evolutionary, perc2017statistical}.

Similarly to the pairwise case, also for multiplayer games multiplexity was first introduced by mean of interdependent fitness functions. ~\textcite{wang2012evolution} suggested to perform the strategy update on both layers of a two-layer multiplex networks based on the following aggregated fitness 
\begin{equation}
\change{f_i\lay{\alpha} = \epsilon f_i\lay{\alpha} + (1 - \epsilon) f_i\lay{\beta}}
\label{games:aggregation2}
\end{equation}

\begin{figure}[t!]
\begin{center}
\includegraphics[width=0.55\textwidth]{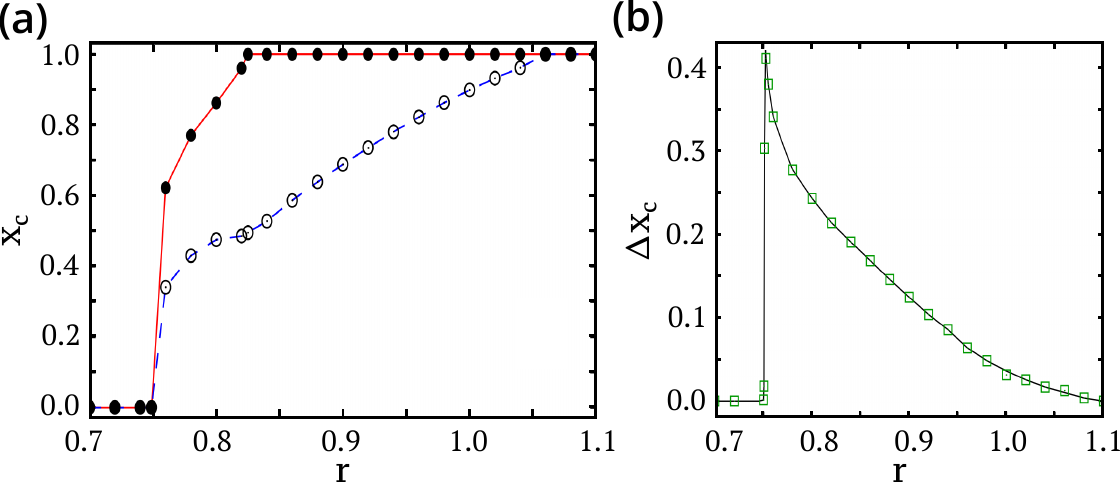}
\caption[]{Enhanced cooperation in the public goods game through biased fitness functions.
(a) Average fraction of cooperators in the first (black full dots) and second (blue empty dots) layer as a function of the reduced synergy factor in a system described by the biased fitness function of Eq.~\ref{games:aggregation2} for $\alpha=0.4$. \change{Symmetry breaking emerges, and the (b) difference in the fraction of cooperators in the whole systems against what the cooperation of a single network in isolation.} 
Figures adapted from~\textcite{wang2012evolution}
 }
\label{fig:games4}
\end{center}
\end{figure}

\change{where $\epsilon$ is a bias parameter that couples the dynamics of the strategy at layer $\alpha$ with the success of the strategy of the same player in the other layer $\beta$. It has been shown that the stronger such bias in the
utility function, the higher the level of public cooperation.
Spontaneous symmetry breaking leads to unequal levels of cooperation, [Fig.~\ref{fig:games4}(a)]. Yet, the aggregate density of cooperators on both networks is higher
than the one attainable on an isolated network [Fig.~\ref{fig:games4}(b)].}

Similar biased fitness functions have been studied by~\textcite{liu2019individual}, where a subset of individuals is assigned a greater influence and wider ability to spread their strategy. 

\begin{figure}[t!]
    \begin{center}
        \includegraphics[width=0.55\textwidth]{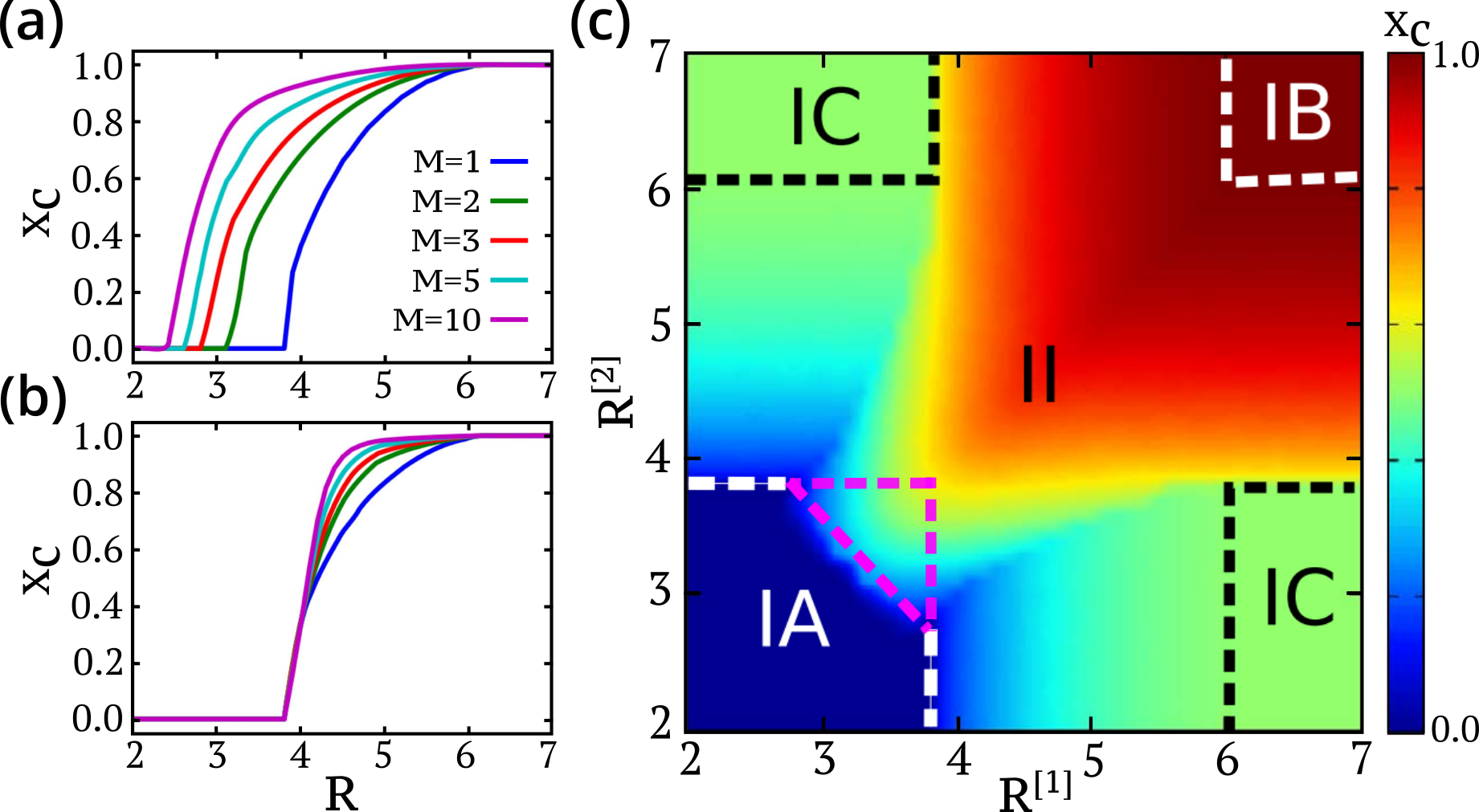}
\caption[]{Structural correlations are required for multiplex reciprocity to enhance public cooperation.
Average fraction of cooperators in the public goods game as a function of the non-reduced synergy factor for multiplex networks with maximal (a) or no (b) edge overlap for different number of layers $M$ for two regular random graphs with $k=4$. For $M=2$, the emergence of cooperators into the system is pushed from $R_c=3.75$ ($r_c=0.75$) in the case of isolated networks to $R_c=3.25$ ($r_c=0.65$) for a multiplex networks with maximal edge overlap. (c) Fraction of cooperators in a multiplex network with different synergy factors $R\lay \alpha, R\lay \beta$ and maximal edge overlap. cooperators survive in the population as long as $\frac{R\lay{1} + R\lay{2}}{2} > 3.25$ (purple dashed line), as long as the two synergy factors do not go beyond the original condition $R_c=3.75$ for the single networks.
Figures adapted from~\textcite{battiston2017determinants}.
 }
\label{fig:games5}
\end{center}
\end{figure}

~\textcite{battiston2017determinants} investigated how different multiplex structures impact collective behavior. Using the fitness function scheme of Eq.~\ref{games:aggregation1}, \change{they found that multiplexity further promotes prosocial behavior only when edge overlap is present in the system, and increases monotonically with it (Figs.~\ref{fig:games5}(a, b))}. 
They also explored the case of different synergy factors at different layers. In such asymmetric case, cooperation can survive in all layers of the system even for extremely low values of the synergy factor in one of them. However, this is only possible if the average non-reduced synergy factor across all layers is greater or equal than the critical threshold obtained for the symmetric case, where all layers have the same game parameter [Fig.~\ref{fig:games5}(c)].

\subsection{Coupling different games}
\label{sec:coupledgames}

\begin{figure}[t!]
    \begin{center}
        \includegraphics[width=0.35\textwidth]{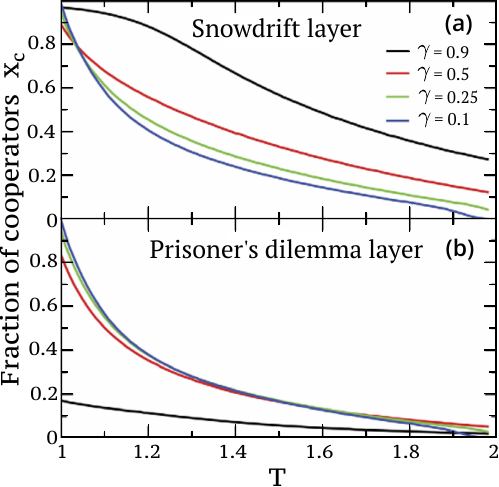}
        \caption[]{\change{Fraction of cooperators in the coupled snowdrift (top) and prisoner's dilemma (bottom) as a function of the temptation T. The games are coupled by a coupling function, where a parameter $\gamma$ indicates the probability to select payoffs from the opposite layer.} Figures adapted from~\textcite{santos2014biased}}
        \label{fig:games6}
    \end{center}
\end{figure}

The multiplex framework has also allowed to investigate the behaviors emerging by coupling different games which take place on different layers. ~\textcite{santos2014biased} investigated the dynamics of a multiplex network where the prisoner's dilemma and the snowdrift game are played on the two layers. They considered a biased imitation mechanism to describe the system dynamics, so that when updating its strategy $\sigma \lay \alpha$, a node selects with probability $\gamma$ a neighbour on the same layer, and with probability $1-\gamma$ a neighbor on the other. By exploring how cooperation varies as a function of the probability $\gamma$, it was found that even a slight deviation from the limit case when transfer strategy across layers is never allowed ($\gamma = 1$) promotes cooperation in the layer where the prisoner's dilemma is played, at the same time hindering cooperative behavior in the snowdrift layer (Figs.~\ref{fig:games6}). A similar setting was considered by~\textcite{xia2018doubly}, showing that strategy sharing generates a greater increase in cooperation for the prisoner's dilemma than the corresponding cooperation loss in the snowdrift game, leading to an overall gain in prosocial behavior. In particular, it exists a critical intermediate value of $\gamma$ for which the growth rates of
cooperation in the system is maximized~\cite{liu2020evolution}. The choice of different selection rules, such as imitation based on majority rule, produce qualitatively similar results~\cite{jiang2015multiple}.

\change{To summarize, the multiplex framework has mainly been exploited to introduce dynamical coupling among payoffs earned at different layers among the different replica-nodes of the same agent. Such a scheme is flexible, has it allows to couple across different layers the same game, the same game but characterised by different parameter, or even entirely different games. 
A crucial finding is that multiplexity naturally leads to the emergence of spontaneous symmetric breaking, and support cooperation beyond the limit of the same game played in isolation in a single network. Yet, such enhancement is not to be taken for granted, but is modulated by the presence of structural correlations in the multiplex structure. Primary examples are the presence of structural correlations such as the edge overlap promoting prosocial behavior, or inter-layer degree correlations giving rise to multiplex hubs which dominate system dynamics regardless of the exact game parameters. Exploiting rich structural and dynamical interdependencies among layers, these findings suggest that multiplexity may serve as a further mechanism to explain the survival of cooperation in adverse conditions.}


\section{Intertwining different types of dynamics} 
\label{sec:intertwined} 


%
We now turn our attention to novel phenomena emerging when two or more    
dynamical processes of entirely different nature take place over the 
different layers of a multiplex network and are mutually coupled. 
\newtext{
In real-world systems, different dynamical processes rarely
evolve in isolation, but often proceed 
simultaneously interacting with and influencing each others. 
%
%
In this sense, multiplex networks are a natural setting to study 
the interaction of two or more dynamical processes. 
Cascading failures in interdependent 
networks of Section \ref{perc1}, coevolution of epidemic spreading and awareness diffusion of Section \ref{sec:coupledgames}, and coupled games on
multiplex networks of Section 
\ref{sec:coupledgames} are examples of coupled 
dynamics across layers.
However, all these cases involve homogeneous processes.  In this
section, we focus instead on the coupling of multiple processes that
are more heterogeneous in nature, with typical scenarios including the
coupling of neural synchronization with metabolic transport or the
interplay between opinion dynamics and information spreading in
society.
}

\newtext{
  A general framework for modeling intertwined dynamics on
  multiplex networks, as in the example with two layers shown  
  in Fig. \ref{fig:figure_section2_dynamics}(c),  
  is to consider the following coupled equations:  
}
\begin{equation}
   \left\{ \begin{array}{c}
       { d \sigma_i\lay1  }/{dt} = F_{\omega_i} ( {\bm \sigma}\lay1 , A\lay1 )   \\
       { d \sigma_i\lay2}/{dt} = G_{\chi_i}  ( {\bm \sigma}\lay2 , A\lay2 )     \\
\end{array}
\right.
\qquad 
i=1,2,\ldots N
\label{eq:intertwine}
\end{equation}
In this case, 
${\bm \sigma}\lay1 = [ \sigma_1\lay1,\sigma_2\lay1,\ldots, \sigma_N\lay1 ]^T \in
\mathbb{R}^N$ and ${\bm \sigma}\lay2 = [ \sigma_1\lay2, \sigma_2\lay2,
\ldots, \sigma_N\lay2 ]^T \in \mathbb{R}^N$ are the states of the systems
respectively at the first and second layer,
and the two dynamics are governed by the functions $F_{\omega}$ and
$G_{\chi}$, which depend on the parameters $\omega$ and $\chi$.  
The two dynamical processes are connected through an appropriate
choice of these parameters. Namely, the value of $\omega_i$ in function 
$F_{\omega_i}$ at the first layer is set to change in time depending on the
dynamical state $\sigma_i\lay2$ of node $i$ at the second layer as 
$\dot{\omega}_i = f ( \omega_i, \sigma_i\lay2)$.
Analogously, the evolution of the
parameter $\chi_i$ at the second layer is ruled by
$\dot{\chi}_i = g (\chi_i, \sigma_i\lay1)$, 
which depends on the state $\sigma_i\lay1$ of node $i$ at the first layer.

\newtext{ The parameter coupling in Eqs.~\eqref{eq:intertwine} offers
  a framework for representing a variety of intertwined processes.
  For instance, in coupled opinion and information spreading, the rate
  at which agents update their opinions ($\omega_i$) may increase with
  exposure to external information on another layer, while the
  dissemination of information ($\chi_i$) can in turn be enhanced or
  suppressed depending on the local consensus of opinions.  Similarly,
  in neural networks, the synchronization dynamics of neurons can be
  coupled to metabolic transport, where energy supply parameters
  evolve in response to neural activity and, conversely, constrain the
  stability of oscillatory states.}
\begin{figure*}
\begin{center}
\includegraphics[width=0.75\linewidth]{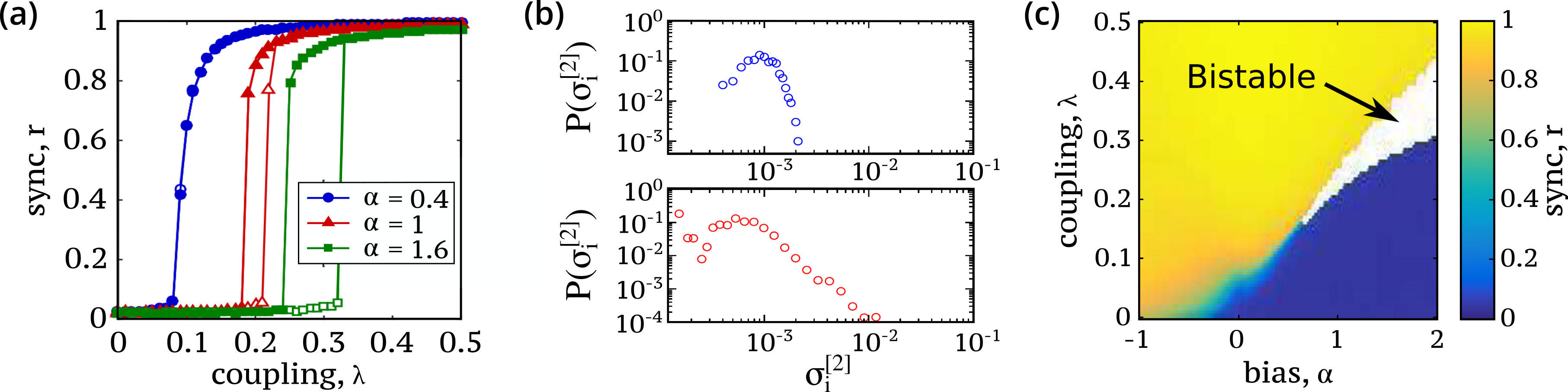}
\caption{Collective behaviors induced by intertwining synchronization and
  transport processes. (a) The level of synchronization $r$ at layer 1
  is shown as a function of $\lambda$, for bias exponents $\alpha=0.4$, $1.0$
  and $1.6$ (blue, red and green, respectively).  (b)
  Stationary distribution $P(\sigma\lay{2}_i)$ of random
    walkers at layer 2 for $\alpha=1.0$, when the
    oscillators at layer 1 are incoherent ($\lambda=0.1$, top, blue)
    and synchronized ($\lambda=0.4$ bottom, red).  (c) Synchronization
    phase diagram showing $r$ as a function of coupling $\lambda$ and
    bias exponent $\alpha$. The bistable region is colored in
    white. Multiplex networks with $N=1000$ nodes, a scale-free graph
    with $\gamma=3$ at layer 1, and a ER random graph
    at layer 2. Figures adapted from \textcite{nicosia2017collective}.
}
  \label{fig:inter_nsal}
\end{center}
\end{figure*}
\newtext{
%
More specifically   
\textcite{nicosia2017collective} studied coupled dynamics consisting
of Kuramoto oscillators modelling neuronal 
activity on one layer of a multiplex network,
and biased random walks for transport 
dynamics of metabolic resources on the other layer.
The intrinsic frequency of each Kuramoto oscillator 
is not fixed but relaxes over time toward a value proportional 
to the fraction of random walkers present at the corresponding 
node, reflecting the idea that neuronal activity depends on the 
local availability of metabolic resources. At the same time, 
each node has a bias variable that evolves toward its dynamic 
synchronization strength. Both numerical and analytical 
results (Fig.~\ref{fig:inter_nsal}) show that this mutual 
feedback induces explosive synchronization in the oscillator 
layer, while, at the same time,
the distribution of random walkers in the transport 
network undergoes a transition from a homogeneous 
to a heterogeneous state.
}

\newtext{
  Using the same framework,
\textcite{li2020double} considered a system of 
intertwined synchronization and game dynamics.
In this study, only cooperating
oscillators contribute to synchrony, and the noise in the
strategy updating procedure of the evolutionary game is governed by
the order parameter in the neighborhood of each oscillator. Similarly
to the previous case, the mutual coupling gives rise to a double explosive
transition both in the Kuramoto and in the Prisoner's Dilemma dynamics, both 
systems normally displaying continuous phase transitions. 
}

In addition, ~\textcite{mikaberidze2022sandpile} investigated the mutual feedback
between Kuramoto oscillators and a model of sandpile cascades model,
where the more out-of-sync a node is with its neighbors the lower its
load-carrying capacity, and where the phase of toppling nodes are
reset at random. While the system is typically trapped in a
synchronized state where load builds up with minimal cascades, it
eventually reaches a tipping point where large cascades of cascades
are triggered. After that, cyclic behavior emerges, with the system
going back to the synchronous buildup phase, preparing for a new rare
sequence of cascading failures.

\newtext{
In another study, \textcite{danziger2019dynamic} introduced  
the concepts of dynamic interdependence and competition between layers, 
providing a general framework to study how cooperative or 
antagonistic interactions between dynamical processes reshape 
collective behavior in multiplex networks.
By implementing this framework in models of coupled oscillators 
to compare interdependent and competitive synchronization 
and in spreading processes to contrast cooperative versus 
competing contagions, they show that dynamic coupling can 
give rise to abrupt transitions, hysteresis, multistability, 
and even chaotic behaviours that do not emerge when the processes 
evolve independently. 
}

\newtext{
Social systems also provide prominent examples where coupled dynamics naturally emerge.}
~\textcite{amato2017interplay} mutually coupled a biased voter model with different game 
dynamics to model the interplay of strategic choices and social influence taking place at
 different layers of a multiplex network. The authors find that such intertwined process 
can significantly increase prosocial behavior, sustainining partial cooperation in the 
prisoner’s dilemma and even full cooperation in the stag hunt game for parameter regions 
where isolated game dynamics leads to full defection, as well as leading to local clusters 
mimicking polarization in social systems.

~\textcite{velasquez2017interacting} investigated the interplay between opinion formation
and disease spreading. In particular, they considered a voter model intertwined with a 
contact process describing disease spreading over multiplex network, with the agent 
propability to update its states depending on both the opinion and disease states of 
the neighbors. A mean-field analysis reveals that beyond a critical value of social 
influence the transition from healthy to endemic state in the disease becomes explosive. 
Moreover, consensus time in the opinion layer behaves non-monotonically as a function of
 the edge overlap, with either full or no overlap associated to the quickest consensus.

\textcite{iacopini2020multilayer} introduced a model of intertwined spreading on 
a two-layer network with a dynamical recovery mechanism. In particular, they studied
 an SIS-like dynamics whose transitions are decoupled across two different network 
layers: while infections follow the standard simple contagion paradigm, a social influence
 mechanism acts on the recovery rule, defining what they call 'complex recovery'---alike 
complex contagion. Numerical simulations and analytical treatments on synthetic and 
real-world networks showed that this change of perspective might lead to explosive adoption 
dynamics and sensitivity to initial conditions. This is especially pronounced in spatial 
systems, where clusters of early adopters help sustain the epidemics.

Finally, ~\textcite{wu2017influence} considered a simplified setting, where the coupling 
among processes is not mutual but unidirectional, and investigated the influence of trust
on information spreading. Spreading is described by a threshold model, where an individual 
becomes informed when a sufficiently high fraction of its neighbours are. Sources which 
have proven themselves reliable are given more weight in the spreading process, with the 
trustability of individuals modelled by the prisoner's dilemma. For a fixed value of the 
threshold,  an analytical solution for the fraction of active spreaders can be computed 
by assuming a locally-tree network structure. \newtext{Intermediate values of temptation
in the prisoner’s dilemma, which balance the number of trustable and untrustable 
individuals in the population, have been identified as the most 
favorable conditions for information spreading.}

\section{Coevolution of networks and processes}
\label{sec:coevolution}

\newtext{
The dynamics of nodal states evolve on top of the network 
structure \cite{boccaletti2006complex}, while the structure itself evolves time 
as shown in studies of temporal and growing networks \cite{holme2012temporal,barabasi1999emergence}. 
In many complex systems, these two levels of dynamics are not independent 
but coevolve: the network adapts in response to the states of its nodes, 
while the dynamics of the nodes are shaped by the evolving topology.
Such coevolution can be observed, for example, in epidemics, where individuals 
not only change their own state from susceptible to infected but also rewire 
connections to reduce the likelihood of exposure \cite{gross2006epidemic}, 
and in social dynamics, where opinions evolve through interaction between 
connected peers while homophily reshape the underlying 
ties \cite{holme2006nonequilibrium,vazquez2008generic}.
From this perspective, coevolution has been recognized as a central feature of 
complex systems~\cite{thurner2018introduction}, and has received much attention in 
many studies~\cite{gross2008adaptive,berner2023adaptive}.
}

When systems comprises different layers of interactions, it is natural to 
consider the case of multiplex coevolution, as formally described 
by Eqs.~\ref{eq:coev_cont} and \ref{eq:coev_discr} for continuous and 
\newtext{discrete} dynamics and illustrated in Fig. \ref{fig:figure_section2_dynamics}(d). 
\newtext{
In this context, coevolution may proceed simultaneously across layers, and structural 
changes in one layer can interact with the dynamics in another in a variety of ways. 
This section discusses representative examples of coevolution on multiplex networks, 
including coevolving voter dynamics, adaptive epidemic models, synchronization 
phenomena, and game-theoretical frameworks.
}

\newtext{
A representative example of these coupled dynamics is the coevolving voter model. 
On single-layer networks, the coevolving voter model is typically defined as follows,
noting that many variants of the model exist in the 
literature~\cite{holme2006nonequilibrium,vazquez2008generic,durrett2012graph}. 
At each update step, a node $i$ is chosen uniformly at random, and one of 
its neighbors $j$ is selected. If their states are equal, $\sigma_i = \sigma_j$, 
nothing happens. If their states differ, $\sigma_i \neq \sigma_j$, then with 
probability $1-p$ node $i$ adopts the state of $j$, while with probability $p$, 
so called plasticity, it cuts the link to $j$ and rewires it to another node $\ell$ that is not already
its neighbor. The model shows an absorbing phase transition between an active phase 
and a frozen phase. The active phase is characterized by a finite density of active 
links, defined as the fraction of links that connect a pair of nodes in different 
states, while the frozen phase corresponds to an absorbing consensus state 
where this density vanishes.
}

\textcite{diakonova2014absorbing} considered one of the first
coevolving multiplex models that studies this effect with a given value 
of multiplexity $q$, where nodes could sever and rewire edges leading to 
neighbours with different opinions. 
\newtext{
Concretely, the system consists of two layers of regular random graphs. 
A fraction $q$ of the nodes are connected by interlayer links; that is, with 
probability $q$ a node has replicas across the two layers. 
Whenever such an interlayer link exists, the corresponding replicas are 
required to share the same state, $\sigma\lay{1}_{i}=\sigma\lay{2}_i$. 
Therefore, the parameter $q$ serves as a control parameter that tunes between 
the dynamics of single-layer networks ($q=0$) and multiplex networks ($q=1$).
Another ingredient of multiplexity is layer-dependent plasticity: each layer 
$\alpha$ is characterized by its own plasticity $p^{\alpha}$. 
The update rule is then as follows: At each step, a layer $\alpha$ is chosen 
uniformly at random, and then a node $i^{[\alpha]}$ and one of its neighbors 
$j^{[\alpha]}$ within that layer are selected.
If $\sigma_i^{[\alpha]} \neq \sigma_j^{[\alpha]}$, 
imitation occurs with probability $1-p^{\alpha}$ and rewiring with probability $p^{\alpha}$. 
In addition, whenever node $i$ has replicas across layers, their states are updated 
simultaneously to remain identical.
}

\begin{figure}[t!]
    \includegraphics[width=0.6\textwidth]{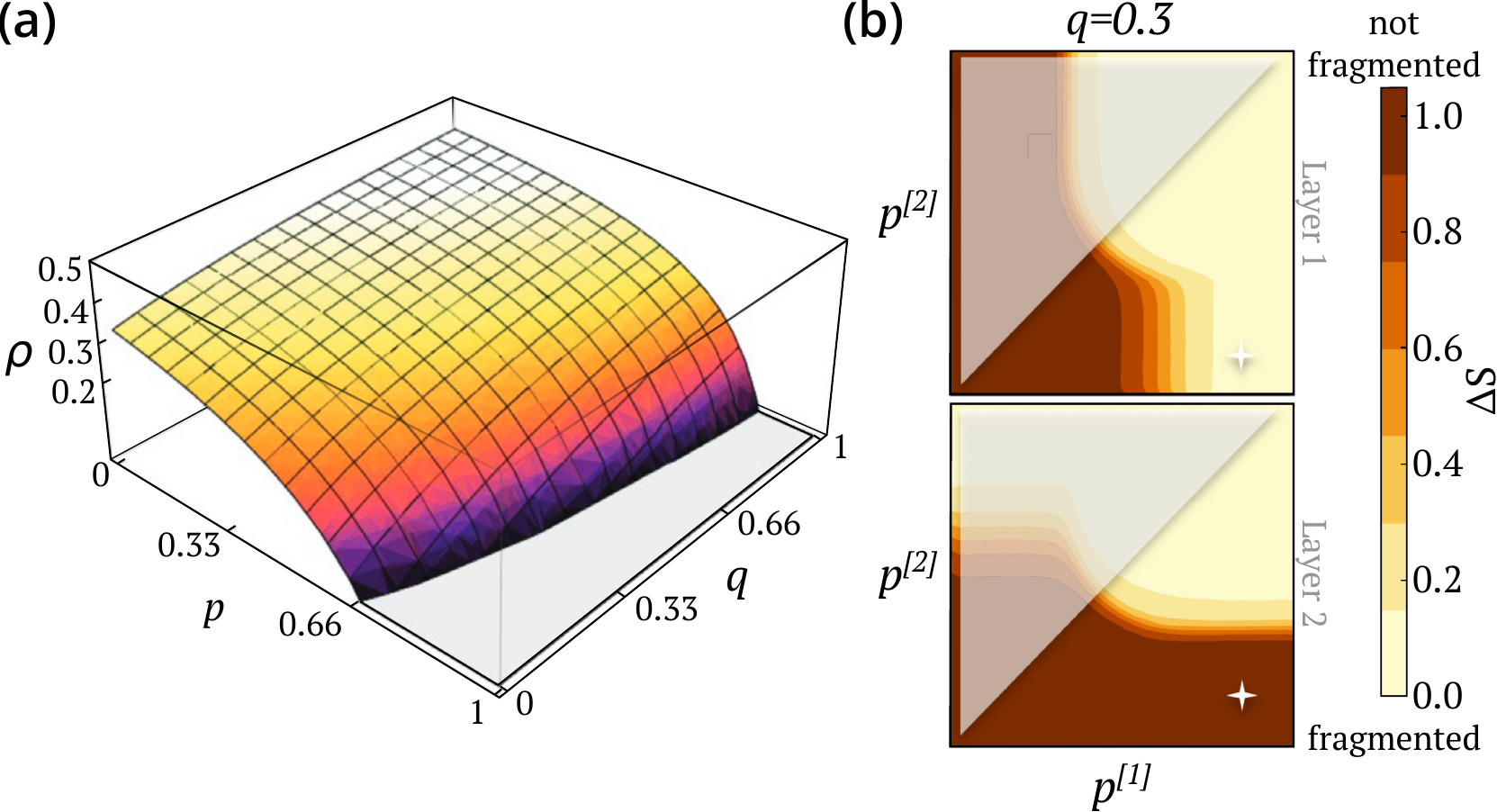}
    \caption[]{Dynamical behavior of the coevolving multiplex voter model. 
    (a) Interface density measuring the pairs of connected nodes with different opinions 
as a function of the plasticity $p=p\lay{1}=p\lay{2}$ and the multiplexity $q$. Higher 
values of plasticity sustains the active phase of the system for a given value of
multiplexity. (b) Difference $\Delta S$ between the two largest clusters of opinions 
in layer 1 and in layer 2 respectively, for $p\lay{1} \neq p\lay{2}$. When the system 
is asymmetric, a new shuttered transition emerges,  with a different number of  components
 across layers. Figures adapted from \textcite{diakonova2014absorbing}.}
    \label{fig:coevolution_diakonova}
\end{figure}

The model has a rich behaviour. As shown in
Fig.~\ref{fig:coevolution_diakonova}(a), in the
symmetric case, i.e., when both layers have the same plasticity
$p\lay{1}=p\lay{2}$, the system can remain active and avoid the
absorbing consensus phase for longer through an adequate increase of
multiplexity $q$. Conversely, Fig.~\ref{fig:coevolution_diakonova}(b)
illustrates that, when the system is
asymmetric, i.e. $p\lay{1} \neq p\lay{2}$, 
%
a new shattered fragmentation transition emerges:
for increasing values of $q$ the layer with higher plasticity
exhibits an increasing number of isolated components
and a giant component whose size depends on $q$.  
In the extreme scenario, when only one of the two layers is active, 
fragmentation is unavoidable, at difference with the classical
coevolving voter model on single-layer networks.

~\textcite{klimek2016dynamical} introduced a particular type of multiplex coevolving 
voter model with triadic closure, where newly created links tend to connect to nodes 
that already share common neighbors. The model exhibits an anomalous fragmentation 
transition, where one layer fragments from one large component into many small components. 
The community structure of the system is shaped by the different link rewiring 
probabilities at each layer, mimicking the heterogeneity in the community size 
distributions observed across the different layers of real-world systems.

A more nuanced model of coevolving multiplex voter dynamics is studied
by~\textcite{min2019votercoevol}, where a non-linear voter dynamics is
implemented. The model is similar to the one studied
in~\cite{diakonova2014absorbing}, but in this case the effect of local
majorities is tuned by a nonlinearity parameter $\chi$. In practice, the 
probability for a sampled node $i$ on layer $\alpha$ to take any action, either 
by copying the state of a neighbour or by rewiring an edge, is equal to
$\left(\frac{a_i}{k_i}\right)^{\chi}$, where $a_i$ is the number of
active links of $i$, i.e., the number of its neighbours on layer
$\alpha$ whose opinion is different from $\sigma\lay{\alpha}_i$. For
$\chi=1$ we recover the coevolving multiplex voter model
in~\cite{diakonova2014absorbing}, but for $\chi>1$ nodes are less
inclined to take any action and local majorities tend to crystallize.
The phase diagram in the parameter space $(p\lay{1}, p\lay{2}, \chi, q)$
is very rich, and comprises consensus, active, shattered, asymmetric
fragmented, and dynamically active shattered phases, thus reflecting the 
interplay of the multiplexity and nonlinearity in coevolutionary dynamics.


Multiplex coevolution, and in particular competition between layers and its effects on the
structural and dynamical features of the system, has also been studied for a population of 
Kuramoto oscillators in \cite{pitsik2018inter}.
\newtext{Adaptive Kuramoto models on single-layer networks 
have been widely investigated~\citep{zhou2006dynamical},
where link weights or interactions evolve in response to the phase coherence among 
oscillators,  leading to the self-organization of synchronized clusters and enhanced 
global coherence.}
In the multiplex coevolution, the multiplex consists of $M$ layers of Kuramoto
oscillators described by:
\begin{equation}
\label{eq:pitsikCompetitive}
\frac{d \sigma_i\lay{\alpha}}{dt}  =   \omega_i + g \sum_{j=1}^N
a_{ij}\lay{\alpha}(t) \sin(\sigma_j\lay{\alpha}-\sigma_i\lay{\alpha})
\end{equation}

\noindent with $\alpha=1,\ldots,M$, $i=1,\ldots,N$, and
$\sigma_i\lay{\alpha}$ representing phase variables. Here, the link
weights in each layer are time-varying quantities, adaptively changed
taking into account homophily (connections between synchronous nodes
tend to be enhanced) and homeostasis (the available resources at each
node to form links with other nodes of the structure are limited). The
dynamics of their evolution read:
\begin{equation}
\label{eq:pitsikUpdatingRule}
\frac{d a_{ij}\lay{\alpha}}{dt}  =  p_{ij}\lay{\alpha}
-a_{ij}\lay{\alpha}(t) \sum_{k\in N_i\lay{\alpha}}
p_{ik}\lay{\alpha}-a_{ij}\lay{\alpha}(t) \sum_{\beta \neq \alpha}
p_{ij}\lay{\beta}
\end{equation}
\noindent where
$p_{ij}\lay{\alpha}(t)=\frac{1}{T}\int\limits_{t-T}^Te^{\mathrm{i}  (\sigma_j\lay{\alpha}(\tau)-\sigma_i\lay{\alpha}(\tau))}d\tau$
measures the average phase coherence between nodes $i$ and $j$ at layer
$\alpha$, where $T$ is the \emph{adaptation time}.
The first term in the right side of Eq.~(\ref{eq:pitsikUpdatingRule})
accounts for homophily, as the \newtext{intra}-layer weight is increased by a high
level of coherence between $\sigma_i\lay{\alpha}$ and
$\sigma_j\lay{\alpha}$. The second term implements homeostasis as the
weight is decreased by a high level of the coherence among all the
neighbors at layer $\alpha$. Finally, the third term accounts for
competition between the layers, as an increase of phase coherence
between nodes $i$ and $j$ at layer $\alpha$ yields a decrease of the
weight of the corresponding link in the other layers.

The effects of the considered mechanisms of competition mainly depend
on two parameters, the intra-layer coupling $g$ and the adaptation time
$T$, and can be illustrated with reference to a structure with two
layers. For small values of $g$ and $T$, adaptation leads to a topology
with weakly marked structural clusters and weights that are distributed
according to a power-law weight distribution and are similar in the two
layers. For increasing values of $g$ and $T$, a strongly modular
structure forms, with the two layers showing more marked
dissimilarities. Finally, for larger values of $g$ and $T$, clusters
mostly disappear and in both layers a homogeneous topology is observed.
At variance with what observed for low values of $g$ and $T$, however,
in this case, the layers evolve towards configurations that are largely
dissimilar.

\newtext{
Concerning epidemic spreading, one of the earliest coevolving models on single-layer networks
was introduced by~\textcite{gross2006epidemic}, where epidemic spreading coevolves 
with the underlying contact network. In their adaptive SIS model, susceptible
individuals may cut connections to infected neighbors and rewire them to other 
susceptibles, thus capturing the feedback between disease dynamics and network topology.
In parallel, in the multiplex setting, \textcite{granell2013dynamical} studied the coupled 
spreading of epidemics and awareness, showing how information diffusion can mitigate disease 
prevalence (see Sec.~\ref{sec:awareness}). In their model, however, the network structures
remain static and the coevolution occurs only at the dynamical level. 
Later, \textcite{peng2021contagion} combined the epidemic–awareness 
framework of~\textcite{granell2013dynamical} with the adaptive rewiring mechanism 
of~\textcite{gross2006epidemic}: when a susceptible becomes 
aware, it may cut a link to an infected neighbor and reconnect it to a randomly 
chosen susceptible, while the awareness itself spreads on a parallel information layer.
Their analysis showed that this structural adaptation raises the epidemic threshold
and reduces the overall prevalence.
}


Coevolutionary dynamics have also received significant attention in game theory, where network structure may be modified as a consequence of the strategic interactions among agents~\citep{perc2010coevolutionary}.
\newtext{On single-layer networks, various adaptive rules such as rewiring or partner selection have been shown to influence the evolution of cooperation, often promoting it under conditions where defection would otherwise dominate, and a comprehensive review is provided in~\textcite{perc2010coevolutionary}.}

\textcite{wang2014self} considered the case of a two-layer network where only a fraction of agents the playing prisoner's dilemma is endowed with an interlayer link. In particular, if a player often wins against its opponent, it is allowed to form an interlayer link to obtain additional earnings, following a scheme of biased coupling functions already illustrated in ~\citep{wang2012evolution}. For all other players, instead, the payoff only depends on their earning at that same layer. The system naturally self-organizes towards a configuration where roughly half of the population created interlayer links, typically prosocial agents which allow cooperation to survive even in adverse scenarios where full defection would arise in isolated layers ~\citep{chu2019self}. A similar setting is investigated in \citep{shen2018coevolutionary}, where this time agents play coupled public goods games, and where interlayer link weight depends on the agents' performance. Also in such a case cooperation naturally evolves towards intermediate values of connectivity among the best performing agents, with high heterogeneity in interlayer link weight among players. 

Finally, ~\textcite{burghardt2018partial} study the impact of economic and exogenous shocks on cooperative multiplex networks formed by several layers, such as trade an alliance layers. In particular, the authors considered an agent based model where, after a shock, agents can make both utility-maximizing decisions and randomly rewire ties to explore the utility landscape under different tie-formation incentives.
Randomly rewired links are found to increase the utility of agents, but only when agents’ incentives for tie-formation are sufficiently high.

\section{Conclusions}

Over the last two decades, complex networks have emerged as the main
framework to model the intricate pattern of interactions in complex
systems from the real world.  Graphs have been succesfully used to
understand how complex dynamical behaviors, such as synchronization and other
collective phenomena, emerge from simple mechanisms in networked systems.
%
%
Stimulated by new and richer data 
from the natural and social sciences, the complex networks community has
only recently recognized the importance of considering more realistic models
of networked systems.  
In particular, going beyond traditional mathematical descriptions,
multiplex networks have become in last years the new 
framework to capture the complex interdependencies
of real-world systems when interactions of different types
and nature coexist.  

What we have presented here is a comprehensive and structured  
review of \vito{many of} the novel dynamical behaviours that emerge 
when the multiplex nature of a complex system is duly taken into
account. As a key takeaway, the following general mechanisms leading 
to truly-multiplex collective phenomena can be identified:


\medskip
{\em -- Structurally correlated layers.}   
Novel multiplex collective phenomena can simply emerge from the
structural properties of the different layers of a system.
Typical examples are the presence of correlations between the layers, e.g.,
non-vanishing overlap of the links at the different
layers, or the sign and intensity of
inter-layer degree correlations. Link overlaps are for instance 
responsible for the emergence of 
catastrophic cascading failures in percolation \cite{cellai2013percolation},
and of multiculturality
in models of cultural diffusion in social systems~\cite{battiston2017layered}.
Differences in the average degree of the mobility layers of 
activators and inhibitors can trigger a novel type of
Turing patterns \cite{kouvaris2015pattern}, while
inter-layer degree correlations can induce topological enslavement in
in evolutionary games \cite{kleineberg2018topological}.

\medskip
{\em -- Dynamical interplay of inter- and intra-layer interactions.} 
A rich variety of non-trivial multiplex phenomena are due to the
dynamical coupling between the interaction layers.  Typically, the
interplay between inter- and intra-layer interactions has been studied
by tuning the relative weight of the inter-layer vs the intra-layer
links, a feature that can be interpreted as changing the cost or time
to switch from a layer to another.
This has led to the discovery of several 
new phenomena: from the emergence of multiplex
superdiffusion \cite{gomez2013diffusion} to the change of the nature of the 
percolation transition \cite{bashan2011percolation}; from layer-based disease localization in multimodal 
epidemic spreading \cite{ferraz2017disease}  
to novel forms of complete \cite{sole2013spectral},
intra- and inter-layer synchronization \cite{sevilla2016inter}.
  

\medskip
{\em  -- Dynamically correlated processes.}
Lastly, even when inter-layer dynamics is not taken into account, 
the heterogeneity between two or more dynamical processes unfolding over
the different layers of a multiplex network can induce novel collective
phenomena. In the simplest possible case, the heterogeneity arises
from considering the same dynamical process on two layers, but
with different parameters, associated for instance to different
velocities in the propagation of information or in the mobility of
moving agents. This leads to a multiplex
``slower is faster'' effect,    
when the agents of a multimodal tranportation system can change mode (layer) 
to avoid congestion and minimize their travel time
\cite{manfredi2018mobility}. 
In more complicated set-ups, the heterogenity comes from 
considering two different processes that 
belong to the same class, such as for instance two different
evolutionary games (snowdrift and prisoner's dilemma) 
\cite{santos2014biased}, or two different 
epidemic spreadings (disease and awareness) \cite{granell2013dynamical}.
Finally, in the most general case, dynamical processes of entirely
different nature, such as a Kuramoto model of synchronization and a
biased random walk \cite{nicosia2017collective}, when mutually
coupled, can trigger
collective phenomena impossible to observe otherwise.

\bigskip

\vito{The different mechanisms outlined above highlight the various ways in
which multiplexity can affect the dynamics of a complex system, giving
rise to collective behaviors that can not emerge on the corresponding
aggregated networks, or when the different layers of a system are
considered in isolation. The study of dynamical processes on multiplex
networks is a vibrant research area that is finding more and more
applications in an increasingly broad range of domains. The framework
we have presented in this review can be expanded in different
directions.  In particular, networks and multiplex networks rely on
the assumption that all the interactions can be captured by dyadic
relationships. Recent works have shown the presence and importance of
group interactions in real-world complex systems. Higher-order
networks, such as simplicial complexes and hypergraphs, are the
natural way to describe interactions in groups of three or more
nodes ~\cite{battiston2020networks, battiston2021physics, bianconi2021higher, bick2023higher}.
However, higher-order networks alone, do not allow to consider 
interactions of different types and nature, prompting for a
generalization to the multiplex case.  
The study of the dynamics of multiplex simplicial complexes 
\cite{sun2021higher,krishnagopal2023homultiplex} and of multiplex
hypergraphs \cite{lotito2024multiplex}, in which interactions of different
order and of different type can be considered at
the same time, is still an unexplored field and a promising direction 
for future research.}


\begin{acknowledgments}

Many of the ideas illustrated in this review would have not presented,
formulated or developed without the many scientific interactions with
the network science community. In particular, we are indebted to: Alex
Arenas, Andrea Baronchelli, Marc Barthelemy, Ginestra Bianconi,
Stefano Boccaletti, Marian Boguna, Arturo Buscarino, Guido Caldarelli,
Timoteo Carletti, Adrian Carro, Giulia Cencetti, Mario Chavez,
Emanuele Cozzo, Tiziana Di Matteo, Fabio D'Ercole, 
Caterina De Bacco, Manlio De Domenico, Pietro De Lellis, Fabrizio De
Vico Fallani, Mario Di Bernardo, Marina Diakonova, Albert
Diaz-Guilera, Ernesto Estrada, Duccio Fanelli, 
Luigi Fortuna, Luca Gallo, Riccardo Gallotti, Lucia 
Valentina Gambuzza, Clara Granell, Jacopo Iacovacci, Iacopo Iacopini,
Gerardo Iniguez, Marton Karsai, J\'anos Kert\'esz, 
Mikko Kivela, Kaj Kolja Kleineberg, Peter Klimek, Nikos
Kouvaris, Philipp Hoevel, Lucas Lacasa, Renaud Lambiotte, 
Cecilia Mascolo, Sandro Meloni, Giulia
Menichetti, Ludovico Minati, Yamir Moreno, 
Adilson Motter, Mirco Musolesi, Vincenzo Nicosia,  Matjaz Perc, Nicola Perra, Mason Porter, Marton Posfai, 
Alessandro Rizzo, Martin Rosvall, Maxi San Miguel,
Mariangeles Serrano, Per Sebastian Skardal, Francesco Sorrentino,
Stefan Thurner.

%

%
%
%
%
F.B. acknowledges support from the Austrian Science Fund (FWF) through projects 10.55776/PAT1052824 and 10.55776/PAT1652425.
M.F. acknowledges support from the Italian Ministry for Research and
Education (MIUR) through Research Program PRIN 2017 under Grant 2017CWMF93.
%
J.G.-G. acknowledges support from the Departamento de Industria e
Innovaci\'on del Gobierno de Arag\'o n y Fondo Social Europeo
(FENOL group E36-20R), and from grant PID2020-113582GB-I00
funded by MCIN/AEI/10.13039/501100011033.
B.M. acknowledges support by the IITP (Institute of Information \& Coummunications Technology Planning \& Evaluation)
-ITRC (Information Technology Research Center) grant funded by 
the Korea government (Ministry of Science and ICT) (IITP-2025-RS-2024-00437284).
F.R. acknowledges support by the Air Force Office of Scientific Research
(FA9550-21-1-0446) and the Army Research Office (W911NF-21-1-0194) 
(the funders had no role in study design, data collection and analysis, decision to publish, or  any opinions, findings, and conclusions or
recommendations expressed in the manuscript). 
A.S. acknowledges support from the European Union’s Horizon Europe research and innovation programme under the Marie Sk\l{}odowska-Curie grant agreement No. 101208090 (Project temporalHOI). 
For the initial stages of this work F.B, V.N. and V.L. acknowledge support from the Project LASAGNE, Contract No.318132 (STREP), funded by the
European Commission. 
\end{acknowledgments}

\bibliographystyle{unsrt_short_et_al}
\bibliography{colored_bibreview.bib}

\end{document}